\documentstyle[11pt,aaspp]{article}

\def\ov{\over\displaystyle}
\def\di{\displaystyle}
\def\ltsima{$\; \buildrel < \over \sim \;$}
\def\simlt{\lower.5ex\hbox{\ltsima}}	
\def\gtsima{$\; \buildrel > \over \sim \;$}
\def\simgt{\lower.5ex\hbox{\gtsima}}	
\font\sevenrm=cmr7
\def\OIII{[O~{\sevenrm III}]}

\def\OII{[O~{\sevenrm II}]}
\def\MGII{Mg~{\sevenrm II}}
\def\ea{et al.\ }
\def\l{{\cal L}}
\def\sm{\mbox{$\cal M_{\odot}$}}

\begin{document}
\begin{flushleft}
Invited Review Paper for PASP
\end{flushleft}
\bigskip\bigskip

\title{Unified Schemes for Radio-Loud Active Galactic Nuclei}
\date{}
\author{C. Megan Urry}
\affil{Space Telescope Science Institute, 3700 San Martin
Drive, Baltimore, MD, 21218\\Electronic mail: cmu@stsci.edu}
\and
\author{Paolo Padovani}
\affil{Dipartimento di Fisica, II Universit\`a di Roma ``Tor
Vergata''\\Via della Ricerca Scientifica 1, I-00133 Roma,
Italy\\Electronic mail:
padovani@roma2.infn.it}
\begin{abstract}

The appearance of active galactic nuclei (AGN) depends so strongly
on orientation that our current classification schemes are dominated by random
pointing directions instead of more interesting physical properties.
Light from the centers of
many AGN is obscured by optically thick circumnuclear matter, particularly at
optical and ultraviolet wavelengths. In radio-loud AGN, bipolar jets emanating
from the nucleus emit radio through gamma-ray light that is relativistically
beamed along the jet axes. Understanding the origin and magnitude of radiation
anisotropies in AGN allows us to unify different classes of AGN; that is, to
identify each single, underlying AGN type that gives rise to different classes
through different orientations.

This review describes the unification of radio-loud AGN, which include
radio galaxies, quasars, and blazars. We describe the classification and
general
properties of AGN. We summarize the evidence for anisotropic emission
caused by circumnuclear obscuration and relativistic beaming. We outline
the two most plausible unified schemes for radio-loud AGN,
one linking the high-luminosity sources (quasars and luminous radio galaxies)
and one the low-luminosity sources (BL~Lac objects and less luminous radio
galaxies). Using the formalism appropriate to samples biased by relativistic
beaming, we show the population statistics for two schemes are in accordance
with available data.

We analyze the possible connections between low- and high-luminosity
radio-loud AGN and conclude they probably are powered by similar physical
processes, at least within the relativistic jet. We review potential
difficulties with unification and conclude that none currently constitutes a
serious problem. We discuss likely complications to unified schemes that are
suggested by realistic physical considerations; these will be important to
consider when more comprehensive data for larger complete samples become
available. We conclude with a list of the ten questions we believe are the
most pressing in this field.

\end{abstract}

\bigskip\bigskip
\centerline{PASP in press, September 1995 issue}

\newpage
\tableofcontents
\newpage
\section{Introduction}
\label{sec:Introduction}

The mystery of active galactic nuclei (AGN) is that they
produce very high luminosities in a very concentrated
volume, probably through physical processes other than the
nuclear fusion that powers stars. AGN are thus special
laboratories for extreme physics which we would like to
understand. They are also our principal probes of the
Universe on large scales, so understanding them is essential
to studying the formation and evolution of the Universe.

At present the approximate structure of AGN is known but
much of the detailed physics is literally hidden from view
because of their strongly anisotropic radiation patterns.
The prevailing (but not necessarily correct) picture of
the physical structure of AGN is illustrated in Fig.~\ref{fig:cartoon}
(Holt \ea 1992).
At the center is a supermassive black
hole whose gravitational potential energy is the ultimate
source of the AGN luminosity. Matter pulled toward the black
hole loses angular momentum through viscous or turbulent
processes in an accretion disk, which glows brightly at
ultraviolet and perhaps soft X-ray wavelengths. Hard X-ray
emission is also produced very near the black hole, perhaps
in connection with a pervasive sea of hot electrons above the disk.
If the black hole is spinning, energy may be extracted
electromagnetically from the black hole itself.

Strong optical and ultraviolet emission lines are produced
in clouds of gas moving rapidly in the potential of the
black hole, the so-called ``broad-line clouds''
(dark blobs in Fig.~\ref{fig:cartoon}). The optical and ultraviolet
radiation is obscured along some lines of sight by a torus
(as shown in Fig.~\ref{fig:cartoon})
or warped disk of gas and dust well outside the accretion
disk and broad-line region. Beyond the torus\footnote{For convenience, we
tend to refer
to the obscuring matter as a torus but to date there is little to
indicate whether it is actually a torus, a warped disk, or some other
distribution (cf. \S~\ref{sec:torus}).},
slower moving clouds of gas produce
emission lines with narrower widths (grey blobs in
Fig.~\ref{fig:cartoon}). Outflows of energetic particles occur
along the poles of the disk or torus, escaping and forming
collimated radio-emitting jets and sometimes giant radio
sources when the host galaxy is an elliptical, but forming
only very weak radio sources when the host is a gas-rich
spiral. The plasma in the jets, at least on the smallest
scales, streams outward at very high velocities, beaming
radiation relativistically in the forward direction.

This inherently axisymmetric model of AGN implies a radically
different AGN appearance at different aspect angles. In
practice, AGN of different orientations will therefore be
assigned to different classes. Unification of these
fundamentally identical but apparently disparate classes is
an essential precursor to studying the underlying physical
properties of AGN. The ultimate goal is to discover which
are the fundamentally important characteristics of AGN ---
e.g., black hole mass, black hole spin, accretion rate, host
galaxy type, interaction with neighboring galaxies --- and
how they govern the accretion of matter, the formation of
jets, and the production of radiation in these bizarre
objects.

This review covers the unification of radio-loud AGN, i.e.,
those with prominent radio jet and/or lobe emission.
Comparable unification schemes for radio-quiet objects
(Rowan-Robinson 1977; Lawrence and Elvis 1982; Antonucci and Miller 1985),
which have not been explored using the same statistical techniques, have
recently been
reviewed by Antonucci (1993; his review includes radio-loud AGN as well)
and are not discussed here. In the following sections, we describe
current AGN classification schemes (\S~2) and the two
principal causes of anisotropic radiation, obscuration
(\S~3) and relativistic beaming (\S~4). We establish the
motivation for current unification schemes for high- and
low-luminosity\footnote{In all that follows, we compute {\it observed}
luminosities assuming spherical symmetry, i.e., we assume uniform emission
into $4\pi$ steradians.
If an AGN radiates anisotropically, it may be called a ``high luminosity''
source even though its intrinsic luminosity is low.}
radio-loud AGN (\S~5) and then discuss them
quantitatively (\S~6). We discuss the possible connections
among high- and low-luminosity AGN and other aspects of the
unification paradigm (\S~7), including potential problems,
complications and future tests (\S~8).
In the final section (\S~9), we briefly summarize
the status of unification and pose what we believe are
the ten most important questions at the current time.
In the Appendices,
we present equations governing the various beaming
parameters (A), the Doppler enhancement (B), and the ratio
of core to extended flux (C), and a glossary of acronyms used in the
paper (D). Throughout this review the
values $H_0 = 50$ km s$^{-1}$ Mpc$^{-1}$ and $q_0 = 0$ have
been adopted (unless otherwise stated) and the spectral index $\alpha$ is
defined such that $F_{\nu} \propto \nu^{-\alpha}$.

\section{Observed Properties and Empirical Classification of AGN} 
\label{sec:AGNprop}

The full complement of active galactic nuclei constitutes a
zoo of different names, detection criteria, and spectral,
polarization, and variability characteristics. As in
biology, however, taxonomy derived from empirical
observations can impose some order on the chaos. Table~1 shows
the principal classes of AGN (adapted from Lawrence 1987, 1993),
organized according to their radio-loudness and their
optical spectra, i.e., whether they have broad emission
lines (Type~1), only narrow lines (Type~2), or weak or
unusual line emission. Within each of the groupings,
different types of AGN are listed by increasing luminosity.
We now explain Table~1 in more detail.

Roughly 15 -- 20\% of AGN are radio-loud, meaning they have
ratios of radio (5~GHz) to optical (B-band) flux $F_5 /F_{\rm B}
\simgt 10$ (Kellermann \ea 1989), although this
fraction increases with optical (Padovani 1993; La Franca \ea
1994) and X-ray (Della Ceca \ea 1994) luminosities, reaching
for example $\sim 50\%$ at $M_{\rm B} \simlt -24.5$. With few exceptions, the
optical and ultraviolet emission-line spectra and the infrared to soft X-ray
continuum of most radio-loud and radio-quiet AGN are very similar (Sanders \ea
1989) and so must be produced in more or less the same way. The characteristic
of radio-loudness itself may be related in some way to host
galaxy type (Smith \ea 1986) or to black hole spin
(Blandford 1990; Wilson and Colbert 1995), which might enable the formation
of powerful relativistic jets.

Based on the characteristics of their optical and ultraviolet
spectra, AGN can be separated into the three broad types
shown in Table~1.

\begin{itemize}
\begin{enumerate}
\item
Those with bright continua and broad emission lines from
hot, high-velocity gas, presumably located deep in the
gravitational well of the central black hole, are known as
Type~1 AGN. In the radio-quiet group, these include the
Seyfert~1 galaxies, which have relatively low-luminosities
and therefore are seen only nearby, where the host galaxy
can be resolved, and the higher-luminosity radio-quiet
quasars (QSO), which are typically seen at greater distances
because of their relative rarity locally and thus rarely
show an obvious galaxy surrounding the bright central
source. The radio-loud Type~1 AGN are called Broad-Line
Radio Galaxies (BLRG) at low luminosities and radio-loud quasars
at high luminosities, either Steep Spectrum Radio Quasars
(SSRQ) or Flat Spectrum Radio Quasars (FSRQ) depending on
radio continuum shape, with the dividing line set at $\alpha_{\rm r} = 0.5$
(where the radio spectrum is measured at a few GHz). Other than luminosity,
little distinguishes Seyfert~1s from radio-quiet quasars, or BLRG from radio
quasars.

\noindent
\item
Type~2 AGN have weak continua and only narrow emission lines, meaning either
that they have no high velocity gas or, as we now believe, the line of sight
to such gas is obscured by a thick wall of absorbing material. Radio-quiet
Type~2 AGN include Seyfert~2 galaxies at low luminosities, as well as
the narrow-emission-line X-ray galaxies (NELG; Mushotzky 1982).
The high-luminosity counterparts are not clearly identified at this point but
likely candidates are the infrared-luminous {\it IRAS} AGN
(Sanders \ea 1988; Hough \ea 1991; Wills \ea 1992b), which may show a
predominance of Type~2 optical spectra (Lawrence \ea 1995, in preparation).

Radio-loud Type~2 AGN, often called Narrow-Line Radio
Galaxies (NLRG), include two distinct morphological types: the low-luminosity
Fanaroff-Riley type~I radio galaxies (Fanaroff and Riley 1974), which have
often-symmetric radio jets whose intensity falls away from the nucleus,
and the high-luminosity Fanaroff-Riley type~II radio galaxies, which have more
highly collimated jets leading to well-defined lobes with prominent hot spots
(see \S~\ref{sec:distinction}). Examples of FR~I and FR~II radio morphologies
are shown in Fig.~\ref{fig:rad_image}.

\noindent
\item
A small number of AGN have very unusual spectral characteristics. Inventing a
term, we call these Type~0 AGN and speculate that they are related by a small
angle to the line of sight (``near 0 degrees''). These include the BL~Lacertae
(BL~Lac) objects, which are radio-loud AGN that lack strong emission or
absorption
features (typical equivalent width limits are set at $W_\lambda < 5$ \AA).
In addition,
roughly 10\% of radio-quiet AGN have unusually broad P-Cygni-like
absorption features in their optical and ultraviolet spectra, and so are
known as BAL (Broad Absorption Line) quasars (Turnshek 1984). If BAL
spectral features are caused by polar outflows at small angles to the
line of sight, they too are Type~0 AGN as indicated in Table~1;
alternatively, they may have edge-on disks with winds instead (Turnshek 1988).
There are no known radio-quiet BL~Lacs.

A subset of Type~1 quasars, including those defined variously as Optically
Violently Variable (OVV) quasars,
Highly Polarized Quasars (HPQ)\footnote{We refer here only to radio-loud HPQ.
While a few radio-quiet quasars also have highly polarized optical emission,
and thus fit the HPQ definition, their polarization is almost certainly caused
by scattering rather than intrinsic emission processes.},
Core-Dominated Quasars (CDQ) or FSRQ, are probably also found at a
small angle to the line of sight. Their continuum emission strongly
resembles that of BL~Lac objects (apart from the presence of a blue ``bump''
in a few cases) and, like BL~Lac objects, they are
characterized by very rapid variability, unusually high and variable
polarization, high brightness temperatures (often in excess of the Compton
limit $T\sim10^{12}$~K; Quirrenbach \ea 1992), and superluminal velocities of
compact radio cores (\S~\ref{sec:anis_rel}). Although the names OVV, HPQ,
CDQ, and FSRQ reflect different empirical definitions, evidence is
accumulating that they are all more or less the same thing --- that is, the
majority of flat-spectrum radio quasars tend to show rapid variability, high
polarization, and radio structures dominated by compact radio cores, and
vice-versa (Fugmann 1989; Impey \ea 1991; Valtaoja \ea 1992; Wills \ea 1992a)
--- so hereafter we refer to them simply as FSRQ. Collectively, BL~Lacs and
FSRQ are called blazars. Even though the FSRQ have strong broad emission lines
like Type~1 objects, they are noted in the ``Type~0'' column in Table~1 because
they have the same blazar-like continuum emission as BL~Lac objects.

\end{enumerate}
\end{itemize}

We have described an empirical division of AGN according to
radio and optical/ultraviolet properties. Table~1 is analogous to the
Periodic Table of the Elements developed by chemists a
century or so ago, when many chemical elements had been
discovered and studied but the relations among them were not
entirely clear. Where in chemistry it was eventually
recognized that valence electrons dominate the horizontal
relations and nuclear mass the vertical relations, we will
argue that the categories in this ``Periodic Table of the
AGN'' are distinguished primarily by orientation effects
along the horizontal direction and by as-yet unknown physics
in the vertical direction.

Whether AGN are classified Type~1
or Type~2 depends on obscuration of the luminous nucleus,
and whether a radio-loud AGN is a blazar or a radio galaxy
depends on the alignment of the relativistic jet with the
line of sight. These two causes of anisotropy, obscuration
of infrared through ultraviolet light by optically thick gas
and dust, and relativistic beaming of radio jets, both
important in radio-loud objects, are discussed in turn
in the next two sections, which can be skipped by readers
interested only in the actual unification schemes. Note that
the unification of quasars with high-luminosity radio galaxies requires
both kinds of anisotropy while the unification of BL~Lac objects
with low-luminosity radio galaxies requires (at present) only relativistic
beaming.

\section{Anisotropic Radiation from Obscuration} 
\label{sec:anis_obs}

The central regions of many AGN appear to contain obscuring
material, probably in the form of dust, that prevents
infrared through ultraviolet light from penetrating some
lines of sight (Rowan-Robinson 1977). This dust may
be distributed in a torus (Pier and Krolik 1992, 1993) or in
a warped disk (Sanders \ea 1989). In either case, it causes
AGN to look markedly different from different aspect angles.
Direct evidence for obscuration has been found in many
Type~2 AGN, although mostly in the more numerous radio-quiet
objects. Those NLRG in which obscuration has been detected because of
scattered broad emission lines are all FR~IIs; detection of such hidden lines
in FR~Is would provide an intriguing connection between the high- and
low-luminosity unification schemes.

\subsection{Polarimetric Evidence for Hidden Nuclear Regions}
\label{sec:polarim}

The most direct evidence for circumnuclear obscuration comes
from spectropolarimetry of Type~2 objects, particularly the
nearby Seyfert~2 NGC~1068 (Antonucci and Miller 1985) and
the radio galaxy 3C~234 (Antonucci 1984). Some fraction of the light from
these objects is highly polarized, and their polarized
spectra have strong broad lines like Type~1 rather than Type~2
objects. Much of the polarization is probably caused by
electron scattering since it is wavelength independent; some
scattering by dust clouds has also been observed (Miller \ea
1991). In 3C~234, the plane of polarization is perpendicular
to the radio jet axis, as expected if a Type~1 nucleus is at least
partially obscured by a thick wall of gas
and dust whose axis coincides with the radio jet axis.\footnote{Antonucci
(1984) identified 3C~234
as a narrow-line radio galaxy even though it has broad H$\alpha$
(Grandi and Osterbrock 1978). His point was that 3C~234 is equivalent
to a narrow-line object like NGC~1068 because its broad lines are
very highly polarized and therefore must be scattered. The broad emission
lines of 3C~234 were not really ``hidden'' prior to the
spectropolarimetric observations (although they were not
strong broad lines that would clearly mark it as a quasar), but
spectropolarimetry indicated that a direct view to the broad-line
region was obscured. The fuzzy definition of ``broad-line'' (Type~1) objects
is discussed further in the footnote in \S~\ref{sec:content}}
The continuum luminosity inferred from the strength of the scattered broad
lines is quasar-like.

Subsequent polarization observations generally support the picture of
scattered light from a luminous, hidden continuum source in NLRG. Like 3C~234,
the radio galaxy IC~5063 has polarized broad lines, indicative of scattered
light from a hidden broad-line region (Inglis et al. 1993). Spectropolarimetry
of three powerful radio galaxies at $z \sim 1$ has shown broad polarized
\MGII~lines, with equivalent widths typical of those observed in radio-loud
quasars (di Serego Alighieri \ea 1994a).
Multiwavelength polarimetry of 3C~109 (Goodrich and Cohen 1992) suggests
polarization by transmission through dust, in which case the intrinsic
luminosity is also quasar-like.\footnote{Like 3C~234, 3C~109 has strong
broad lines observable in total flux, in this case H$\alpha$. While the
spectral shapes of the
polarized continuum in 3C~109 and 3C~234 are similar, the polarization in
3C~234 is too high to be caused by dust transmission (Goodrich, private
communication).}

Recent {\it HST} spectra of Cygnus~A show a broad \MGII~emission line which
appears to be reflected from the southeast knot (a region of extended
featureless optical continuum; Pierce and Stockton 1986). This is
a direct signature of the hidden quasar in this luminous FR~II galaxy
(Antonucci \ea 1994); that broad emission lines have not been detected with
optical spectropolarimetry (Jackson and Tadhunter 1993) may be because of
dilution by the local optical continuum at the scattering site
(Antonucci \ea 1994; \S~\ref{sec:ext_cont}).

Imaging polarimetry of the radio galaxies PKS~2152$-$69 (di Serego Alighieri
\ea 1988), 3C~277.2 (di
Serego Alighieri \ea 1989), Cygnus~A (Tadhunter \ea 1990), 3C~368 (di
Serego Alighieri \ea 1989; Scarrott \ea
1990), and additional radio galaxies (Cimatti \ea 1993, and references therein)
reveals extended regions of polarized continuum emission which appear to be
nuclear continuum scattered by ambient dust and/or electrons.
Tadhunter \ea (1989; see also Cimatti \ea
1993 and di Serego Alighieri \ea 1994a) have suggested that
dust-scattering of quasar radiation could also explain the so-called
``alignment effect'' observed in high-redshift radio galaxies (McCarthy \ea
1987; Chambers \ea 1987), i.e., the correlation between the direction of the
radio axis and the optical structure (extended emission line regions plus
continuum).

\subsection{Infrared and X-Ray Evidence for Hidden Nuclear Regions}
\label{sec:infr}

Since optical depth decreases at wavelengths longer than the optical, infrared
observations are potentially much deeper probes of the nuclear regions of
Type~2 AGN. Compact, bright, infrared cores and/or wavelength-independent
perpendicular polarization as in 3C~234 have been found in about a
dozen narrow-line radio galaxies, of both high and low luminosities (Bailey
\ea 1986; Fabbiano \ea 1986; Hough \ea 1987; Antonucci and Barvainis 1990;
McCarthy \ea 1990; Djorgovski \ea 1991). In a few cases, infrared spectroscopy
has revealed broad wings on the Paschen lines (Fabbiano \ea 1986;
Hill \ea 1995), at least partially revealing the infrared
high-velocity gas that is completely obscured at optical wavelengths. In
others the hidden infrared-optical continuum source and broad-line region
remain stubbornly hidden suggesting the expected optical depths are quite
large;
for example, up to $25-50$ magnitudes of visual extinction to the nucleus are
estimated in the case of Cygnus~A (Ward et al. 1991; Djorgovski et al. 1991).

The weakness of the X-ray continuum from Type~2 AGN relative to Type~1s is
also consistent with the idea of an obscured nucleus (Lawrence and Elvis 1982;
Fabbiano \ea 1984; Fabbiano \ea 1986). The X-ray spectrum of Cygnus~A, a
classic
high-luminosity narrow-line radio galaxy, is commensurate with a typical quasar
spectrum absorbed by a high column density of cold gas along the line of sight
(Arnaud \ea 1987; Ward \ea 1991; Ueno \ea 1994). Similar high column densities
are deduced from X-ray-measurements of a number Type~2 AGN, exactly as expected
if the central source in Type~2s is obscured from direct view
(Mulchaey \ea 1992).

\subsection{Anisotropic Illumination of Narrow Emission Line Gas}
\label{sec:anisNLR}

The presence of anisotropic continuum emission can be inferred from the
extended narrow line regions, sometimes tracing ionizing light cones, seen in
direct imaging of many nearby Type~2 AGN. High-resolution {\it HST} images of
NGC~1068 in the light of \OIII~(Evans \ea 1991) confirm in exciting detail
earlier ground-based evidence for an ionization cone with apex at an obscured
nucleus (Pogge 1988). Fig.~\ref{fig:ion_cones} shows the {\it HST}
image of another nearby Seyfert~2 galaxy, NGC~5728 (Wilson \ea 1993);
the conical shape of the ionized gas and its filamentary structure
are both apparent.
In these and other cases, the bi-conical structure suggests that an obscured
nuclear source with quasar-like luminosity is photoionizing gas in the
extended narrow-line region (Robinson \ea 1987; Baum and
Heckman 1989; Tadhunter and Tsvetanov 1989; Wilson \ea 1993).

At present, there is no direct evidence (i.e., from spectropolarimetry)
for obscuration in low-luminosity, FR~I radio galaxies,
although there are strong indications of anisotropic continuum emission
in some individual objects. (Indirect evidence for obscuration in FR~Is
as a class is discussed in \S~\ref{sec:iso_low}.)
The FR~I radio galaxy PKS~2152$-$69
shows optical line emission from a gas cloud at a projected distance
of 8~kpc from the nucleus which, if due to excitation from a beamed nuclear
source, implies an ionizing beam power well within the range for BL~Lacs
(di~Serego Alighieri \ea 1988).

Similarly, in the FR~I radio galaxy Cen~A, optical emission-line filaments
subtend a
small projected solid angle in the northwest part of the galaxy, roughly
aligned with the position of the jet-like structures found in the X-ray and
radio bands (Morganti \ea 1991).
A detailed analysis of the line ratios in the Cen~A filaments suggests
they are photoionized by strongly anisotropic radiation, about two orders of
magnitude larger in the X-ray band than that observed directly
(Morganti \ea 1992). Were this due to simple obscuration of an isotropic
continuum by a thick torus, the obscured radiation would be reradiated
in the infrared, contrary to observation.
At least some of the continuum emission must be obscured, however,
as the variable (i.e., nuclear) X-ray source in Cen~A is heavily absorbed
(Mushotzky \ea 1978).
The continuum anisotropy in Centaurus~A could be caused by relativistic
beaming (see \S~\ref{sec:anis_rel}) --- its inferred beam power is similar to
that of BL~Lacertae (Morganti et al. 1991) --- or the optical filaments could
be excited by local shocks where the northern radio jet interacts with a dense
cloud of material (Sutherland \ea 1993).
We note that illuminated clouds like those seen in Cen~A and PKS~2152-69
are expected if FR~Is found in cooling flow clusters contain hidden BL~Lacs
(Sarazin and Wise 1993).

\section{Anisotropic Radiation from Relativistic Beaming} 
\label{sec:anis_rel}

When an emitting plasma has a bulk relativistic motion relative to a fixed
observer, its emission is beamed in the forward direction (in the fixed frame),
a direct consequence of the
transformation of angles in special relativity. An observer
located in or near the path of such a plasma sees much more
intense emission than if the same plasma were at rest. Time
scales for variability are also shorter, and this can cause the
emission region to appear to move superluminally in the
transverse direction (Appendix~A). Strong relativistic beaming is thought
to explain the rapid variations, high polarization, and high
luminosities that characterize blazars (Blandford and Rees
1978), and if present in blazars, it must also be present in other radio-loud
AGN. The consequences of the anisotropic beamed radiation pattern are
considerable (Appendices B and C),
introducing significant selection effects in almost any flux-limited sample.

\subsection{Evidence for Relativistically Beamed Gamma-Rays}
\label{sec:gammaray}

More than forty blazars have now been detected with the EGRET high
energy experiment on the {\it Compton} Gamma-Ray Observatory (von Montigny \ea
1995). Without exception, all EGRET-detected extragalactic objects are
radio-loud blazars, either FSRQ or BL~Lac objects (primarily the former). Not
only are blazars bright at $E \ge 100$ MeV, in many cases their observed
gamma-ray luminosity dominates the luminosity in other wavebands by a factor
between 1 and 1000.
Multiwavelength spectra of the superluminal quasar 3C~279
(Fig.~\ref{fig:3c279_spec}) show that the
ratio of gamma-ray to bolometric luminosity increases with overall luminosity
(Maraschi \ea 1994a).

In several blazars, the observed high-energy gamma-rays are highly variable,
on time scales of a few days.\footnote{The variability time scale can
be defined in several ways, including the doubling time
($t_d = \langle F \rangle (dF/dt)^{-1}$) or the e-folding time
($t_e = (d\ln F / dt)^{-1}$). Either is fine for the rough
estimates here as long as it is derived from a substantial change
in flux ($\simgt 30$\%).}
For example, the intensity of 3C~279
($z=0.538$) declined by a factor of 4-5 in less than 3 days
(Fig.~\ref{fig:3c279_lc}; Kniffen \ea 1993),
and the blazar PKS~0528+134 ($z=2.06$) more than doubled
its intensity in 5 days (Hunter \ea 1993).

This rapid variability leads to a largely model-independent argument that the
gamma-rays, at least, must be relativistically beamed (Maraschi \ea 1992).
This argument does not depend on which physical mechanism gives rise to the
gamma-ray emission, simply on the observed luminosity and variability time
scales at high energies. Specifically, in order for gamma-rays to escape
the source, the optical depth to pair production, $\tau_{\gamma\gamma}$, must
be of order unity or less, which is equivalent to saying the compactness, a
convenient dimensionless parameter that represents source luminosity divided
by dimension, must be less than about 40 at the threshold for pair-production.
That is, $\tau_{\gamma\gamma} \sim
{{\ell} \over {40}} \ll 1$, where $\ell = (L/r) (\sigma_T/m_ec^3)$ is the
compactness, $L$ and $r$ are the source luminosity and dimension,
respectively, and $m_e$ and $c$ are the usual constants (electron mass and
speed of light). The Thomson cross section, $\sigma_T$, is appropriate because
most pairs will be produced by X-$\gamma$ interactions.

For 3C~279 and PKS~0528+134, the inferred values for the compactness are 5000
to 15,000,
well in excess of the optical depth limit. In order that we
observe gamma-rays from these blazars, the true gamma-ray luminosity, $\cal
L$, must be much smaller than observed and the true dimension much larger.
Relativistic beaming has the effect that $L_{\rm obs} = \delta^4 \cal{L}$
(Eq.~\ref{eq:boost4}), where $\delta$ is the Doppler beaming factor
(Appendix~A). If $r$ is estimated from the variability time scale
($r\sim \delta^{-1} c\Delta t$), then
\begin{equation}
\ell = \delta^{-5} {{L_{\rm obs}} \over {\Delta t_{\rm obs}}}
{{\sigma_T}\over{m_e c^4}}.
\end{equation}
The limit $\ell \simlt 40$ then translates to $\delta \simgt
6$ for 3C~279 and $\delta \simgt 7$ for PKS~0528+134, where
$L_{\rm obs}$ has been evaluated at X-ray energies under
reasonable spectral assumptions (Maraschi \ea 1992);
similar limits are obtained for all the gamma-ray blazars
(Dondi and Ghisellini 1995).
For comparison, the values derived in an entirely independent
way from a synchrotron self-Compton calculation using radio
and X-ray data (see \S~\ref{sec:SSC}) are $\simgt 18$ for 3C~279 and
3 for PKS~0528+134 (Ghisellini \ea 1993).

Note that no radio-quiet AGN have been detected so far with
EGRET. One cannot rule out that, since gamma-ray power
appears to correlate with radio power (Padovani \ea 1993),
radio-quiet AGN might still have blazar-like spectra with
high-energy ($>50$~MeV) emission well below the sensitivity
of EGRET. This is unlikely, however, because OSSE
observations of Seyfert~1 galaxies show steep cutoffs above
$\sim 50$ keV, in sharp contrast to the hard spectra of
radio-loud objects (Kurfess \ea 1994). The difference in gamma-ray
properties may then be related to radio-loudness, which in turn must be closely
associated with relativistic beaming.

\subsection{Superluminal Motion in Radio Jets} 
\label{sec:superluminal}

The term ``superluminal motion'' describes proper motion of source structure
(traditionally mapped at radio wavelengths) that, when converted to an
apparent speed $v_{\rm a}$, gives $v_{\rm a} > c$. This phenomenon
occurs for emitting regions moving at very high (but still
subluminal) speeds at small angles to the line of sight
(Rees 1966). Relativistically moving sources ``run after''
the photons they emit, strongly reducing the time interval
separating any two events in the observer's frame and giving
the impression of faster than light motion (Appendix~A).

Typical proper motions observed with Very Long Baseline
Interferometry (VLBI) are in the range 0.1 to 1~milliarcsecond~yr$^{-1}$
and imply apparent velocities up to
$\sim 30c/(H_0/50)$ (Vermeulen and Cohen 1994). The majority
of superluminal sources are FSRQ and BL~Lacs, although this
is in part a selection effect since these objects have the
brightest cores and so are more easily observed with VLBI.
Blazars do tend to have the largest apparent velocities
(Ghisellini \ea 1993; Vermeulen and Cohen 1994), in
agreement with the idea that their jets are more aligned
with the line of sight than other classes of radio-loud AGN.

Detection of superluminal motion does not necessarily imply
that the source of radiation is moving at near-relativistic
speeds. For example, the tip of a rotating beam of light
moves faster than light at a distance $r > c/\omega$, where
$\omega$ is the angular speed. If the beam ionizes
material that re-emits the radiation, the observer will
detect superluminal motion even though the ionized material
is not moving at all. (Of course, a simple rotating-beam model would
predict some cases of superluminal contractions,
i.e., negative values of $v_{\rm a}$,
contrary to observation; Vermeulen and Cohen 1994).
Apparent superluminal motion requires only a relativistic
phase or pattern speed, which could be different from the bulk velocity of the
radiating plasma itself.

Whether the pattern speed inferred from superluminal motion
differs from the bulk motion of the plasma, as predicted by
jet models that include relativistic shocks (Lind and
Blandford 1985), is a matter of current debate (Ghisellini
\ea 1993; Vermeulen and Cohen 1994; Kollgaard 1994). There
is an observed correlation between the apparent superluminal
velocity and the Doppler beaming factor (Ghisellini \ea 1993), thus
connecting superluminal motion directly to bulk relativistic motion
(\S~\ref{sec:independent}). A particularly nice local laboratory has been found
by Mirabel and Rodr\'\i guez (1994), who discovered a Galactic superluminal
source for which both jet and counter-jet are seen and the two expansion
velocities and luminosities are measured. In this case the
pattern and bulk velocity are probably quite similar, with
$\beta_{\rm bulk} / \beta_{\rm pattern} \sim 0.8$
(Appendix~B; Bodo and Ghisellini 1995). A second galactic superluminal
source has since been discovered, with similar bulk velocity and
lying roughly in the plane of the sky, but with significant intrinsic
jet asymmetries (Tingay \ea 1995; Hjellming and Rupen 1995).

\subsection{Observed Radio Jet Asymmetries}
\label{sec:asymm}

One of the intriguing characteristics of jets imaged on the sky is that they
are often one-sided, almost exclusively so for the high-luminosity AGN (Bridle
and Perley 1984), for the parsec scale jets measured with VLBI (Cawthorne
1991), and even at the base of the symmetric jets in low-luminosity objects
(Bridle and Perley 1984; Parma \ea 1987). This raises the key question of
whether jets are intrinsically one-sided or have one of two intrinsically equal
jets brightened by relativistic beaming.
If observed superluminal velocities on
parsec scales are due to bulk relativistic motion of the emitting plasma, the
enormous intensity enhancement from relativistic aberration would inevitably
cause jet one-sidedness (Eq.~\ref{eq:jcj}). The continuity of sidedness between
parsec and kiloparsec scales would then argue for large-scale superluminal
motions, which in turn implies kinetic energies many orders of magnitude
larger than required by the parsec-scale jets.

The evidence for relativistic speeds on the larger scales is largely
circumstantial
but not easily dismissed. Wherever one-sided jets are seen on the small scale,
they are on the same side as those seen on the large scale, strongly
suggesting a connection between the two (Bridle 1992). Studies of the nearby
FR~I radio galaxy M87 support the hypothesis that its jets are intrinsically
two-sided: first, modest superluminal motion has been
detected on kiloparsec scales (Biretta \ea 1995) and second, an invisible
counter-jet has been inferred from the optically polarized hot spot in the
extended radio lobe on the counter-jet side (Stiavelli \ea 1992; Sparks
\ea 1992). Similarly, optical evidence for an invisible counter-jet was seen
in the superluminal BLRG 3C~120 (Axon \ea 1989).

More general evidence for Doppler enhancement of large-scale jets is found in
the depolarization asymmetry in radio lobes (Laing 1988; Garrington and Conway
1991). If
jet one-sidedness indicates an oncoming high-velocity jet, then the extended
radio lobe on the jet side should be closer to us than the lobe on the opposite
side. In 49 of the 69 sources studied by Garrington and Conway (1991), the
lobe on the counter-jet side is more depolarized than the one on the jet side,
exactly as expected if it were further away and had more depolarizing material
along the line of sight. The differential depolarization is explained
well by a hot gaseous halo surrounding the radio source (Tribble 1992).
It may be possible to ascertain the characteristics of the depolarizing
gas independently through sensitive X-ray observations with {\it AXAF},
which would confirm that the one-sidedness of most kpc-scale jets is
caused by relativistic beaming. We note that in some cases asymmetric
radio jet emission correlates spatially with the extended optical emission
line gas, which is clearly not beamed, indicating that intrinsic asymmetries
are also present at some level in radio sources (McCarthy \ea 1991).

Finally, supporting evidence for the preponderance of relativistic jets
comes from large surveys of radio-loud AGN (Hough and Readhead 1989; Impey \ea
1991). Just as one would expect if all jets were relativistic, the ratio of
core radio emission (presumably relativistically beamed) to extended (clearly
unbeamed) radio emission is correlated with optical polarization, optical
power-law fraction, degree/rapidity of variability, jet curvature, and
superluminal motion, and is inversely correlated with linear
size.\footnote{Lister \ea (1994a) found no correlation between the ratio
of core-flux to total-flux and other beaming indicators, but this may be in
part because for that particular ratio, core-dominated objects
(those with $R>1$; Appendix C) are restricted to the limited range
$0.5 \leq f_{\rm core} / f_{\rm total} \leq 1$.}

\subsection{Brightness Temperature Calculations and SSC Models}
\label{sec:SSC}

The smooth nonthermal radio-through-infrared continuum
emission in radio-loud AGN is probably synchrotron
radiation, i.e., emission from relativistic electrons moving
in a magnetic field. Some of the synchrotron photons will be
inverse Compton scattered to higher energies by the
relativistic electrons, which is known as the synchrotron
self-Compton (SSC) process. For some radio-loud AGN the
synchrotron radiation density inferred from the observed
radio power and angular size predicts SSC X-rays well in
excess of the observed X-ray flux (Marscher \ea 1979; Ghisellini
et al. 1993), which is called the ``Compton catastrophe''
(Hoyle \ea 1966). A related (but not equivalent) statement
is that extremely rapid radio variability in some blazars
(Quirrenbach \ea 1992) implies brightness temperatures,
$T_{\rm B} = I_{\nu}c^2/(2k\nu^2)$, where $k$ is Boltzman's
constant, larger than the $10^{12}$~K limit (Kellermann and
Pauliny-Toth 1969; see also Singal and Gopal-Krishna 1985 and
Readhead 1994, who suggest a limit $T_{\rm B} \simlt 10^{11}$~K appropriate to
the equipartition of magnetic field and relativistic electron energy
densities).

It follows that the true synchrotron photon density must be
lower than observers infer by assuming isotropy. The strong
anisotropy and shortened time scales caused by relativistic
beaming can account naturally for the Compton catastrophe
(or non-catastrophe, as it happens). A lower limit to
the Doppler factor, $\delta$, which characterizes these effects (Appendix~B)
can be estimated from the ratio of predicted to
observed SSC flux (Jones \ea 1974; Marscher \ea 1979). In
the case of a spherical emission region of observed angular
diameter $\phi_{\rm d}$, moving with Doppler
factor $\delta_{\rm sphere}$, the limit is (Ghisellini \ea 1993):
\begin{equation}
\delta_{\rm sphere} > f(\alpha) \, F_{\rm m} \left(
{{\ln(\nu_{\rm b}/\nu_{\rm m})}\over {F_{\rm x}\,
\phi_{\rm d}^{6+4\alpha} \, E_{\rm x}^{\alpha} \,
\nu_{\rm m}^{5+3\alpha}}} \right) ^{{1\over {(4+2\alpha)}}}
(1+z) \,,
\label{eqn:ssc}
\end{equation}
where $\phi_{\rm d}$ is in milliarcseconds,
$\nu_{\rm m}$ is the observed self-absorption
frequency of the synchrotron spectrum in GHz, $F_{\rm m}$ is
the observed radio flux at $\nu_{\rm m}$ in Jy, $E_{\rm
x}$ and $F_{\rm x}$ are the observed X-ray energy and
flux in keV and Jy respectively, and $\nu_{\rm b}$ is the observed synchrotron
high frequency cut--off. The function $f(\alpha)$, where $\alpha$ is
the spectral index of the optically thin synchrotron
emission, depends only weakly on the various assumptions
used by different authors (see discussion in Urry 1984) and
has the approximate value $f(\alpha) \simeq 0.08 \alpha +
0.14$ (Ghisellini 1987). If the radio source is a continuous
jet, which is perhaps more realistic (Appendix~B, case
$p=2+\alpha$), then (Ghisellini \ea 1993):
\begin{equation}
\delta_{\rm jet} \, =\, \delta_{\rm sphere}^{(4+2\alpha)/(3+2\alpha)}.
\end{equation}
For a continuous jet compared to a single blob, therefore, the same
observed quantities imply a higher Doppler beaming factor (for $\delta >1$).

The limit in Eq.~(\ref{eqn:ssc}) has been calculated for many
radio-loud AGN (Marscher \ea 1979; Madejski and Schwartz
1983; Madau \ea 1987), with the result that $\delta$ has
lower limits both larger and smaller than unity, depending
on the AGN. One complication is that the angular size
($\phi_{\rm d}$) is a function of observation frequency and so
is to some extent arbitrary. A self-consistent approach is
to use, in Eq.~(\ref{eqn:ssc}), the observing frequency as
$\nu_m$ and the flux and angular size (preferably measured
with VLBI) at that frequency as $F_m$ and $\phi_d$,
respectively.

For $\sim100$ radio sources for which the VLBI size of the
radio-emitting core is published, Eq.~(\ref{eqn:ssc}) gives
$\delta >1$ for a large fraction of BL~Lacs and essentially
all FSRQ (Ghisellini \ea 1993). That is, blazars for
which appropriate data exist do appear to have
relativistically beamed emission. Similarly, using
variability time scales to infer Doppler factors from the
condition $T_{\rm B,max} <10^{12}$~K gives $\delta >1$ for a
number of blazars (Ter\"asranta and Valtaoja 1994). The
latter values for BL~Lacs are somewhat low
compared to the SSC calculation, but if the equipartition
brightness temperature is more appropriate (Singal and Gopal-Krishna 1985;
Readhead 1994), the derived Doppler factors increase by a factor of $2-3$,
since $\delta = [T_{\rm B, obs}/T_{\rm B, max}]^{1/3}$.

\section{Basis for Unification Schemes} 
\label{sec:basis}

The previous two sections described the abundant evidence
for strongly anisotropic radiation from radio-loud AGN.
Their appearance therefore depends strongly on
orientation, and unification schemes are inevitable. The
light from some AGN must be directed toward the observer,
whether by relativistic beaming or by obscuration, and the
remaining misaligned AGN must constitute the so-called
``parent population.''\footnote{There is a
difference between what should really be called the parent
population, which is to say all intrinsically identical
objects regardless of orientation, and the misaligned
subclass, which is the parents minus the aligned objects. In
most cases involving relativistic beaming, however, the relative number of
aligned objects is small and there is little difference between the two.}
The known properties of radio-loud AGN --- radio
galaxies, quasars, and BL~Lac objects --- are key to
identifying the correct parent and beamed populations.

Here we describe the basic unification schemes that have been proposed to date
(\S~\ref{sec:history}), and how the isotropic properties of
radio-loud AGN support or undermine these schemes. Both FR~I
and FR~II radio galaxies tend to have pairs of jets more or
less in the plane of the sky and thus are candidates for
misoriented blazars. We discuss the FR~I/FR~II distinction
(\S~\ref{sec:distinction}) and then pursue the two most viable unification
schemes, linking quasars with FR~II galaxies (\S~\ref{sec:iso_high}) and BL
Lac objects with FR~I galaxies (\S~\ref{sec:iso_low}). Specifically, we
summarize the principal matching criteria, which include the
isotropic properties of extended radio power and
morphologies, optical narrow line emission, far-infrared
continuum emission, host galaxies, environments external
to the AGN, and evolution over cosmic time. Finally, we argue that statistical
treatment of
proper unbiased samples is not currently possible (\S~\ref{sec:myth})
and we review the method for incorporating the selection
effects introduced by relativistic beaming (\S~\ref{sec:effect}).

We note that all of the comparisons of FR~II galaxies and radio-loud quasars
ignored the existence of a separate class of FR~IIs with low-excitation optical
emission lines (Hine and Longair 1979; Laing \ea 1994).
Since the low-excitation FR~IIs tend to have lower radio luminosities than the
high-excitation FR~IIs (Laing \ea 1995), as well as more complex radio
morphologies (Laing \ea 1994), the comparisons of
isotropic properties described below could be affected in a systematic way.
In particular, the low-excitation FR~IIs may be more closely
associated with BL~Lac objects than with quasars.
Re-analyzing existing data in the light of this alternative classification
would be very interesting.

\subsection{History of Radio-Loud Unification Schemes} 
\label{sec:history}

The suggestion that radio-loud quasars were an aligned
version of radio-quiet quasars (Scheuer and Readhead 1979)
was an early attempt to unify quasars, but was
ruled out by the lack of strong large-scale, diffuse radio
emission in the radio-quiet sources. An alternative
suggestion was that flat-spectrum radio quasars were aligned
versions of steep-spectrum radio quasars (Orr and Browne
1982). This satisfies the requirement for equivalent
large-scale radio structures and has not been ruled out.
However, a third suggestion, that the SSRQ and FSRQ are
increasingly aligned versions of FR~II galaxies (Peacock
1987; Scheuer 1987; Barthel 1989) appears to fit the data better (Padovani
and Urry 1992).

Since BL~Lac objects are also blazars but have much lower
equivalent width lines than FSRQ, it was originally
suggested that they could be still more aligned versions of
quasars (Blandford and Rees 1978). The intrinsic BL~Lac
luminosities are much smaller than quasar luminosities (for well defined
samples), however, and crude estimates of the number densities of
parents required are more consistent with low-luminosity
radio galaxies (Schwartz and Ku 1983; Browne 1983; P\'erez-Fournon and
Biermann 1984), as are
quantitative estimates of the number densities and
luminosities of BL~Lacs and FR~I galaxies (Ulrich 1989; Browne 1989;
Padovani and Urry 1990; Urry \ea 1991a).

\subsection{Distinction between FR~I and FR~II Radio Galaxies} 
\label{sec:distinction}

Two decades ago, Fanaroff and Riley (1974) recognized that radio galaxies
separate into two distinct luminosity classes, each with its own morphology.
As described in \S~\ref{sec:AGNprop}, radio emission in the low-luminosity
FR~Is peaks near the nucleus, while the high-luminosity FR~IIs have radio
lobes with prominent hot spots and bright outer edges. Jets, when seen, tend
to be more collimated in the FR~IIs. Examples of both morphological types are
shown in Fig.~\ref{fig:rad_image}.

The luminosity distinction is fairly sharp at 178 MHz, with
FR~Is and FR~IIs lying below and above, respectively, the
fiducial luminosity $L_{178} \approx 2 \times 10^{25}$ W
Hz$^{-1}$. The separation is cleanest in the two-dimensional
optical-radio luminosity plane, where each class follows a
separate, approximately linear, $L_{\rm r}$ - $L_{\rm o}$
correlation (Owen and White 1991; Owen and Ledlow 1994),
implying the FR~I/FR~II break depends on optical as well as
radio luminosity. At a given radio luminosity, FR~Is are more luminous
optically than FR~IIs although both classes span the same range in optical
luminosity overall (Owen and Ledlow 1994).

At higher radio frequencies, the
luminosity ranges for each class do overlap by as much as
two orders of magnitude, with FR~Is having 2.7~GHz
luminosities as high as $L_{2.7}\sim 6 \times
10^{26}$~W~Hz$^{-1}$ while FR~IIs have luminosities as low
as $L_{2.7}\sim 2 \times 10^{25}$~W~Hz$^{-1}$ for the 3CR
sample (Laing \ea 1983) and even $L_{2.7}\sim 3 \times
10^{24}$~W~Hz$^{-1}$ for the 2~Jy sample (Morganti \ea
1993).

FR~I and FR~II galaxies differ systematically in several ways in addition
to host galaxy magnitude. At the same radio luminosity or host galaxy
magnitude,
FR~IIs have optical emission lines about an order of magnitude stronger
than those of FR~Is
(Rawlings \ea 1989; Baum and Heckman 1989; Zirbel and Baum 1995).
FR~Is tend to inhabit moderately
rich cluster environments (Prestage and Peacock 1988; Hill and Lilly 1991)
in which they are sometimes the first-ranked ellipticals (Owen and Laing 1989),
while FR~IIs are more isolated.

Classification of radio galaxies is surely more complicated
than a simple bifurcation. While FR~II morphologies are well defined
by their clear outer hotspots and/or bright edges,
the FR~I class includes many disturbed
and atypical radio structures (Parma \ea 1992). At the same
time, optical spectra of FR~IIs are quite heterogeneous,
with some having strong emission lines and others having
only very weak, low-excitation emission lines (Hine and Longair 1979;
Laing \ea 1994). An
alternative classification of radio galaxies by optical
spectra might group the low-excitation FR~IIs with the
(weak-lined) FR~Is. The physical connection between FR~I and FR~II radio
galaxies is a clue to the formation of jets and the extraction of energy
from the black hole (Baum \ea 1995).

The blurry line between FR types does affect the proposed AGN unification
schemes, but only as a perturbation on the basic picture. In terms of
statistical analysis, it affects the details of the low-luminosity end of the
FR~II luminosity function and the high-luminosity end of the FR~I luminosity
function
but has no major impact because what matters in a quantitative analysis
(\S~\ref{sec:statistical}) is the knee of the parent luminosity function.

\subsection{Isotropic Properties of Quasars and FR~II Galaxies} 
\label{sec:iso_high}

\subsubsection{Extended Radio Emission} 
\label{sec:ext_high}

FR~II radio galaxies are dominated by emission from large,
diffuse radio lobes which is almost certainly unbeamed,
especially at low frequencies where the contribution from
flat-spectrum hot spots is minimal. The radio luminosities
of the extended structures of FR~II galaxies and quasars
should therefore be comparable if the unified scheme is right.
The fact that most compact
radio sources have substantial extended radio power has been
known for more than a decade (Ulvestad \ea 1981) and
the distributions of $P_{\rm ext}$ for compact radio sources
(largely quasars)
and 3CR/B2 radio galaxies (at 1.5~GHz) do overlap
(Antonucci and Ulvestad 1985).

In terms of the quasar -- FR~II unification scheme in particular,
the extended radio luminosities at a few GHz of complete samples of FSRQ and
SSRQ are typical of those for complete samples of FR~II
radio galaxies, as shown in Fig.~\ref{fig:pext_high}
for the 2 Jy sample. (All radio powers
have been de-evolved using the best-fit evolution
appropriate for that class; see \S~\ref{sec:unif_high}) We note that
it is not appropriate to use the Kolmogorov-Smirnov
(KS) statistic to test whether these distributions are
compatible --- or similarly, whether the distributions of
redshift or narrow-line luminosity are comparable for
blazars and radio galaxies --- because
the sample identification
itself involves strong selection effects from relativistic
beaming and thus one should not expect the {\it shape} of the
distributions to be the same. For example, since beamed
objects can be detected to higher redshifts, in a
flux-limited sample they will be intrinsically more luminous
as well. The histograms in Fig.~\ref{fig:pext_high}
demonstrate only that the range in unbeamed radio luminosity
overlaps completely, not that the distributions have the
same shape.

\subsubsection{Narrow Emission Lines} 
\label{sec:narrow_high}

Assuming the narrow emission lines are emitted isotropically, unification
schemes predict similar narrow-line luminosities in quasars and FR~II radio
galaxies. Due to the intrinsic correlation between narrow-line luminosity and
extended radio emission (Baum and Heckman 1989; Rawlings and Saunders 1991;
Zirbel and Baum 1995), the comparison has to be done matching objects of
similar extended radio luminosity. For flux-limited samples, this is roughly
equivalent to matching in redshift (although the scatter in the
redshift-$P_{\rm ext}$ correlation can be large for core-dominated objects
since it is an induced correlation). Also, matching the objects in redshift
will exclude
the effects of cosmological evolution, as well as observational selection
effects due to the bandpass used.

Quasars do have systematically higher [O~{\sevenrm III}] $\lambda 5007$
luminosities than radio galaxies (Jackson and Browne 1990),
but this line
appears to be emitted anisotropically when compared to [O~{\sevenrm II}]
$\lambda 3727$ (McCarthy 1989; Hes \ea 1993), possibly due to contribution to
[O~{\sevenrm III}] from the broad-line region (although there is no sign of
broad wings on the [O~{\sevenrm III}] line) and/or partial obscuration by
the thick torus. (Excluding the low-excitation FR~II galaxies also reduces the
discrepancy in [O~{\sevenrm III}] luminosities: Laing \ea 1994.) As shown in
Fig.~\ref{fig:oii}, the [O~{\sevenrm II}] $\lambda 3727$ line luminosities of
3CR steep-spectrum quasars and FR~II radio galaxies matched in radio power
are completely overlapping (Hes \ea 1993), as expected from the unification
scheme.\footnote{Any trend with redshift is induced through the correlation of
luminosity and redshift in flux-limited samples, and should therefore be
ignored in this and similar plots.}

\subsubsection{Infrared Properties} 
\label{sec:infrared_high}

Models for the obscuring torus suggest it becomes
transparent in the far-infrared (Pier and Krolik 1993),
where the radiation, possibly thermal emission by dust
grains (Sanders \ea 1989), therefore becomes
isotropic. To date, the only bulk measurements of AGN at
far-infrared wavelengths come from {\it IRAS}, which was not sensitive enough
to detect most radio galaxies individually. Separate co-adds
of {\it IRAS} data for radio galaxies and quasars, including many
non-detections, do show that
quasars are systematically more luminous at all observed
wavelengths from 12~$\mu$m to 100~$\mu$m. This suggests
that there is a fundamental difference between the two populations,
or there is a significant contribution from beamed far-infrared continuum
in the quasars,
or the torus is still radiating anisotropically at
$\sim50~\mu$m in the rest frame (Heckman \ea 1994). {\it
ISO} observations will be fundamental in
addressing this issue further.

In AGN for which the optical depth to the nucleus
is not too high, detection of broad Paschen lines may be possible.
To date, Pa$\alpha$ has been seen in some broad-line radio galaxies
and also in a few narrow-line radio galaxies, and in all cases the
reddening to the broad-line region was larger than the reddening to
the narrow-line region (Hill \ea 1995). With
more sensitive instruments, the expectation is that infrared broad
lines will be found in a number of Type~2 radio galaxies. The torus
is likely quite thick in the equatorial plane, so it is only at
intermediate angles that the extinction would be small enough for Pa$\alpha$
photons to escape. Interestingly, radio galaxies appear to have
nuclear sources stronger in the infrared than in the optical, exactly
as expected for a reddened, hidden nucleus (Dunlop \ea 1993).

\subsubsection{Host Galaxies} 
\label{sec:host_high}

A simple test of unification is that the morphology and magnitude of
radio galaxies and quasar host galaxies be comparable. At present the
observational situation for quasars and FR~II radio galaxies is not
entirely clear, as comparisons of their mean host galaxy luminosities
have led to conflicting conclusions.
Some studies have found radio galaxies to be fainter by
$\sim0.5-1$ magnitude (Hutchings 1987; Smith and Heckman 1989), while others
concluded the mean magnitudes (and dispersions) were comparable (V\'eron-Cetty
and Woltjer 1990; Lehnert \ea 1992).
In part, determining quasar host galaxy
magnitudes is difficult, and in part the differences between the conflicting
results may be explained by systematic effects (Abraham \ea 1992) and/or by
incomplete or heterogeneous samples.
For example, in the present unification scheme the
parent population consists of luminous narrow-line radio galaxies, yet samples
of ``powerful radio galaxies'' used to test unification sometimes include
broad-line radio galaxies (the local counterpart of quasars) and FR~Is.

With high-resolution ground-based imaging, host galaxies have been detected
in quasars out to a couple of gigaparsecs.
Figure~\ref{fig:qhost} shows a CFHT image of the radio-loud quasar 2135$-$147,
which has redshift $z=0.2$. Even though the quasar nucleus is very bright,
the host galaxy is clearly visible, as are companion objects $\sim2$~arcsec
and $\sim5$~arcsec to the east (Hutchings and Neff 1992). Most quasars are at
higher redshift, so determining whether they are in ellipticals requires high
resolution imaging of the kind now available with {\it HST}
(Hutchings \ea 1994; Hutchings and Morris 1995;
Bahcall \ea 1995).
Such studies of the galaxy morphologies of large, complete samples of quasars
and radio galaxies will be important for testing unified schemes.

Most host galaxy studies have been done at optical wavelengths,
where the ratio of nuclear to galaxy flux is relatively high.
Diffuse stellar light surrounding bright AGN is more easily detected
in the infrared; for example, the K-band images of 2135$-$147, the quasar in
Fig.~\ref{fig:qhost}, also show the diffuse galaxy
emission and companion objects quite clearly, if not at the same
resolution (Dunlop \ea 1993). A recent infrared comparison of
radio galaxies and radio-loud quasars at $z\simlt0.4$,
matched in radio power and redshift, shows that the host galaxies of both
classes appear to be luminous ellipticals, have the same average half-light
radii, and have the same mean absolute magnitudes (Dunlop \ea 1993;
Taylor \ea 1995). For both the radio galaxies and quasars,
infrared galaxy surface brightness and scale length are well correlated
(Fig.~\ref{fig:host_prop}), with the same slope and normalization as for
brightest cluster galaxies (Schneider \ea 1983).
This gives strong support to the unified scheme, at least for relatively
nearby quasars.

\subsubsection{Environments} 
\label{sec:env_high}

For a unification scheme to be correct, the environments of the unified
classes --- here, FR~IIs and radio quasars --- have to match.\footnote{Even
within the
unified scheme, environment may have an effect on the relative numbers of
radio galaxies and blazars. For example, the opening angle of the torus
might depend on gas density external to the active nucleus. In that case,
the number ratio of blazars to radio galaxies could differ between cluster and
field. If the opening angle were smaller in clusters due to extra quenching,
for example, radio galaxies would be found more often in clusters than are
blazars. This is a matter of degree, however; the fact that both radio galaxies
and blazars should be found in clusters and in the field would not change.}
The observational
situation here is unclear. In an oft-quoted study, Prestage and Peacock (1988)
found that compact/flat-spectrum radio sources appear to lie in regions of
galaxy density a factor of $\sim 2$ lower than those typical of FR~Is and
possibly even lower than for FR~II sources. However, this was based on small
samples of only $\sim 10$ compact flat-spectrum sources, which included
BL~Lacs,
local FSRQ, and even some radio galaxies. The Prestage and Peacock (1988)
data on FSRQ alone are too limited to derive a
spatial correlation amplitude for comparison to the FR~IIs.

Other studies at low redshift showed either no significant difference in
environment (Smith and Heckman 1990), or only a marginal difference (Yates \ea
1989) that disappears when the radio galaxies without definite FR~II structure
are excluded. At higher redshift ($0.35 \simlt z \simlt 0.5$), the environments
of powerful radio galaxies are as rich as those observed around radio-loud
quasars in the same redshift interval (Yates \ea 1989).

Thus there is no evidence for concluding that the
environments of FR~IIs and radio quasars differ. Obviously, additional data
for larger samples covering a range in redshift are needed to confirm that the
environments are similar.

\subsubsection{Cosmic Evolution} 
\label{sec:evol_high}

If radio galaxies and quasars are related strictly through aspect, then
the relative numbers and luminosities of each class should be broadly
comparable
at all cosmological epochs. Indeed, the cosmological evolution of
steep-spectrum
radio quasars, flat-spectrum radio quasars, and radio galaxies can be described
in each case by luminosity increasing with redshift roughly as $(1+z)^3$ out to
$z \sim 2$, followed by a comparable decline in co-moving density at higher
redshifts (Dunlop and Peacock 1990). This means current samples are broadly
consistent with unification independent of redshift.

Over cosmic time, changes in the relative strengths of beamed and unbeamed
components in individual sources, or in the opening angle of an obscuring torus
(Lawrence 1991), might be expected. In either of those cases, the
observed evolution would be different depending on orientation angle.
The magnitude of this effect is clearly model-dependent but if the unification
of radio galaxies and quasars is correct, it can not be large given the
observed similarity in their evolutionary properties.

\subsection{Isotropic Properties of BL~Lac Objects and FR~I Galaxies} 
\label{sec:iso_low}

\subsubsection{Extended Radio Emission} 
\label{sec:ext_low}

More than ten years ago it was noted that the extended radio emission of
BL~Lacs was comparable to that of low-luminosity (i.e., FR~I) radio galaxies
(Browne 1983; Wardle \ea 1984; Antonucci and Ulvestad 1985), although these
early studies were hampered by the lack of redshift information and small and
incomplete samples. Now that sizeable complete BL~Lac samples are available,
indeed the mean extended radio powers at 5~GHz for the 1~Jy radio-selected
BL~Lac sample (Stickel \ea 1991) and a subsample of the 3CR FR~Is are
essentially the same, $\langle P_{\rm ext} \rangle \sim 10^{25}$~W~Hz$^{-1}$
(Padovani 1992a). Figure~\ref{fig:pext_low} shows the overlapping histograms
of $P_{\rm ext}$ observed for the 1~Jy BL~Lacs and the 2~Jy FR~Is (Wall and
Peacock 1985; di Serego Alighieri \ea 1994b), which have similar flux
limits and selection frequencies. It is not clear whether the BL~Lac
morphologies are consistent with FR~I morphologies seen at small angles;
if the angle separating the two classes is large enough, one might expect
to see more double-jet radio structures in BL~Lacs, although the FR~I
morphologies are quite varied (Parma \ea 1992).

There is similar agreement between B2 FR~Is and X-ray selected BL~Lac objects
(XBL) from the {\it Einstein} Medium Sensitivity Survey (EMSS; Perlman and
Stocke 1993) or 3CR and B2 FR~Is and XBL from the {\it HEAO-1} Large Area Sky
Survey (LASS; Laurent-Muehleisen \ea 1993). Figure~\ref{fig:pext_low} includes
the distribution of $P_{\rm ext}$ for the EMSS XBL;
both the 3CR and B2 samples have slightly lower fluxes at 5 GHz than the 2 Jy
sample, which results in their extended powers reaching smaller values.

It is notable that the mean extended radio power of XBL is more than an order
of magnitude lower than for radio-selected BL~Lacs (RBL; see also
Laurent-Muehleisen \ea 1993). Moreover, the weak
radio cores of XBL appear relatively unbeamed, as if they are ``off-axis'' in
the radio band (Padovani 1992a; Perlman and Stocke 1993) or intrinsically less
luminous (Giommi and Padovani 1994; Padovani and Giommi 1995). The large
difference in extended power supports the latter interpretation (see
\S\S~\ref{sec:rel_xr}, \ref{sec:terms}).

Some of the 1~Jy RBL, which extend to higher redshifts and thus to higher
luminosities, appear to have radio morphologies more consistent
with FR~IIs than with FR~Is (or rather, with what FR~IIs would look like at a
small angle with the line of sight; Kollgaard \ea 1992; Murphy \ea 1993). This
is not a serious challenge to the idea of FR~Is being the parent population of
BL~Lac objects (see also Maraschi and Rovetti 1994, who consider the
continuity explicitly), as it affects only a relatively few objects at the high
end of the FR~I luminosity function, but it reinforces the blurriness of the
distinction between FR~Is and FR~IIs in the overlapping luminosity range. Just
as the luminosities overlap at high frequencies, perhaps the clear
morphological separation breaks down in higher frequency maps (which are
needed to get high spatial resolution in the more luminous, higher redshift
objects). Also, none of the BL~Lacs observed by Kollgaard \ea (1992) displays
FR~II characteristics as clearly as some core-dominated quasars observed by
the same group (albeit with somewhat better resolution; Kollgaard \ea 1990).
Finally, a substantial number of 3CR FR~II sources
have low-excitation optical spectra, i.e., with lines such as \OIII~that
are very weak
(or undetectable) in comparison to the hydrogen lines (Laing \ea 1994).
Since the
narrow-line spectra of these objects are clearly different from those of
classical FR~II radio galaxies and radio quasars, instead resembling those of
FR~Is, it is possible that the low-excitation sources, despite their FR~II
morphology, could well belong to the low-power unification scheme. It would
then follow that some BL~Lacs are {\it expected} to display FR~II morphology
and also that that does not necessarily associate them with all the high-power
sources.

\subsubsection{Narrow Emission Lines} 
\label{sec:narrow_low}

It is not yet clear whether the narrow-line luminosities of BL~Lac objects, in
the few cases when such lines can be observed, are consistent with those of
FR~I galaxies. \OIII~luminosities are available for eight of the eleven 1~Jy
BL~Lac objects with certain redshift $z\simlt 0.5$ (Stickel \ea 1993). The
mean emission line luminosity is $L_{\rm O III} = 10^{41.30 \pm 0.11}$
erg s$^{-1}$,
with a spread of less than an order of magnitude. Figure~\ref{fig:oiii}
shows these values compared to the available
\OIII~luminosities for the FR~I galaxies in the 2 Jy sample (Tadhunter \ea
1993), plotted versus redshift to compensate
simultaneously for any trends with observed wavelength range or
luminosity (which correlates well with redshift in flux-limited samples).

For $z\simlt 0.2$, the maximum redshift of the FR~Is, and excluding
non-detections, the mean \OIII~luminosities are $L_{\rm O III} = 10^{41.10 \pm
0.07}$ and $10^{40.49 \pm 0.24}$ erg s$^{-1}$ for five BL~Lacs and five FR~Is,
respectively. The difference is a factor of $\sim4$, significant at the $\sim
94\%$ level, and is more than an order of
magnitude if one includes the upper limits.
However, the FR~I line fluxes may be underestimated due to
small slits and/or contamination from a strong stellar continuum
(Tadhunter \ea 1993). Also,
a comparable set of uniform upper limits for BL~Lac objects is not available
for
comparison; possibly the \OIII~luminosity in most BL~Lac objects is well
below the few detections available.
High signal-to-noise-ratio spectra through large apertures
of a reasonably large sample of FR~I radio galaxies (and BL~Lac objects)
are needed to determine whether $L_{\rm O III}$ in FR~Is is really
smaller than in BL~Lacs.

The apparent difference in [O~{\sevenrm III}] luminosities for BL~Lacs and
FR~Is could also arise if [O~{\sevenrm III}] were emitted anisotropically, as
for radio-quasars and FR~II radio galaxies (\S~\ref{sec:iso_high}). The fact
that broad \MGII~emission lines are seen in some BL~Lac objects (Stickel \ea
1993) means that for unification to be correct, some FR~Is (albeit
high-redshift
ones) must have obscured broad line regions. By analogy to quasars
(\S~\ref{sec:narrow_high}), the obscuring matter could affect the
\OIII~emission more than the \OII, thus explaining the apparent discrepancy.
While there is no direct evidence for an obscuring torus surrounding the
[O~{\sevenrm III}]-emitting clouds in BL~Lac objects or FR~Is, neither is
there evidence against it.

Recent {\it HST} spectra of the nucleus of M87, a nearby FR~I galaxy of modest
luminosity (Harms \ea 1994), do show relatively broad emission lines, ranging
from FWHM $\sim
1400$~km~s$^{-1}$ for H$\beta$ up to $\sim 1900$~km~s$^{-1}$ for [O~{\sevenrm
I}]. These lines are broader than the typical widths for Seyfert~2 galaxies
(FWHM $\sim 250-1200$~km~s$^{-1}$; Osterbrock 1989)
and the [O~{\sevenrm III}]/$H\beta$ ratio is much larger than 1,
unlike the typical case for Seyfert~1 galaxies. If M87 has a hidden nuclear
broad-line region, further spectroscopic investigation of this and other FR~Is
is obviously of great interest.

Unlike the quasar - FR~II unification scheme, where SSRQ appear at
intermediate angles, the BL~Lac - FR~I scheme has essentially no transition
objects (the low-luminosity BLRG 3C~120 might be an exception).
These would be FR~I-like radio sources, with steep radio spectra
and FR~I-like morphologies because they are oriented outside the
relativistic beaming cone, but would show BL~Lac-like emission lines
(possibly broad lines) because they are oriented inside any
obscuration cone. It may be that BL~Lacs just do not have any appreciable
emission, lines or continuum, beyond that relativistically beamed from the jet.
Alternatively, broad lines could be hidden in FR~Is, a phenomenon implied
by unification with the high-redshift BL~Lac objects, in which case there
should be (as for quasars) broad-line FR~Is. The lack of such objects
could be explained if the opening angle of the obscuring torus is narrower
at lower powers (Lawrence 1991; Falcke \ea 1995; see also Baum \ea 1995).
Far infrared observations
of BL~Lacs and FR~Is with {\it ISO} should indicate whether or not there
is obscuring matter in low-luminosity radio sources.

\subsubsection{Host Galaxies} 
\label{sec:host_low}

There is good observational evidence that low-redshift BL~Lacs reside in giant
ellipticals as FR~Is do.
The host galaxy of the BL~Lac object PKS~0548$-$322
(Fig.~\ref{fig:bll_host}, center) has a de~Vaucouleurs
$r^{1/4}$~law profile (Falomo \ea 1995),
as do the point-source-subtracted optical profiles
of many low-redshift BL~Lacs (Ulrich 1989; Stickel \ea 1991, 1993).
The distribution of host galaxy absolute magnitudes peaks
around $M_{\rm V} \sim -23$, which is comparable to the brightest galaxies in
clusters (Kristian \ea 1978). The mean host galaxy magnitude for the seven
1~Jy BL~Lac objects with $z < 0.2$ is $\langle M_{\rm V} \rangle = -22.9 \pm
0.3$ (Stickel \ea 1991, 1993), in excellent agreement with the mean value for
FR~Is, $\langle M_{\rm V} \rangle = -23.1 \pm 0.1$ (derived from Smith and
Heckman 1989, converting to $H_{\rm 0} = 50$ km s$^{-1}$ Mpc$^{-1}$). There
is no useful imaging information available for the BL~Lacs at higher redshifts
($z \simgt 0.2$), a project now underway with {\it HST}.

A few of the 20 or so BL~Lac host galaxies classified in the literature show
evidence for disks but this is far from conclusive. The surface
brightness profile of PKS~1413+135 appears to be better fitted by an
exponential law than by a de Vaucouleurs law (Abraham \ea 1991), suggesting a
disk host, while for OQ~530 the results are mixed (Abraham \ea 1991; Stickel
\ea 1991, 1993). Stocke \ea (1992; see also Perlman \ea 1994)
have suggested that PKS~1413+135, an
unusually reddened BL~Lac object, may lie behind a spiral galaxy rather than
within it (cf. McHardy \ea 1994). Finally, while Halpern \ea (1986) reported a
disk
host in 1E~1415.6+2257, Romanishin (1992) found the brightness profile to be
consistent with an elliptical galaxy, with the possible presence of a jet-like
feature that went unnoticed by Halpern et al. About 10\% of FR~Is are known to
have some sort of morphological peculiarity in the optical (Smith and Heckman
1989). This could complicate the surface brightness profile analysis for BL~Lac
host galaxies, although to our
knowledge there is no FR~I radio galaxy whose optical surface brightness
profile is better fitted by an exponential law than a de Vaucouleurs law. It
would be useful to have higher resolution imaging of both BL~Lacs and FR~Is to
explore this further.

\subsubsection{Environments} 
\label{sec:env_low}

The systematic study of the environment of BL~Lacs is relatively recent. The
work of Prestage and Peacock (1988), discussed in \S~\ref{sec:iso_high}, is
usually quoted (wrongly) as evidence that BL~Lacs avoid rich environments; it
actually shows that (within large uncertainties) the spatial correlation
amplitude of (4) BL~Lacs is consistent with that of FR~I radio galaxies.

Fields of low-redshift BL~Lac objects that have been studied intensively tend
to show statistically significant excesses of galaxies, compatible with
clusters of galaxies of Abell richness class 0 and 1 (Falomo \ea 1993;
Pesce \ea 1994; Fried \ea 1993; Wurtz \ea 1993; Smith \ea 1995),
in good agreement with
the environments of FR~I type radio galaxies (Hill and Lilly 1991).
The environment of PKS~0548$-$322 (Fig.~\ref{fig:bll_host}) is clearly
rich in galaxies, including two companion elliptical galaxies within
$\sim40$~kpc projected distance (if at the redshift of the BL~Lac, $z=0.069$)
and numerous other galaxies with brightnesses appropriate for a cluster
at $z=0.069$. Some nearby BL~Lacs lie in somewhat poorer clusters,
as do some FR~Is (Wurtz \ea 1993; Laurent-Muehleisen \ea 1993).
In short, BL~Lacs and FR~Is do appear to have similar environments.

Reversing the question leads to an unexplained discrepancy. A spectroscopic
survey of 193 radio sources found in Abell clusters revealed 186 FR~I galaxies
but no BL~Lac objects, when according to beaming models roughly 8 would have
been expected in the observed luminosity range, a discrepancy significant at
the $>99\%$ level (Owen \ea 1995). That is, this unbiased cluster-selected
survey for BL~Lac objects suggests they occur preferentially outside of normal
FR~I environments.
We note that the FR~I luminosities observed by Owen \ea (1995),
$L_{1.4}<10^{26}$~W~Hz$^{-1}$, and the observed BL~Lac radio
luminosity function,
for which $L_5>10^{25}$~W~Hz$^{-1}$ (Fig.~\ref{fig:rlf_low};
Urry \ea 1991a), overlap by less than one decade. In this interval
there are only 18 FR~Is and the absence of BL~Lacs is only
marginally significant (95\% confidence; Owen \ea 1995). Furthermore, the
conflict in number density is actually between the {\it observed} BL~Lac and
FR~I luminosity functions rather than with the predictions of the beaming
model. It is possible that since the redshift limit of the cluster sample is
low, $z < 0.09$, recognition effects (Browne and March\~a 1993) could
be important.

At higher redshifts ($z \sim 1$) there is no indication of a density
enhancement near the BL~Lacs, as expected because cluster galaxies at these
redshifts would be well below the typical detection limits (Fried \ea 1993).
It is important to study blazar environments at high redshift because it is
there that quasar environments appear to undergo significant evolution. At
$z\sim0.6$, radio quasars are often found in clusters of galaxies while
locally they are in regions of lower galaxy density (Yee and Ellingson 1993).
Furthermore, if there were a connection between the more luminous BL~Lac
objects and radio quasars (see \S~\ref{sec:connection}), it would only be
apparent at higher redshift.

\subsection{The Myth of Unbiased Selection} 
\label{sec:myth}

Given the strong selection effects introduced by relativistic beaming and
obscuration, it is very difficult to find useful unbiased samples. In the
ideal case, surveys would be designed to select AGN by some completely
isotropic property. Since surveys are generally flux-limited, this effectively
means selecting sources at a wavelength not strongly affected by
obscuration or beaming, such as in the low-frequency radio, far-infrared,
or hard X-ray bands.

At present, however, it is not possible to select a useful AGN sample in an
unbiased way. The 3CR radio sample, selected at 178~MHz where steep-spectrum
emission from the lobes dominates, contains only a few core-dominated objects
(i.e., blazars); in the 3CR catalog (Spinrad \ea 1985), only 2/298 (0.7\%)
objects are BL~Lac objects (0\% in the complete subsample of Laing \ea 1983).
That is, the 3CR catalog may be (essentially)
unbiased by beaming but it does not include enough blazars to test their
participation in unified schemes. In contrast, in the 1~Jy radio source
catalog (Stickel \ea 1994), selected at 5 GHz, there are
37/527 BL~Lacs (7\%; there are 34 in the complete sample, with $V \leq
20$~mag), and 214/527 (41\%)
FSRQ, i.e., about 50\% of the sources are blazars. Low-frequency-selected
samples that are larger than the 3CR, like the Molonglo or B2 surveys, are
still not completely identified: a complete sample of 550 sources selected
from the Molonglo catalog is in the process of being identified (Kapahi \ea
1994), while two relatively small but completely identified sub-samples of the
B2 catalog exist (Fanti \ea 1987) but are defined by a cut in optical
magnitude, which is certainly not an isotropic property. Finally, the emission
at low radio frequencies need not be entirely isotropic, as it can be
affected somewhat by beaming in the cores and radio-lobe hot spots.

Far-infrared selection does avoid the bias introduced by dust obscuration of
optical/ultraviolet emission but introduces a bias against dust-free AGN.
Furthermore, the selection effects at far-infrared wavelengths depend on the
properties of the obscuring torus or other dusty structures, which may not
generate isotropic emission (Pier and Krolik 1992) and which may differ from
source to source by more than just orientation. The far-infrared may also be
affected by beaming insofar as there is a contribution from an infrared
nonthermal source. Finally, the sample identification for the {\it IRAS}
survey is a
major burden, not to mention that AGN are not found efficiently (less than
1\% of the high-latitude sources are AGN).

Hard-X-ray surveys are theoretically useful but no deep large-area surveys are
currently planned. Also, X-ray spatial resolution at 10 keV is not as good as
radio or infrared resolution, making the confusion limit correspond to an
effectively higher flux; i.e., hard X-ray surveys can not go as deep. The
lower resolution also makes optical identification of the sample more
difficult. In the future, an all-sky survey with arcsecond resolution at
$10-20$ keV would be very valuable for evaluating unified schemes in a
model-independent way.

In short, to address the key question of unified schemes --- whether the
numbers and luminosities of objects in the classes one proposes to unify are
as expected by the scheme --- means using well-defined (i.e., flux-limited)
samples with known biases, and accounting for those selection effects
quantitatively. This allows us to comment on whether a specific unified scheme
is plausible and what parameter values are needed to make it so. It also
avoids the pitfall of defining samples {\it post facto}, based on ``isotropic
quantities'' like narrow-line emission or host galaxy type, which may well
be incomplete because the detection threshold is not related to the sample
definition.
Fortunately, allowing for selection bias head-on, at least in the case of
relativistic beaming, is a robust process which is not terribly sensitive to
the details of the beaming model.

\subsection{Effect of Relativistic Beaming on Number Statistics} 
\label{sec:effect}

AGN with anisotropic emission patterns will be boosted into and shifted out of
flux-limited samples according to their orientation. For some radiation
patterns, including those caused by relativistic beaming (Urry and Shafer
1984) and thick accretion disks (Urry \ea 1991b), a narrow distribution in
intrinsic luminosity is broadened into a flat distribution over a wide range
of observed luminosities (which follows from the probability function for flux
enhancement.) This leads to a distortion in the measured shape of the
luminosity function of the boosted AGN relative to their intrinsic luminosity
function (Urry and Shafer 1984; Urry and Padovani 1991).

More specifically, under the basic assumption that AGN are randomly oriented
on the sky, and assuming a radiation pattern, we can predict the exact numbers
of AGN with a given observed luminosity relative to their intrinsic luminosity
(summed over all angles). Given the luminosity function (LF) of the misaligned
AGN, then, one can predict the LF of the aligned AGN, subject to the form of
the radiation pattern. In practice, the pattern for relativistic beaming
depends on the assumed distribution of Lorentz factors and on whether one
assumes a uni-directional jet or a fan beam; the pattern for obscuration
depends on the size and optical depth of the torus; the pattern for thick
disks depends primarily on the funnel geometry.

To evaluate the number statistics of radio-loud unification schemes we
incorporate the effect of relativistic beaming on the observed LFs (Urry and
Shafer 1984). Consider an ensemble of emitters all having the same intrinsic
luminosity (${\cal L}$) and all moving with the same relativistic bulk speed
($\beta$) at random angles ($\theta$) to the line of sight.
Given that $L = \delta^p \l$ (Eq.~\ref{eq:lj_lu}), the probability of
having a particular Doppler factor $\delta = [\gamma(1 -\beta
\cos\theta)]^{-1}$ (Appendix~A) is $P(\delta) = d(\cos\theta)/ d\delta
= (\beta\gamma\delta^2)^{-1}$. The probability of observing
luminosity $L$ given intrinsic (emitted) luminosity ${\cal L}$ is
\begin{equation}
{P(L\vert{\cal L})} = { P(\delta) ~ {{d\delta} \over {dL}}}
={ {{1} \over {\beta\gamma\delta}} ~ {\cal L}^{1/p} ~
L^{-(p+1)/p}~.}
\label{eq:prob}
\end{equation}
For fixed {$\cal L$}, the distribution of observed luminosities
is a flat power law (with index in the range $1 - 1.5$ for likely values of
$p$; Appendix~B) extending from $L\sim (2\gamma)^{-p} {\cal L}$ to $L\sim
(2\gamma)^p {\cal L}$. The observed luminosity distributions are illustrated
in Fig.~\ref{fig:fake_lf}a (thick dashed lines) for three different intrinsic
luminosities (thin dashed lines). The low-luminosity cutoff corresponds to an
emitter moving directly away from us ($\theta = 180^\circ$) and the
high-luminosity cutoff to an approaching emitter perfectly aligned ($\theta =
0^\circ$). The normalization of this flat power law decreases with increasing
beaming (higher $\beta$, $\gamma$, or $p$) because the beaming cone angle gets
smaller.

For a distribution of intrinsic luminosities (i.e., a luminosity function),
the observed LF is just the integral of the intrinsic luminosity distribution
times the conditional probability function in Eq.~(\ref{eq:prob}):
\begin{equation}
\Phi_{\rm obs} (L) = \int d{\cal L} ~ P(L\vert{\cal L}) ~
\Phi_{\rm intr} ({\cal L}) ~.
\label{eq:int}
\end{equation}
This integral (thick solid line in Fig.~\ref{fig:fake_lf}a)
is basically the envelope of the beamed LFs for fixed intrinsic
luminosities (the intrinsic LF is shown as a thin solid line). For simple power
law luminosity functions, Eq.~(\ref{eq:int}) can be integrated analytically
(Urry and Shafer 1984).

In practice, one expects an unbeamed component (e.g., radio lobes) to be
present in addition to a beamed component (jet). We use the simple
parametrization that the intrinsic luminosity in the jet is a fixed fraction
of the unbeamed luminosity, ${\cal L}_{\rm j} = f {\cal L}_{\rm u}$, so the
total observed luminosity is
\begin{equation}
L_{\rm T} = {\cal L}_{\rm u} + L_{\rm j}
 = {\cal L}_{\rm u} (1 + f \delta^p)~ .
\label{eq:ltot}
\end{equation}
This takes account of the approaching jet only; it can be shown that if jets
come in oppositely directed pairs of similar intrinsic power,
the receding beamed component (with
Doppler factor equal to $\delta = [\gamma (1 +\beta \cos \theta)]^{-1}$) will
have a negligible contribution for likely values of the Lorentz factor and
approaching jets within $\sim 60^\circ$ of the line of sight (Appendix~C).
As before, the
conditional probability is derived from Eq.~(\ref{eq:ltot}) but in this case
Eq.~(\ref{eq:int}) is integrated numerically. A (likely) distribution of
Lorentz factors can also be included (Urry and Padovani 1991).
If we define the critical angle,
$\theta_{\rm c}$, to be where the beamed and unbeamed luminosities are
comparable (i.e., $f\delta^p = 1$), then for $\theta < \theta_{\rm c}$ the
luminosity will be dominated by beamed emission and we can identify these
sources as blazars. The observed parent and beamed LFs for the case of a
single-power-law intrinsic LF are shown in Fig.~\ref{fig:fake_lf}b, for four
values of $f$.

The key point is that the luminosity function of the beamed population has a
characteristic broken-power-law form, flat at the low luminosity end and steep
at the high luminosity end (Urry and Shafer 1984).
This remains approximately the case even when the
intrinsic luminosity function has a more complicated form, without sharp
cutoffs (Urry and Padovani 1991). This means that the comparison of the number
densities of beamed and parent populations is a strong function of luminosity,
and for samples biased by relativistic beaming
can be evaluated only by measuring the luminosity functions for each.

Since radio maps of blazars show the presence of a diffuse component we
consider the two--component model for which $L_{\rm T} = {\cal L}
(1+f\delta^p)$ (Eq.~\ref{eq:ltot}). The assumption that the intrinsic jet
power is linearly proportional to the extended power may not be consistent
with observations. Observed core power and total radio power are well
correlated but apparently not linearly (Feretti \ea 1984; Giovannini \ea 1988;
Jones \ea 1994). Although results vary, all seem to find that observed core
power has a less than linear dependence on total radio power. (Since their
samples include mainly radio galaxies, total power is essentially the same as
extended power.) For example, Jones \ea (1994) find that $P_{\rm core}
\propto P_{\rm total}^{0.8}$. The range of observed slopes, generally
calculated taking upper limits into account, is $0.4-0.8$. Calculating the
regression with $P_{\rm core}$ as the dependent parameter will lead to a
systematically different slope and indeed treating the variables symmetrically
would be the appropriate approach (Isobe \ea 1990).

We note that the observed nonlinearity does not immediately conflict with the
assumption that ${\cal L}_{\rm jet} = f {\cal L}_{\rm ext}$. First, beaming
will cause large scatter in the observed $P_{\rm c} - P_{\rm ext}$ plane due
to the spread in viewing angles and possibly Lorentz factors. Second,
selection effects could influence the slope of the correlation, although we do
not immediately see why it should be flatter than unity. If the intrinsic
relation is nonlinear (e.g., ${\cal L}_{\rm jet} = f {\cal L}_{\rm ext}^x$),
then the calculation described here (represented by Eqs.~\ref{eq:int} and
\ref{eq:ltot}) would need to be modified. This would change the
derived parameter values but as long as $x$ is reasonably close to 1, it
should not be a major effect.

Radio-loud unification schemes, which involve primarily relativistic beaming,
have been verified with the luminosity function approach just described.
Radio-quiet schemes have not been tested this way because the radiation
patterns are still very uncertain and the radiation anisotropy is probably much
more Draconian at the usual (blue) selection wavelengths (Pier and Krolik
1992; Ward et al. 1991; Djorgovski et al. 1991). For the
high-power radio-loud scheme, obscuration is needed to explain the optical
properties (namely, to allow for the fact that we do not have a direct view of
the broad line region in FR II radio galaxies)
but it is not important for statistical arguments based on
radio luminosities alone. In the next section we outline the quantitative
evaluation of the observed and beamed luminosity functions for radio-loud
AGN.

\section{Statistical Unification of Radio-Loud AGN} 
\label{sec:statistical}

A fundamental test of the proposed unification of blazars and radio
galaxies is whether the number statistics of the populations agree
with the relativistic beaming hypothesis. The total number of beamed
objects (here, blazars) must be small compared to the number of parent
objects (radio galaxies), as they are presumably oriented at small
angles to the line of sight. This ratio depends only on the critical
angle dividing blazars and radio galaxies, which in turn depends on
the amount of beaming (essentially, the Lorentz factor and the relative
luminosities of beamed and unbeamed components). The critical angle is
therefore central to unification.

In recent years it has become possible to test unified schemes via the
number statistics of complete samples of blazars and radio galaxies.
This section describes separately the statistics of the unification of
quasars and FR~II radio galaxies (\S~\ref{sec:unif_high}) and of BL~Lac objects
and FR~I radio galaxies (\S~\ref{sec:unif_low}). Since relativistic beaming is
best constrained by radio observations (thanks to VLBI), we compare in
\S~\ref{sec:independent} the Lorentz factors from superluminal motion, SSC
calculations, and jet/counter-jet ratios with those derived from number
statistics.

\subsection{Unification of Radio Quasars and FR~II Galaxies} 
\label{sec:unif_high}

We start by deriving the luminosity function of FR~II radio galaxies, then we
beam it according to the prescription outlined in \S~\ref{sec:effect} and
compare the beamed and
observed radio luminosity functions for quasars. The free parameters
are $\gamma$,
the bulk Lorentz factor of the jet, and $f$, the fractional luminosity of the
jet. One can further constrain these two parameters from $R$, the ratio
of beamed to unbeamed flux (core to extended flux in high resolution radio
maps; Appendix~C).

\subsubsection{Content of the 2 Jy Sample of Radio Sources} 
\label{sec:content}

As in Padovani and Urry (1992), we derive quasar and radio galaxy luminosity
functions from the 2~Jy sample (Wall and Peacock 1985), a complete
flux-limited sample of 233 sources with flux at 2.7~GHz $F_{2.7} \geq 2$ Jy.
Redshifts and optical spectroscopic identifications have been updated using
the latest version of the 1~Jy catalog (Stickel \ea 1994) and new optical
spectra for sources with declination $d< 10^{\circ}$ (di Serego Alighieri \ea
1994b).

Quasars (and BLRG) are distinguished from radio galaxies (NLRG) by whether
they have broad optical/ultraviolet emission lines or not. This criterion
might not be as straightforward to apply as it sounds: recent papers (Laing
\ea 1994; Economou \ea 1995; Hill \ea 1995) have pointed out that some objects
classified as narrow-line galaxies actually show broad H$\alpha$ emission.
This is particularly likely for high-redshift sources classified on the basis
of an H$\beta$ line, for which the broad wings can be extinguished by
modest reddening, thus introducing a redshift dependence (in practice)
to the definition of Type~1 versus Type~2 AGN.\footnote{The
precise meaning of ``narrow-line'' AGN,
particularly if every Type~2 AGN harbors a hidden Type~1 nucleus,
is not well-defined. For some astronomers it means AGN completely devoid
of high-velocity ionized gas, while for others it means that
broad lines are simply not detectable in the best spectrum to date.
To rule out the presence of a hidden broad-line region in every Type~2
AGN would require spectropolarimetry more sensitive than that currently
available, not to mention more observing time.
It is also not particularly useful to have objects
changing type based on the latest best observations. Thus we favor a
definition based on fixed observational criteria; for example, a given
signal-to-noise ratio in a given wavelength range (perhaps out to
$H\alpha$ in the rest frame but excluding the Paschen lines).
Based on spectra meeting these criteria, all AGN could be categorized
definitely as Type~1 or Type~2. One would then have to devise other
names to represent objects with similar intrinsic (as opposed to
observed) properties. For example, 3C~234 would be called a Type~1 object
because it has strong broad lines but at the same time it would belong
with NGC~1068 (a Type~2 AGN) in some new category because both have
hidden broad-line regions.}

Only 3 of the 2~Jy sources (1\% of the sample) have no optical
counterpart, while fourteen more (6\%) have no redshift information. The great
majority of the objects without redshifts are classified as galaxies, i.e.,
they appear extended, and based on their visual magnitudes they have estimated
redshifts larger than 0.7.

Morphological classifications as FR~I or FR~II for the
galaxies in the 2~Jy sample have been updated according to new radio maps for
objects with $d < 10^{\circ}$ and $z < 0.7$ (Morganti \ea 1993), and using
additional information as available (Zirbel and Baum 1995, and references
therein). Fifteen galaxies are classified as compact or unresolved while 7 more
are, to the best of our knowledge, unclassified. The division between steep-
and flat-spectrum quasars was made at a radio spectral index between 2.7 and 5
GHz of $\alpha = 0.5$.

These updates have essentially no effect on the FSRQ and SSRQ luminosity
functions. For FR~II radio galaxies, the luminosity function was originally
derived (Padovani and Urry 1992) from a different sample, the 3CR
catalog (Laing \ea 1983), because at the time a large fraction (nearly
one-third) of FR~II galaxies in the 2~Jy sample had uncertain redshift
estimates
and their estimated evolution differed from that of radio galaxies
in other samples.\footnote{The evolutionary
properties of a sample can be characterized by the mean value of the ratio
between $V$, the volume enclosed by an object, and $V_{\rm max}$,
the maximum accessible volume within which the object
could have been detected above the flux limit of the sample.
In a Euclidean geometry, then, $V/V_{\rm max} =
(F/F_{\rm lim})^{-3/2}$, where $F$ is the observed flux of an
object and $F_{\rm lim}$ is the flux limit of the survey. In
the absence of evolution $V/V_{\rm max}$ has the property of
being uniformly distributed between 0 and 1, with a mean
value of 0.5 (Rowan-Robinson 1968; Schmidt 1968). When the survey is made up of
separate fields with different flux limits, as is the case
for the EMSS, it is more appropriate to use $V_{\rm
e}/V_{\rm a}$, that is the ratio between {\it
enclosed} and {\it available} volume (Avni and Bahcall
1980).}
While neither the different selection frequency nor the
different flux limit of the 3CR appeared to affect the resulting luminosity
function substantially, the 2~Jy FR~II sample is now much better defined
so we revisit the question of using it.

The new 2~Jy FR~II galaxies still exhibit less evolution
than the 3CR FR~II galaxies, although the evolutionary properties of the two
samples are consistent at the $1~\sigma$ level.
Using only the definite FR~II
radio galaxies with $z < 0.7$ (the completeness limit
of the spectroscopic subsample of di Serego Alighieri \ea 1994b), one
obtains $\langle V/V_{\rm max} \rangle = 0.55\pm0.04$, where the quoted
error is $1/\sqrt{12N}$, appropriate for a uniform distribution.
The best-fit
evolution parameter\footnote{Here we characterize evolution in terms of
exponential luminosity evolution, $P(z)=P(0) exp[T(z)/\tau]$, where $T(z)$ is
the look-back time and $\tau$ is the time scale of evolution in units of the
Hubble time.}
is $\tau = 0.26$ and the associated 1~$\sigma$ interval
is [0.16, 1.0]. This evolution is consistent with the earlier estimate
(Padovani and Urry 1992) and is within $1~\sigma$ uncertainty of the
value for the 3CR FR~II galaxies, for which the corresponding number is $\tau
= 0.17$ with 1~$\sigma$ interval of [0.15, 0.19]. If there is significant
contamination of the 3CR NLRG sample with BLRG, as mentioned above, this could
make their observed evolution too high (more like quasars). The two FR~II
luminosity functions, for the 2~Jy and 3CR galaxies, are in good agreement,
although due to the smaller evolution the 2~Jy one is somewhat flatter at
higher powers. For consistency with the FSRQ and SSRQ samples,
we now use the 2~Jy FR~II LF (with $\tau =
0.26$) in the calculations that follow.

The 2~Jy sample includes, as do other radio
samples, a sizeable fraction of compact steep-spectrum (CSS) and gigahertz
peaked-spectrum (GPS) sources ($\sim 20\%$ CSS; Morganti \ea 1993). We exclude
galaxies classified by Morganti \ea (1993) as compact or with unresolved radio
structures, so no CSS or GPS objects should be in our FR~II (or FR~I)
samples. However, since 2~Jy objects with broad optical line emission
and unresolved radio emission are classified as quasars (FSRQ or SSRQ depending
on radio spectral index), some SSRQ could actually be CSS
sources and some FSRQ could actually be GPS sources.
The place of CSS and GPS sources in the unified scheme is discussed
further in \S~\ref{sec:CSS} Even if they should be excluded from the unified
scheme, which is not necessarily the case, they are unlikely to
have ``contaminated'' the quasar sample
because the fraction of such sources is not large; the derived luminosity
functions will be distorted only if there are systematic trends with luminosity
or redshift.

\subsubsection{Observed LFs of High-Luminosity Radio Sources} 
\label{sec:obs_highlf}

Given the 2~Jy sample as defined, we calculate the luminosity functions of
high-power radio sources as follows. First, we
fit an exponential pure luminosity evolution model to each
sample, by finding the evolutionary parameter which makes $\langle V/V_{\rm
max}
\rangle = 0.5$ and then testing the goodness of fit via a KS test.
The best-fit values of $\tau$ for each sample are listed in Table~2, along
with median redshifts and $\langle V/V_{\rm max}\rangle$ values.
Each luminosity function was de-evolved using the appropriate
best-fit evolution. Figure~\ref{fig:rlf_high} shows the resulting local
luminosity functions at 2.7 GHz for FR~IIs, SSRQ, and
FSRQ.\footnote{The FSRQ luminosity function includes PKS~0521$-$365, which was
inadvertently excluded by Padovani and Urry (1992). This is
a low redshift, highly polarized object (Angel and Stockman
1980), classified as a galaxy by Wall and Peacock (1985) and
as a BL~Lac by V\'eron-Cetty and V\'eron (1993) but showing
broad Balmer lines (see \S~\ref{sec:connection}). It was not
included in the 1 Jy BL~Lac sample (Stickel \ea 1991)
because two emission lines exceeded the equivalent width criterion,
but its \OIII~luminosity is more typical of BL~Lacs than of
FSRQ, so it could be a transitional object (see \S~\ref{sec:low_highzbl}).}

The luminosity function of FSRQ LF (filled circles) is flattest and
extends to the highest luminosities.
The luminosity functions of SSRQ (open triangles; Fig.~\ref{fig:rlf_high}) and
FR~II galaxies (open squares) are similar in shape over the common range of
powers, but the SSRQ are lower in number density: the number ratio of the two
classes for $P_{2.7} \simgt 5\times 10^{25}$ W Hz$^{-1}$ is 6.4. This is
larger than in Padovani and Urry (1992), due to the different LF used for the
FR~II galaxies

\subsubsection{Beamed LFs of High-Luminosity Radio Sources} 
\label{sec:pred_highlf}

The beaming calculation is as follows. We start with the derived parent
luminosity function, then calculate the effect of beaming it
(Eqs. \ref{eq:prob} -- \ref{eq:ltot}) adjusting free parameters
$\gamma$ (the Lorentz factor) and $f$ (the fraction of luminosity intrinsic to
the jet) to match the observed luminosity function of FSRQ. It was
not possible to fit the observed FSRQ luminosity function with a single
Lorentz factor; it required instead a distribution in the range $5 \simlt
\gamma \simlt 40$, weighted toward low values: $n(\gamma) \propto
\gamma^{-2.3}$, with a mean value $\langle \gamma \rangle \simeq 11$ and $f
\simeq 5 \times 10^{-3}$.
(Note that this last parameter is fixed by the
largest $\gamma$ and by the maximum value of the ratio between beamed and
unbeamed radio flux, $R$; Eq.~\ref{eq:fraction}.)

Figure~\ref{fig:rlf_high} shows the
beamed (solid line) and observed (filled circles) radio LFs of FSRQ, which
are in very good agreement. The ratio between FSRQ and parents, integrated
over the full luminosity function assuming that of the FR~IIs cuts off at the
low luminosity end, is $\sim 2$\%. (Because of the flat LF slope at low
luminosities, this percentage is not too sensitive to the cutoff.)
The critical angle separating FSRQ from SSRQ and FR~IIs\footnote{There
is actually a range of critical angles, one for each $\gamma$
(Eq.~\ref{eq:thetac}). The larger the Lorentz factor, the smaller the critical
angle;
$\theta_{\rm c} (\gamma_{1}) = 14^\circ$ is the largest angle within
which an FR~II would be identified as an FSRQ.
Using this angle is appropriate for our purposes because
the fitted distribution of Lorentz factors is skewed to low values.}
is $\theta_{\rm c}(\gamma_1) \sim 14^{\circ}$. The fitted parameters of
this beaming model are summarized in Table~3.

According to the beaming hypothesis, the steep-spectrum radio quasars are
supposed to be at intermediate angles and their intrinsic properties --- in
particular, the value(s) of $\gamma$ and $f$ --- must be identical to those of
the FSRQ. Therefore, these parameters are already fixed.
The method of calculation is similar to the one used in the previous section;
however, since SSRQ are supposed to be misaligned objects, $\theta_{\rm min}
\neq 0^{\circ}$. We take $\theta_{\rm min} = 14^{\circ}$, and
the observed value of $R_{\rm min} \sim 0.002$ is
used to constrain $\theta_{\rm max}$ for SSRQ (Eq.~\ref{eq:thetac}).
With no free parameters, then,
we calculate the beamed luminosity function of SSRQ.

The comparison between observations (open triangles) and the
beaming model (dashed line) for SSRQ is shown in
Fig.~\ref{fig:rlf_high}. The agreement is quite good, especially considering
we did not adjust the parameters to optimize the FSRQ and SSRQ fits jointly.
The angle dividing SSRQ from FR~IIs is $\theta \sim 38^{\circ}$ (Table~3).

The same distribution of Lorentz factors produces a good fit to both the
luminosity functions of FSRQ and SSRQ. The value of $\theta \sim 38^{\circ}$
for the angle separating SSRQ from FR~II galaxies is in reasonable agreement
with the angle derived from the number density ratio in the observed range
of overlapping luminosity, $\theta =
\arccos (1 + 1/6.4)^{-1}$ $\simeq 30^{\circ}$.
The latter estimate is valid only when dealing with unbeamed luminosities or
(in
an approximate way) when the effect of beaming is not very strong because the
objects are viewed off-axis, as in this case.
For the 3CR sample, in which beaming is unimportant, Barthel (1989) found
$\theta = 44.\!^\circ 4$ from the ratio of quasars to radio galaxies in the
interval $0.5 < z < 1$.

Table~3 summarizes the beaming parameters for the radio band for different
classes of objects (see \S~\ref{sec:unif_low} for a discussion of the
parameters for the BL~Lac class). Note that we have used $p = 3 +\alpha$ (see
Appendix~B); using instead $p = 2 +\alpha$ results in the Lorentz factors
extending to higher values (Padovani and Urry 1992), and in a slightly larger
critical angle.

{}From the observed values of $R$ we can estimate a lower limit to the maximum
Lorentz factor (Appendix~C). The FR~II galaxy OD~$-$159 is the most
lobe-dominated source in the 2~Jy sample. The most core-dominated FSRQ known
is 0400+258, which does not actually belong to the 2~Jy sample (its 2.7 GHz
flux $\sim 1.5$ Jy); its measured $R$ is comparable to the lower limits (when
no extended emission was detected) for some of the 2~Jy FSRQ. For these two
objects, the $R$-values, $K$-corrected (Eq.~\ref{eq:kcorr}) and extrapolated
(when necessary)
to 2.7 GHz rest frequency assuming $\alpha_{\rm core} - \alpha_{\rm ext} = -1$,
are $R_{\rm min, FR~IIs} < 6 \times 10^{-5}$ (OD~$-$159; Morganti \ea 1993)
and $R_{\rm max, FSRQ} \simeq 1000$ (0400+258; Murphy \ea 1993). Using
Eq.~(\ref{eq:gamma_rmax_rt}),
we find $\gamma_{\rm max} > (1.7 \times 10^7 2^{1-p})^{1/2p} \sim 13$
for $p = 3$.

For these observed $R$ values, small values of $p$ imply high values for the
largest Lorentz factor (here, for $p =2$, $\gamma_{\rm max} \simgt 54$). More
precisely, $\alpha_{\rm r} \simeq -0.3$ for the 2~Jy FSRQ (Padovani and Urry
1992) so $p \simeq 1.7 - 2.7$ if $p$ ranges between $2+\alpha$ and $3+\alpha$.
Then
$\gamma_{\rm max} \simgt 17$ (for $p=2.7$) or $\simgt 120$ (for $p=1.7$). The
need for quite high values of the largest $\gamma$s for smaller values of $p$
was noted by Urry \ea (1991a) and Padovani and Urry (1992) from their fits to
the observed LFs.

While the SSRQ-FSRQ-only scheme (ignoring the radio galaxies; Orr and Browne
1982) can not be ruled out, it is much harder to reconcile with the available
data, mainly because there seem to be too few SSRQ for them to be the parents
of FSRQ (Padovani and Urry 1992). The FR~II-SSRQ-FSRQ scheme, illustrated by
the curves in Fig.~\ref{fig:rlf_high}, can be tested further via the predicted
radio counts of flat- and steep-spectrum quasars, which converge at easily
accessible levels (Padovani and Urry 1992).

\subsection{Unification of BL~Lac Objects and FR~I Galaxies} 
\label{sec:unif_low}

The unification of low-luminosity radio-loud AGN can be tested in the same
way, by comparing the observed number densities of BL~Lac objects and FR~I
radio galaxies. We derive the FR~I luminosity function (Padovani and Urry
1990; Urry \ea 1991a), beam it according to the prescription outlined in
\S~\ref{sec:effect} and fit it to the observed BL~Lac luminosity
functions (Padovani and Urry 1990; Urry \ea 1991a; Celotti \ea 1993). The free
parameters are $\gamma$, the bulk Lorentz factor of the jet, and $f$, the
fractional luminosity of the jet. We further constrain these two
parameters either from $R$, the ratio of beamed to unbeamed flux (that is,
the core to
extended flux in high resolution radio maps), as was done for quasars and
FR~IIs (\S~\ref{sec:unif_high}), or from the ratio of maximum BL~Lac
luminosity to maximum FR~I luminosity. The former approach is used for the
radio comparison (\S~\ref{sec:pop_r}) and the latter approach is used for
the X-ray comparison (\S~\ref{sec:pop_x}) because X-ray spatial resolution is
insufficient to measure $R$. Some separation of beamed and unbeamed
components is now becoming possible for radio galaxies with {\it ROSAT}
(Worrall and Birkinshaw 1994) and in the future, we expect X-ray measurements
of $R$ for large samples of BL~Lacs will be possible with {\it AXAF}.

\subsubsection{X-Ray and Radio Samples of BL~Lac Objects} 
\label{sec:bl_samples}

Sizable well-defined samples of BL~Lac objects are now available. X-ray
surveys have produced a number of samples, including the {\it HEAO-1} A-2
all-sky survey (only 5 BL~Lac objects; Piccinotti \ea 1982), the EMSS (36 BL
Lacs; Stocke \ea 1991), and the {\it EXOSAT} High Galactic Latitude Survey
(HGLS: 11 BL~Lacs; Giommi \ea 1991). The last two go down to fairly low X-ray
fluxes, $\sim 2 \times 10^{-13}$~erg~cm$^{-2}$~s$^{-1}$ in the $0.3 - 3.5$~keV
band. Redshifts are available for most EMSS BL~Lacs.

The 1~Jy sample of BL~Lac objects, a radio-selected sample of size comparable
to the EMSS X-ray-selected sample, has also become available recently (Stickel
\ea 1991). These BL~Lac objects were selected from a catalog of 518 radio
sources with 5 GHz fluxes $\ge 1$ Jy on the basis of their flat radio spectra
($\alpha_{\rm r} \le 0.5$), weak or absent emission lines (rest-frame
equivalent widths $W_\lambda \le 5$ \AA), and optical magnitudes
(brighter than V = 20 mag). Of the 34 1~Jy BL~Lacs in the complete sample, 26
have redshift information from
emission lines (Stickel \ea 1991; Stickel \ea 1994), 4 have lower limits to
their redshifts from absorption lines, while in 4 objects neither emission or
absorption lines have been detected but a lower limit of $z > 0.2$ can be
estimated from their stellar appearance (Stickel \ea 1993).

\subsubsection{Properties of X-Ray-Selected and Radio-Selected BL~Lac Objects}
\label{sec:prop_xr} 

The properties of X-ray-selected and radio-selected BL~Lac objects are
systematically different (Ledden and O'Dell 1985; Stocke \ea 1985; Maraschi
\ea 1986; Padovani 1992a; Laurent-Muehleisen \ea 1993). On average, XBL have
lower polarization, less variability, higher starlight fractions, and are less
luminous and less core-dominated than RBLs
(Stocke \ea 1985; Morris \ea 1991; Perlman and Stocke 1993; Jannuzi \ea 1994).

The two classes occupy different regions on the $\alpha_{\rm ro} - \alpha_{\rm
ox}$ plane, which is indicative of different broad-band energy distributions
(Stocke \ea 1985; Ledden and O'Dell 1985).
This is illustrated in Fig.~\ref{fig:multispec} (Maraschi \ea 1994b),
which shows the multiwavelength spectra of a
radio-selected BL~Lac (PKS~0537$-$441) and an X-ray-selected BL~Lac (Mrk~421).
The overall shape of the RBL spectrum, notably the wavelength of the peak of
the
synchrotron emission and the relative strength of the gamma-ray emission,
is similar to that of the flat-spectrum radio quasar 3C~279
(Fig.~\ref{fig:3c279_spec}), as is true
for RBLs and FSRQ in general (Maraschi \ea 1994b).
The peak of the synchrotron component is typically at
infrared/optical wavelengths for the RBL (and FSRQ) and in the soft X-rays
for the XBL. If the peak wavelength changes smoothly and continuously from
short (ultraviolet/X-ray) to long (infrared/optical) wavelengths for
BL~Lac objects as a whole, the radio/optical/X-ray colors of RBL and XBL
can be reproduced easily (Padovani and Giommi 1995).

The 1~Jy sample of RBL shows a weak {\it
positive} evolution, consistent at the $2~\sigma$ level with no evolution
($\langle V/V_{\rm max} \rangle = 0.60 \pm 0.05 $; Stickel \ea 1991),
while the EMSS XBL
display a {\it negative} evolution, appearing less
abundant and/or less luminous at higher redshifts, ($\langle V_{\rm e}/V_{\rm
a} \rangle = 0.36 \pm 0.05$; Wolter \ea 1994). For luminosity
evolution of the form $L_{\rm x}(z) = L_{\rm x}(0) exp[c_{\rm x}T(z)]$, where
$T(z)$ is the look-back time, the best-fit evolution for the EMSS sample
is $c_{\rm x} = -7.0$, with a $2~\sigma$ range of $-15.9$ to $-1.3$
(Wolter \ea 1994), nearly
consistent (at the $\sim 2~\sigma$ level) with zero evolution.
For the 1~Jy RBL sample, $c_{\rm r} = 3.1$ with a $1~\sigma$
range of 1.7 to 4.2 (Stickel \ea 1991, with $c=1/\tau$), also consistent
with zero evolution at the 2$\sigma$ level.

Both samples are still relatively small (30 and 34 objects
for XBL and RBL, respectively), so it is premature to draw any conclusion
regarding the supposedly different evolutionary behaviors of the two classes.
This is confirmed by two additional results. First, for the fourteen RBL of the
S4 survey ($f_{\rm r} \ge 0.5$~Jy at 5 GHz; Stickel and K\"uhr 1994), $\langle
V/V_{\rm max} \rangle = 0.44 \pm 0.08$ (Padovani and Giommi 1995), more like
the EMSS XBL. Second, for the EMSS,
raising the flux limit only slightly, to $10^{-12}$~erg~cm$^{-2}$~s$^{-1}$,
changes the mean $V_{\rm e}/V_{\rm a}$ from
$0.33\pm0.06$ to $0.48\pm0.06$ (Della Ceca 1993).
These results for RBL and XBL are completely consistent with no
evolution.

\subsubsection{Population Statistics for X-Ray Samples} 
\label{sec:pop_x}

The EMSS and HGLS XBL samples, which went down to fairly low X-ray fluxes,
$\sim 2 \times 10^{-13}$~erg~cm$^{-2}$~s$^{-1}$ in the $0.3 - 3.5$~keV band,
revealed a decided flattening of the counts at low flux, in marked contrast to
the steep X-ray log N - log S curves for other kinds of AGN. Padovani and Urry
(1990) compared predictions of the FR~I unified scheme with the X-ray counts
from these two samples (plus the {\it HEAO-1} A-2 sample) and found good
agreement for a bulk Lorentz factor $\gamma_{\rm x} \sim 3$ and a ratio of
intrinsic jet luminosity to unbeamed luminosity $f \simeq 0.1$. The total
number of BL~Lac objects was $\sim14$\% of the number of FR~I galaxies,
with more
than 90\% of the BL~Lacs at lower X-ray luminosities than currently
observed, and the critical angle separating BL~Lacs from FR~Is was
$\theta_{\rm c}\sim30^\circ$.

Since then, the EMSS-derived X-ray luminosity function of BL~Lac objects has
been published (Morris \ea 1991; Wolter \ea 1994) and now we can fit it
directly. We compare our beaming predictions with the X-ray luminosity
function for the EMSS subsample of 30 BL~Lacs with $f_{\rm x} \ge 2 \times
10^{-13}$ erg cm$^{-2}$ s$^{-1}$, assuming no evolution (which is consistent
at the $\sim 2~\sigma$ level with the observational data; Wolter \ea 1994; see
also \S~\ref{sec:prop_xr}). We follow the method of Avni and Bahcall (1980) to
take into account the fact that the volume surveyed in the EMSS is a function
of limiting flux. Redshifts, fluxes, and sky coverage were taken from Wolter
\ea (1994). Five objects in the sample have no redshift determination; these
were initially excluded from the sample but their presence was taken into
account by multiplying the normalization of the resulting luminosity function
by 30/25.

Figure~\ref{fig:xlf_low} shows the luminosity function (solid line)
from the fitted beaming model of Padovani and Urry (1990) compared to the
observed LF for the EMSS sample for no evolution (filled circles). The
model agrees very well with the data ($\chi^2_{\nu} \simeq 0.3$),
especially considering that the parameters for the model were optimized for
the number counts. If instead we use the anti-evolving X-ray luminosity
function
of Wolter \ea (1994), with $\tau=-0.14$, then by
increasing $R_{\rm max}$ (the ratio between the maximum luminosities of the
parent and beamed populations) from 250 (as in Padovani and Urry 1990) to 1000
we get the dot-dashed line in Fig.~\ref{fig:xlf_low}. The ratio between BL
Lacs and FR~Is increases to about 30\%, $\gamma_{\rm x} \sim 2.9$, $f \sim
1.1$, and $\theta_{\rm c} \sim 47^{\circ}$. This is a much poorer fit to the
data ($\chi^2_{\nu} \simeq 2.5$) but given the uncertainties in the X-ray
luminosity function of FR~Is (see below), one cannot definitely rule out this
case.

The redshift distribution of the EMSS is difficult to reproduce. Despite the
fact that beaming in the zero evolution case fits the observed LF better, it
cannot reproduce the peak in $N(z)$ at $z \sim 0.2 - 0.3$, whereas our fit to
the anti-evolving LF does. Browne and March\~a (1993) have suggested that
recognition problems affecting low-luminosity BL~Lacs whose light is swamped
by the host galaxy, typically a bright elliptical (\S~\ref{sec:iso_low}),
might explain the unusual redshift distribution and anti-evolution of the
EMSS XBL. Padovani and Giommi (1995) have shown through numerical
simulations, however, that although recognition problems have some effect,
they are probably not strong enough to affect these properties so drastically.

A complication in applying beaming models to the X-ray band stems from the
uncertainties in the X-ray luminosity function of FR~Is. These are due to its
bivariate nature, small statistics, and the non-detection in X-rays of $\sim
30\%$ of the complete radio-selected sample (which was already small).
{\it ROSAT} will eventually
produce an X-ray selected sample of FR~Is, or at least complete X-ray data on
a sizeable radio-selected sample of these objects. Already one interesting
{\it ROSAT} result is the identification of resolved (thermal) and unresolved
X-ray emission in radio galaxies, as assumed by Padovani and Urry (1990). The
beaming model discussed in this section predicts ratios of intrinsic-jet to
extended X-ray flux in the range $0.1 - 1$, marginally inconsistent with the
observed ratios of nonthermal to thermal emission for six low-luminosity radio
galaxies of $1 - 5$ (Worrall and Birkinshaw 1994). The disagreement is even
worse because in radio galaxies the jets should be de-amplified (for $\gamma
\sim 3$, $\delta < 1$ for $\theta > 45^{\circ}$; Appendix~A). Further
high-resolution X-ray observations will put more constraints on the
$f$-parameter.

The XBL samples are still relatively small. Larger XBL samples are now being
optically identified, including the {\it HEAO-1} LASS sample (Schwartz \ea
1989; Laurent-Muehleisen \ea 1993) and the {\it Einstein} Slew Survey Sample
(Elvis \ea 1992; Schachter \ea 1993; Perlman \ea 1995). In addition, the {\it
ROSAT} All-Sky Survey should produce large numbers of new BL~Lac objects ---
our beaming model for the X-ray samples predicts roughly two objects
per square degree down to $F_{\rm x} = 10^{-14}$~erg~cm$^{-2}$~s$^{-1}$ in the
$0.3 - 3.5$~keV band --- some of which have already been identified (Bade \ea
1994). With these new data we will be able to address the beaming hypothesis
with better statistics.

\subsubsection{Population Statistics for Radio Samples} 
\label{sec:pop_r}

The radio data are better suited than the X-ray data for testing the beaming
hypothesis for BL~Lac objects. One reason is that the radio sample of FR~I
galaxies is larger than the X-ray sample. In addition, the higher spatial
resolution available with radio interferometry allows estimates of the ratio
between beamed and unbeamed radio flux (Antonucci and Ulvestad 1985), thus
adding an important constraint to the combination of the parameters $\gamma$
and $f$ (Appendix~C).

The radio luminosity function of FR~I radio galaxies was derived by Urry \ea
(1991a) from the 2~Jy catalog (and converted to 5 GHz) and then extended to
lower radio powers using the radio LF of elliptical galaxies of Franceschini
\ea (1988). Here we choose {\it not} to include the low-luminosity radio
ellipticals but restrict ourselves only to sources classified explicitly as
FR~Is (some of the lowest luminosity ellipticals could in fact
have radio emission dominated by thermal emission rather than the nonthermal
BL~Lac nucleus; Phillips \ea 1986). Our results, however, are basically
unchanged, since the beamed LF
for BL~Lacs at $P_{\rm 5} \simgt 5\times 10^{25}$ W Hz$^{-1}$ (i.e., where it
overlaps with the observed LF) is largely unaffected by the behavior of the
FR~I LF at such low powers. The new optical identifications (di Serego
Alighieri \ea 1994b) and radio maps (Morganti \ea 1993) also have little effect
on the FR~I luminosity function;
the fit to the new LF used here is consistent with the one
in Urry \ea (1991a). The evolutionary properties of FR~I radio galaxies are
consistent with no evolution, with $\langle V/V_{\rm max} \rangle =
0.42\pm0.05$.

We can compare our FR~I luminosity function, derived from the 2~Jy sample
(selected at 2.7~GHz), with that of de Ruiter \ea (1990), based on the B2
sample (selected at 408~MHz) plus nearby ($z < 0.2$) 3CR radio galaxies
(selected at 178~MHz). Converting our FR~I LF to $H_0 = 100$ km s$^{-1}$
Mpc$^{-1}$ and $\nu = 408$ MHz (assuming $\alpha_{\rm r} = 0.7$), we find
excellent agreement from $P_{408} \sim 6\times 10^{23}$~W~Hz$^{-1}$ up to the
break at $P_{408} \sim 3 \times 10^{25}$~W~Hz$^{-1}$ (see Fig.~13 of de Ruiter
\ea 1990).
Above the break our LF is steeper, not surprisingly, since they did not
exclude FR~II sources. The two lowest luminosity bins in our LF
($P_5 \simlt 2\times 10^{23}$~W~Hz$^{-1}$), lie a factor of 2--3
above the de Ruiter \ea LF. This disagreement is not surprising, as our two
points are highly uncertain, with only one object in each bin. Also, the two
objects, M82\footnote{Although M82 is classified as an FR~I galaxy, its
strong starburst component makes it an unlikely BL~Lac parent; however, its
inclusion does not change our results significantly.}
 and M84, are nearby ($z = 0.0014$ and 0.0028, respectively) and
so sample the local overdensity, whereas the B2 sample, being deeper than the
2~Jy, averages over a larger volume of space.
(The equivalent flux limit of the B2 survey translated to 2.7~GHz with
$\alpha=0.7$ is 130~mJy.) In any case,
excluding the first two bins does not alter our fitted beamed LF significantly.

Using a two-power-law approximation to our FR~I luminosity function,
we fitted a beamed LF to the observed LF for the 1~Jy BL~Lacs. The latter was
obtained as described in Stickel \ea (1991) with the addition of S5 0454+844,
which had no redshift at the time. It is impossible to fit the data with a
single Lorentz factor; instead, an acceptable fit to the LF and the observed
$R$-values is obtained for Lorentz factors distributed in the range
$5\simlt\gamma_{\rm r}\simlt32$. While the form of the distribution is not
well constrained, it is weighted toward lower values; e.g., for a power law
of the form $N(\gamma_{\rm r}) \propto {\gamma_{\rm r}}^G$, the best-fit index
is $G\sim-4$.

Figure~\ref{fig:rlf_low} shows the excellent agreement between beamed
(solid line) and observed (filled circles) radio luminosity functions for the
1~Jy sample of BL~Lacs. The mean Lorentz factor, which is approximately
independent of the exact shape of the distribution, is $\langle \gamma_{\rm r}
\rangle \sim 7$, corresponding to a ratio between BL~Lacs and FR~Is of $\sim
1:50$, about an order of magnitude smaller than in the X-ray case
(\S~\ref{sec:pop_x}). The total number of BL~Lacs is most sensitive to the
lowest value of $\gamma_{\rm r}$, while the maximum ratio of beamed to unbeamed
flux is sensitive to the highest value. Our fitted model predicts that
radio-selected BL~Lacs are aligned within $\theta_{\rm c} \sim 12^{\circ}$ of
the line of sight. The fitted beaming parameters for radio-selected BL~Lacs
are summarized in Table~3.

As we did for quasars, we can estimate a lower limit to the maximum Lorentz
factor from the observed maximum and minimum values of $R$ for BL~Lacs and
FR~Is, respectively (Appendix~C). OJ~287 is the most
core-dominated BL~Lac in the 2~Jy sample (and in the 1~Jy sample),
excluding those cases where no extended emission has
been detected, and Fornax~A is the most lobe-dominated FR~I. The corresponding
values of $R$, $K$-corrected (Eq.~\ref{eq:kcorr}) and extrapolated (when
necessary) to 5~GHz
rest frequency assuming $\alpha_{\rm core} - \alpha_{\rm ext} = -1$, are
$R_{\rm min, FR~I} \simeq 4 \times 10^{-4}$ (Fornax~A; Morganti \ea 1993) and
$R_{\rm max, BL~Lac} \simeq 780$ (OJ~287; Kollgaard \ea 1992). Using
Eq.~(\ref{eq:gamma_rmax_rt})
as before, $\gamma_{\rm max} \simgt (2.0 \times 10^6 2^{1-p})^{1/2p} \sim 9$
for $p = 3$.

As was the case for the quasars, small values of $p$ imply high values for the
largest Lorentz factor (for $p = 2$, $\gamma_{\rm max} \simgt 30$). More
precisely, since $\alpha_{\rm r} \simeq -0.1$ for the 1~Jy BL~Lacs (Stickel \ea
1991), $p \simeq 1.9 - 2.9$ for $p$ between $2+\alpha$ and $3+\alpha$. Then
$\gamma_{\rm max} \simgt 10$ (for $p=2.9$) or $\simgt 38$ (for $p=1.9$).

The value of $H_0$ does affect the luminosity functions but has no direct
effect on the derived beaming parameters. The fitted Lorentz factor, which is
inversely proportional to the normalization of the beamed luminosity function,
depends only on the ratio between the total numbers of objects (the integrals
of the parent and beamed LFs). $H_0$ enters only because we constrain
$\gamma$ to some extent with the observed values of superluminal motion (for
BL~Lacs and for quasars separately; see \S~\ref{sec:independent}). A higher
$H_0$ would mean lower superluminal velocities, and in general, fits with
lower $\gamma_1$ are possible. We did not explore parameter space
exhaustively but we show an example of a low-$\gamma$ fit for RBL in Table~3.

It is interesting that the luminosity functions of FR~I and FR~II
galaxies overlap smoothly, as do those of BL~Lac objects and FSRQ
(Figs. \ref{fig:rlf_high},~\ref{fig:rlf_low}). Possibly they represent
different manifestations of the same basic central engine, in which case
the different radio morphologies and emission line strengths would have to
be closely linked to observed radio power (Maraschi and Rovetti 1994).

\subsubsection{Relation of X-Ray- and Radio-Selected BL~Lac Objects}
\label{sec:rel_xr}

The difference between XBL and RBL is surely a fundamental issue in
understanding the BL~Lac phenomenon. Our beaming results
indicate that radio-selected BL~Lac objects constitute a smaller fraction of
FR~I radio galaxies than do X-ray-selected BL~Lacs (Table~3). This can lead to
a higher fitted Lorentz factor in the radio calculation, since $\gamma$ is
inversely proportional to the normalization of the beamed luminosity
function. The values from our published calculations are
$\langle \gamma_{\rm r} \rangle \sim 7$ (Urry et al. 1991a) and $\gamma_{\rm
x} \sim 3$ (Padovani and Urry 1990), in effect suggesting the radio emission
is more collimated than the X-ray.

This result was exciting because it suggested XBL and RBL might represent
slightly different orientations of the same underlying relativistic jet.
Problems with this interpretation have recently emerged, as discussed below,
but first we summarize the original argument. The apparent difference
in Lorentz factors was interpreted (Urry \ea 1991a) in terms of an accelerating
jet model (suggested for other reasons by Ghisellini and Maraschi 1989)
in which the X-ray emission from the most compact region would have
a smaller $\gamma$ than the more extended radio-emitting region. In addition
to explaining the apparent difference in number densities of XBL and RBL,
the accelerating jet explained naturally the lower variability, polarization,
and luminosities of XBL (\S~\ref{sec:prop_xr}, and references therein).

The original number density argument was actually independent of the Lorentz
factors obtained from beaming calculations. Because XBL and RBL have similar
X-ray luminosities, X-ray selection should be relatively unbiased (Maraschi
\ea 1986). Since X-ray surveys detected very few known RBL, they must be
relatively rare, exactly as expected if the radio emission is more beamed than
the X-ray. This was supported by the X-ray number counts of RBL and XBL (Urry
\ea 1991a). If, as our beaming results implied, RBL are viewed within $\sim
12^{\circ}$ of the jet axis while XBL are between $\sim 12^{\circ}$ and
$30^{\circ}$, then RBL are a factor of $\sim7$ less numerous than XBL
(Table~3).

An alternative explanation of the same data was that the physical collimation
itself increased along the jet, so that the narrower radio beams made it less
likely that our line of sight would intercept their emission. This was
commensurate with the X-ray and radio luminosity functions of RBL and XBL
(Celotti \ea 1993). For this kind of fan beam, the luminosity function
calculation (Celotti \ea 1993)
implied a Lorentz factor that was quite high, $\gamma = 29$ (with
$\theta_{\rm c,r} = 2/\gamma \simeq 4^{\circ}$, $\theta_{\rm c,x} =
13^{\circ}$, $f_{\rm r} = 6 \times 10^{-3}$ and $f_{\rm x} = 5 \times
10^{-3}$, assuming no evolution for both XBL and RBL) but it could be reduced
if, for example, only 10\% of FR~Is were parents of BL~Lacs ($\gamma = 10$,
$\theta_{\rm c,r} \simeq 11^{\circ}$, $\theta_{\rm c,x} = 41^{\circ}$, $f_{\rm
r} = 0.2$ and $f_{\rm x} = 0.5$).

This simple and satisfying picture based on jet collimation is now in doubt.
Most significantly, the accelerating jet and the fan-beam jet are both
contradicted by analysis of the multiwavelength spectra of complete
samples of XBL and RBL (Sambruna 1994; \S~\ref{sec:multi}).
In addition, the fitted $\gamma$ for RBL can actually be pushed to lower
values, reducing the apparent difference between XBL and RBL
(\S~\ref{sec:independent} and Table~3;
but note that the fraction of beamed objects still remains smaller for
RBL than for XBL). Finally, it is important to remember that the X-ray
unification calculation is more poorly constrained than the radio calculation
because the FR~I samples contain fewer objects with X-ray data and
because of the less extensive imaging information; thus interpreting
any difference in fitted Lorentz factors was premature.

\subsubsection{New Terminology and a New Connection Between BL~Lac Classes}
\label{sec:terms}

There is now a viable interpretation of the observational data that is
unrelated to jet collimation (Giommi and Padovani 1994; Padovani and Giommi
1995). It is based on a single population of BL~Lac objects characterized by a
wide range of multiwavelength spectral shapes, with bolometric luminosities
peaking at infrared/optical wavelengths for most RBL and at ultraviolet/X-ray
for most XBL, as is observed (Giommi et al. 1995; Sambruna 1994). Before
discussing this scenario further, we discuss new terminology that
distinguishes clearly between the selection method used to find BL~Lacs
and the spectral characteristics of the sources themselves.

The new terms are motivated by the fact that the division of
BL~Lac objects into RBL and XBL can get garbled or confused.
Strictly speaking, those terms are based on selection band rather
than intrinsic physical properties, in which case some BL~Lac objects
already qualify as both RBL and XBL (e.g., Mrk~501),
and many more ``double agents'' will appear
as deeper radio and X-ray samples become available (Elvis \ea 1992).
This leads to awkward terminology such as ``XBL-like RBL'' and
``RBL-like XBL'' when discussing
the strong differences between the broad-band spectral properties
of RBL and XBL in the historical samples. Since we wish to explore
these differences --- notably the peak luminosities in the infrared/optical
or ultraviolet/X-ray, respectively ---
it is useful to define categories in terms of the ratio of
X-ray to radio flux (Ledden and O'Dell 1985).

Padovani and Giommi (1995) have suggested dividing RBL-like and XBL-like
objects into LBL and HBL, respectively (for ``low-energy cutoff BL~Lacs'' and
``high-energy cutoff BL~Lacs''),
according to whether $\alpha_{\rm rx}$ (between 5~GHz and 1~keV) is greater
(LBL) than or less (HBL) than 0.75.
These names reflect the actual spectral characteristics of the two types
of BL~Lacs, allowing the terms ``XBL'' and ``RBL'' to be reserved strictly
for sample membership (which is well-defined).
BL~Lac objects like Mrk~501 or OJ~287 which appear in both radio- and
X-ray-selected samples can be uniquely categorized as HBL or LBL,
respectively. Most XBL are HBL (OJ 287 is an exception) and most RBL are LBL
(Mrk 501 is an exception).
The lack of precision of ``high'' and ``low'' reflects the possibility that
the synchrotron peak in BL~Lacs occurs at a wide range of wavelengths, perhaps
not fully probed in current samples, from infrared through X-ray.

The essence of the argument of Padovani and Giommi (1995) is that
X-ray selection favors objects with a peak at ultraviolet/X-ray wavelengths
and thus finds fewer with peak at infrared/optical wavelengths.
If radio rather than X-ray surveys are unbiased (because the radio
emission does not ``know'' the wavelength of the peak luminosity)
then HBL are
relatively rare, about 10\% in the 1~Jy plus S4 plus S5 radio-selected samples
(Padovani and Giommi 1995). In effect, Giommi and Padovani take the opposite
approach from Maraschi \ea (1986), assuming radio selection rather than
X-ray selection is unbiased, and they find the opposite result: in complete
contrast to the accelerating jet picture, they conclude that HBL constitute a
minority of the BL~Lac population.

Specifically, Giommi and Padovani (1994) argue that HBL outnumber LBL at a
given X-ray flux, even though they are
intrinsically less numerous, because the two classes sample different
parts of the BL~Lac radio counts (Giommi and Padovani 1994). As a consequence
of their higher $f_{\rm x}/f_{\rm r}$ ratios, HBL have lower radio fluxes
($\sim 10$ mJy) and since fainter objects are more numerous than brighter ones
(the radio counts are rising), their surface density is higher. Stated
differently, X-ray surveys sample the BL~Lac radio counts at low fluxes and
mostly detect the $\sim 10\%$ of objects with high $f_{\rm x}/f_{\rm r}$
ratios. This holds down to quite faint X-ray fluxes, well below the {\it
ROSAT} deep survey limit, below which the fraction of LBL should increase
slowly and eventually dominate by a factor of 10 (Padovani and Giommi 1995).
The Giommi and Padovani hypothesis explains most of the properties of HBL ---
e.g., X-ray luminosity function, X-ray number counts, and radio flux
distribution --- using only the observed properties of RBL (and no free
parameters).

There are at present insufficient data to determine whether the number ratio
of HBL to LBL (integrated over all luminosities) is $\sim 0.1$,
as Giommi and Padovani conclude, or roughly the opposite!
One persistent objection to the Giommi and Padovani picture is
that the polarization characteristics of LBL
and HBL differ in a way not obviously explained by a change in break
frequency. Specifically, the LBL have higher polarization which varies
more in both degree and position angle. According to Giommi and Padovani,
the HBL are intrinsically less luminous --- less extreme --- so they should
have lower polarization but why the polarization angle is more stable is
not obvious in their picture.

The beaming picture, comparing BL~Lacs and FR~I LFs, suggests that HBL are
more numerous than
LBL by at least a factor of 3. It does seem intuitive that more luminous
objects are more rare, i.e., that luminosity functions rise to low
luminosities (which they do for all known galaxies, active or not). Thus one
could ask how the bolometric luminosity functions of LBL and HBL compare;
i.e., what the relative number densities are, at least in the range of
overlapping luminosities. Using data from Giommi \ea (1995), we formed a close
approximation to the bolometric LF using the luminosities at the peak of the
emission in $\nu L_{\nu}$. Figure~\ref{fig:bol_lf} shows this
pseudo-bolometric LF for the 1~Jy LBL (excluding the two HBL in the 1~Jy RBL
sample) and the EMSS~HBL (this sample includes no LBL).
{\it This is a bivariate LF
so it does not compensate for the selection effects
inherent in the 1~Jy and EMSS samples.}
The HBL appear to be more numerous in the range of overlap with number
densities
systematically above the LBL points.
We note that the EMSS is effectively much deeper than the 1~Jy
survey (its equivalent radio flux limit is $\sim 1$~mJy; Padovani and Giommi
1995). We also note the uncertainties are very large. This kind of figure will
be very useful to re-derive when larger samples of LBL and HBL are available.

Clearly one would like to select samples in an unbiased way. Somewhere between
the wavelengths where LBL and HBL have their spectral peaks, they must have
roughly comparable fluxes. The optical has to be less biased than the X-ray or
radio band, and indeed the optically brightest BL~Lacs in the 1~Jy and Slew
Survey samples have comparable optical magnitudes (Stickel \ea 1991; Perlman
\ea 1995). The relative numbers of LBL and HBL in an optically-selected
sample, then, should reflect their relative numbers globally, at least better
than radio or X-ray surveys do. Unfortunately, the best optically-selected
sample, the complete PG sample, has only 6 BL~Lac objects (Green \ea 1986;
Fleming \ea 1993). Of these, 1/3 are LBL and 2/3 are HBL, but
the statistics are obviously poor. While formally there are more HBL than LBL
in the PG sample,
the observed ratio actually agrees well with what the Giommi and Padovani
scenario would predict in the optical band (Padovani and Giommi, in
preparation). Larger optical
samples of BL Lac objects are important for addressing the fundamental
differences between LBL and HBL.

\subsection{Independent Estimates of Relativistic Beaming Parameters}
\label{sec:independent}

Independent estimates of Lorentz factors and angles to the line of sight are
available, mostly from VLBI data. In VLBI observations of twelve FSRQ
in the 2~Jy sample (Vermeulen and Cohen 1994, and
references therein), all but one FSRQ show evidence of superluminal motion
($\beta_{\rm a}> 1$). The apparent separation of radio components gives a
lower limit to the Lorentz factor, $\gamma_{\rm min}=\sqrt{{\beta_{\rm
a}}^2+1}$
(Appendix~A); the distribution of $\gamma_{\rm min}$ for the fastest
superluminal components in FSRQ covers the range 5 -- 35, roughly the same as
that necessary to explain the observed luminosity function
(\S~\ref{sec:unif_high}).

The observed $\beta_{\rm a}$ distribution can in principle be compared
with that predicted using the LF-derived beaming parameters.
Vermeulen and Cohen (1994) determined that a distribution of Lorentz factors
peaked at high values best fit the observed $\beta_{\rm a}$ distribution
for a small heterogeneous sample defined by a {\it post-facto} flux limit.
(They calculated the expected distribution of $\beta_{\rm a}$ for naked
jets without extended emission, which is appropriate to their sample of
core-dominated objects and which is straightforward to calculate.)
This is quite different from the steep distribution of Lorentz factors
we found from fitting luminosity functions (\S~\ref{sec:pred_highlf}).
Whether these two results can be reconciled remains to be seen but
we note that our calculation is most sensitive to low values
of $\gamma$, which yield the largest number of beamed objects,
while superluminal motion statistics (particularly for samples of
``favorite objects'' studied in the past) are most sensitive to high values.
Alternatively, one can compare the results from
the recent VLBI surveys of well-defined flux limited samples
(Pearson and Readhead 1988; Polatidis \ea 1995; Thakkar \ea 1995;
Taylor \ea 1994)
to the $\beta_{\rm a}$ distribution predicted by a more-difficult
calculation incorporating selection effects appropriate to jets plus
extended emission, but this calculation has not yet been done.

As is the case for emission-line blazars, all thirteen 1 Jy BL~Lac objects
observed more than once with
VLBI techniques appear to show superluminal motion (Vermeulen and Cohen 1994;
Gabuzda \ea 1994). For all but two objects, $\gamma_{\rm min}$ is higher than
$\sim4$, and reaches at least $\sim15$. These results agree
roughly with the distribution in $\gamma_{\rm r}$ required to fit the
luminosity function and the core-halo ratios of radio-selected BL~Lacs
(\S~\ref{sec:unif_low}).

Since two BL~Lac objects have $\gamma_{\rm min}$ as low as 1.5, we tried a fit
to the radio LF of BL~Lacs with $\gamma_1 = 2$, leaving all other parameters
unchanged. The resulting beamed LF is still in good agreement with the
observed one, although the Lorentz factors are now lower, $2 \simlt \gamma_{\rm
r} \simlt 20$ with an average value $\langle \gamma_{\rm r} \rangle \sim 3$,
corresponding to a ratio between BL~Lacs and FR~Is of 5\% (Table~3). This
shows that a scenario where the Lorentz factors are lower for BL~Lacs than for
FSRQ (Gabuzda \ea 1994; Morganti \ea 1995), although not required by our fits,
is consistent with the luminosity function and VLBI data.

The SSC formalism (\S~\ref{sec:SSC}) can be used to estimate the Doppler
factor independent of superluminal motion or luminosity functions, via
comparison of the predicted and
observed X-ray fluxes. For a sample of $\sim 100$ radio sources with published
VLBI measurements of the core angular size, lower limits to the Doppler factor
correlate well with the apparent velocity $\beta_{\rm a}$ obtained from
multi-epoch VLBI maps (Ghisellini \ea 1993; \S~\ref{sec:superluminal}). This
result suggests superluminal motion is related to bulk motion of the emitting
plasma and is not simply an illusion. The Doppler factor, $\delta$, can be
combined with the measured superluminal speed, $\beta_{\rm a}$, where
available, to constrain $\gamma$ and the angle to the line of sight, $\theta$
(Eqs.~\ref{eq:gamma_bd} and \ref{eq:tan_theta});
the mean values of these parameters for the different classes
(BL~Lacs, FSRQ, and SSRQ; Ghisellini \ea 1993) are in good agreement with
those derived from the radio luminosity function studies
(\S\S~\ref{sec:unif_high}, \ref{sec:unif_low}).

Similarly, the jet velocities and orientations of radio galaxies, derived from
the jet to counter-jet brightness ratio (Eq.~\ref{eq:beta_cos_theta}
) and the SSC formalism,
are broadly consistent with those derived from luminosity function studies
(Giovannini \ea 1994). Their Doppler factors are also much lower than for
BL~Lacs and FSRQ, as expected if the radio jets are closer to the plane
of the sky.

\section{Relation of Quasars and BL~Lac Objects} 
\label{sec:q_bl}

\subsection{Low- and High-Redshift BL~Lac Objects}
\label{sec:low_highzbl}

The high-redshift members of any flux-limited sample tend to be systematically
more luminous than the low-redshift members because of the induced
correlation of luminosity and redshift. Among BL~Lac objects, such differences
have been interpreted as evidence for a ``true'' (Type~0) BL~Lac class at low
redshift and a ``quasar-like'' (Type~1) BL~Lac class at high redshift (Burbidge
and Hewitt 1989).
A particular motivation for this division (Antonucci 1993) is the fact that
broad emission lines, occasionally with large equivalent widths, have been seen
in some high-redshift but not
low-redshift BL~Lac objects. (Here ``high-redshift'' essentially means the
distant half of the
1~Jy RBLs, which extend to $z\sim1$; few members of
any complete XBL sample have $z\simgt 0.5$. In this section we review whether
existing data support and/or require a bifurcation via redshift for BL~Lacs.

One of the defining features of BL~Lac objects is their weak or absent
emission lines. For the complete 1~Jy radio-selected sample (Stickel \ea
1991), BL~Lacs are defined in part by rest-frame equivalent widths of optical
emission lines less than 5~\AA.
The EMSS and Slew Surveys are defined in a similar way,
although using the observed equivalent widths (Stocke \ea 1991; Perlman \ea
1995). With this equivalent width limit, there can in principle be cross-over
objects --- for example, for fixed emission-line luminosities, the equivalent
widths can change when the continuum varies.

Emission line types and strengths are central to the question of whether
nearby and distant BL~Lacs constitute distinct populations. The nearby
BL~Lacs, which are associated with nearby ellipticals (typically at $z \simlt
0.2$) with typical galaxy absorption spectra, sometimes have weak narrow
emission lines (typically \OIII), while the more distant BL~Lacs, around which
host galaxies have never been detected, sometimes have weak but broad emission
lines (typically \MGII; Stickel \ea 1991, 1993). Observational limitations
(and sometimes habits), however, militate against uniform information on
emission line properties across the full redshift range. In particular, to
detect broad \MGII~in the nearby BL~Lacs requires high signal-to-noise
ultraviolet spectra, and to detect \OIII~in the distant BL~Lacs requires
fairly red spectra.

With available data we first ask whether BL~Lac objects have line luminosities
comparable to or clearly distinct from FSRQ, for samples selected in the same
way (therefore, the 1~Jy BL~Lacs and the 2~Jy FSRQ, both selected at high
radio frequencies). Figure~\ref{fig:oiii} shows \OIII~line luminosities as a
function of redshift, which removes observational selection effects due to
spectral bandpass and at the same time effectively matches objects for
luminosity in the selection band (which should be unbiased with respect to the
optical line emission). We found \OIII~data in the literature for all the 2~Jy
FSRQ with $z < 0.7$ and for eight of the twelve 1~Jy BL~Lacs sources with
certain $z <
0.5$; the other four BL~Lac objects have much weaker \OIII~lines (Stickel \ea
1993).

Figure~\ref{fig:oiii} shows that the \OIII~luminosities of BL~Lac objects are
systematically lower than those of FSRQ. Inclusion of upper limits for BL~Lacs,
were they available, would exacerbate the difference. There are not enough
data points, however, to determine whether the two classes actually have
distinct narrow-emission-line properties.

The one BL~Lac with quasar-like \OIII~luminosity
is PKS~0521$-$365, a nearby low-luminosity
object which had been previously classified as a BL~Lac (V\'eron-Cetty and
V\'eron 1993) but which was not included in the 1~Jy BL~Lac sample (Stickel
\ea 1991) because the equivalent width of two lines was larger than the 5
\AA~limit (specifically, $W_{\rm H\beta} \simeq 9$~\AA\ and
$W_{\rm [O~{\sevenrm II}]} \simeq 7$~\AA;
this object also has broad H$\alpha$; Ulrich 1981; Stickel and K\"uhr
1993). Figure~\ref{fig:oiii} suggests that PKS~0521$-$365 has BL~Lac-like line
luminosity but qualifies as a quasar because its continuum emission was
unusually faint (exactly as found earlier by Ulrich 1981) when Stickel
\ea classified it.

Broad emission lines are perhaps more crucial to the comparison of
high-redshift BL~Lac objects and quasars. We collected from the literature all
available \MGII~$\lambda2798$ fluxes for the 2~Jy FSRQ and the 1~Jy BL~Lacs,
plotted in Fig.~\ref{fig:mgii} versus redshift to facilitate comparisons
at similar luminosities and observed wavelengths.
Roughly 35\% of the 2~Jy FSRQ and 30\% of the
1~Jy BL~Lacs are represented in the Figure, including six of the seven BL~Lac
objects with $z > 0.7$. Additional data for the FSRQ would be relatively easy
to obtain, but for the BL~Lacs, which by definition have very weak spectral
features (in fact $\sim$1/3 do not have a firm redshift determination),
spectra with higher signal-to-noise ratios are needed. Because of the
low mean redshift of BL~Lac samples, the \MGII~line usually lies in the
ultraviolet, accessible at the required signal-to-noise only with {\it HST}.

In contrast to the \OIII~plot, Fig.~\ref{fig:mgii} shows that a few 1~Jy
BL~Lacs have \MGII~luminosities as high as those of the weaker-lined FSRQ.
In particular, two BL~Lacs have $L_{\rm \MGII} > 3 \times 10^{43}$~erg~s$^{-1}$
(the luminosity for the faintest FSRQ), namely PKS~0537$-$441 and
B2~1308+326, both at $z \simgt 0.7$. The equivalent width of \MGII~in
PKS~0537$-$441 was on at least one occasion larger than 5 \AA\ (Wilkes 1986),
though not when the 1~Jy sample was being defined. This object has also been
suggested as a lensing candidate, meaning it might actually be a quasar,
although recent observations do not confirm this (see \S~\ref{sec:lensing}).
B2~1308+326 may have an FR~II radio morphology (Kollgaard \ea 1992), while
its VLBI polarization properties seem to be more similar to those of quasars
than of BL~Lacs (Gabuzda \ea 1993).

Apart from these two cases, the
\MGII~luminosities of BL~Lac objects are clearly lower than for FSRQ across
the entire redshift range sampled.
Still, we can not exclude a continuous distribution of line luminosities
from BL~Lacs to quasars, which could result simply from the definition of
BL~Lacs in
terms of their low equivalent widths. In particular, the radio selection in
the 1~Jy and 2~Jy samples is similar, so the radio luminosity distributions of
these BL~Lacs and FSRQ at a given redshift must be similar, suggesting their
total luminosities may be comparable. Division in terms of equivalent width
would then translate into the observed division in line luminosity.
It remains to be seen if this explanation is viable quantitatively
when more line luminosities become available.

The apparent difference between the evolutionary properties of radio-selected
BL~Lacs ($V/V_{\rm max} = 0.60 \pm 0.05$) and X-ray-selected BL~Lacs ($V_{\rm
e}/V_{\rm a} = 0.36 \pm 0.05$) has been ascribed to contamination of the RBL
sample from
strongly evolving FSRQ (Morris \ea 1991). The comparison of line luminosities
in Figs.~\ref{fig:oiii} and \ref{fig:mgii} shows that RBL are clearly not
quasar-like for $z \simlt 0.5$. If we divide the 24 1~Jy BL~Lac objects having
certain redshift information at $z=0.5$, there is no evidence that the
$V/V_{\rm max}$ distributions or mean values for the low- and high-redshift
subsamples are different, as shown by a KS and Student's t-test, respectively.
More simply, there is no significant correlation between $V/V_{\rm max}$ and
redshift in the 1~Jy sample. For the five objects with FR~II or FR~I/II
morphology (Kollgaard \ea 1992), which might be considered the most
quasar-like, $\langle V/V_{\rm max} \rangle$ is also not significantly
different
from the $\langle V/V_{\rm max} \rangle$ for the rest of the sample. In short,
the evolutionary properties of BL~Lac objects do not depend on redshift and do
not indicate contamination by strongly evolving quasars.

A number of other results suggest that low- and high-redshift BL~Lacs are more
alike than the latter are like FSRQ. In particular:
\begin{itemize}
\item
The distributions of extended radio power for low- and high-redshift BL~Lacs
are not significantly different, while those of high-redshift BL~Lacs (the
1~Jy with $z>0.3$; the largest measured redshift is $z=1.048$; Stickel \ea
1991) and FSRQ (the 2~Jy sample from Wall and Peacock 1985) differ at the
99.9\% confidence level (Padovani 1992b).
\item
The ratio between compact and extended radio emission is not significantly
different for low- and high-redshift BL~Lac objects. This is in spite of the
possibility that some high-redshift BL~Lacs are FR~IIs with low-excitation
optical spectra (\S~\ref{sec:iso_low}), and therefore might have
systematically different radio properties.
\item
The X-ray spectral shapes of quasars and of BL~Lac objects observed
with the {\it
Einstein} Observatory (Worrall and Wilkes 1990) and with {\it ROSAT} (Urry \ea
1995) are systematically different, with the FSRQ having flatter spectral
indices. Even BL~Lacs with faint but broad emission lines have X-ray spectra
more similar to the other BL~Lacs than to the FSRQ (Worrall and Wilkes 1990),
although those with the broad lines tend to have flatter-than-average
(i.e., more FSRQ-like) X-ray spectra
(Urry \ea 1995).
\item
The characteristic VLBI polarization structure of BL~Lac objects, which
implies a magnetic field perpendicular to the jet axis, is quite different
from that of quasars, which have magnetic field parallel to the jet (Gabuzda
\ea 1992), with one possible exception (the 1~Jy BL~Lac B2~1308+326; Gabuzda
\ea 1993). Furthermore, the milliarcsecond polarization structures of low- and
high-redshift BL~Lacs, divided at $z=0.3$ as suggested by Burbidge and Hewitt
(1989), are indistinguishable. It is perhaps possible to create the distinction
between quasars and BL~Lacs if the latter are systematically more aligned
with the line of sight because of the sensitive dependence of polarization on
aspect at small angles (Gopal-Krishna and Wiita 1993).
\end{itemize}

We have asked two simple questions: (1) whether there is any evidence
requiring separate high- and low-redshift BL Lac
populations, and (2) whether the
evidence shows that high-redshift BL~Lacs are similar to quasars. We believe
that, based on presently available data, the answer to both questions is
``no.'' BL~Lacs selected on the basis of their equivalent width represent a
fairly homogeneous class, with no strong differences between objects with
redshift below and above $z\sim 0.3$, $z\sim 0.5$ or other, or between objects
with broad and with narrow emission lines. There may be a few intermediate
objects,
characterized by FR~II radio morphology, high emission-line luminosities,
unusually weak continuum or quasar-like VLBI polarization, but their small
number is unlikely to affect unified schemes significantly.

\subsection{Possible Connections between BL~Lac Objects and FSRQ} 
\label{sec:connection}

Because of their similar continuum properties, BL~Lac objects and FSRQ are
collectively called blazars, yet we saw in the previous section that there are
also considerable differences between them. What is the relationship between
the two blazar classes? Three suggestions --- that FSRQ evolve into BL~Lacs,
that they are different manifestations of the same physical process, and that
BL~Lacs are gravitationally micro-lensed FSRQ --- are discussed in this
section (some of these points were originally discussed by Padovani 1992b).

\subsubsection{The Evolutionary Connection} 
\label{sec:econnection}

The de-evolved number densities of FSRQ and BL~Lacs are similar (correcting
for evolution, FSRQ are about twice as numerous at $z=0$; Padovani 1992b), but
the FSRQ are distributed to much higher redshift. It is a well-known problem
that high-redshift
quasars must disappear and/or become considerably less luminous by the present
epoch (Schmidt 1968). Here we consider the suggestion that FSRQ evolve into
BL~Lac objects, becoming weak-lined objects by virtue of increased beaming of
the continuum, that is, a Lorentz factor increasing with cosmic time
(decreasing with redshift; Vagnetti \ea 1991). The merit of this approach is
that it includes evolution more directly than usual and it attempts to link
the (at present) separate unified schemes for high- and low-luminosity
radio-loud AGN. It also predicts the evolution of the critical angle
separating blazars from radio galaxies (Vagnetti and Spera 1994). Specific
predictions of distributions of jet/counter-jet ratios and superluminal
velocities are also possible (Vagnetti, in preparation).

The results in \S~\ref{sec:low_highzbl}, however, demonstrate the lack of
continuity in redshift between FSRQ and BL~Lac objects. For example, their
extended radio powers and line luminosities are very different --- by $1-2$
orders of magnitude at comparable redshifts. Figures~\ref{fig:oiii} and
\ref{fig:mgii} show that BL~Lacs and FSRQ occupy separate, approximately
parallel, regions in the $L_{\rm \OIII} - z$ and $L_{\rm \MGII} - z$ planes,
which is more suggestive of related but distinct histories than of evolution
from high-luminosity to low-luminosity blazar.

The strongest objection to the suggestion that strong-lined objects
evolve into weak-lined ones because of increased beaming is the fact that
BL~Lacs have intrinsically weak lines. That is, the equivalent widths are not
small because the optical continuum is stronger but because the lines are
weaker. There is no evidence that the Lorentz factor, and to first order the
Doppler factor, is larger in BL~Lacs than in FSRQ. If anything, the limits
on Doppler
factors derived from SSC arguments are smaller for BL~Lacs than for FSRQ
(\S~\ref{sec:independent}). Furthermore, superluminal motion data might
actually support an increase with redshift of the Lorentz factor for low
values of $q_0$ (Vermeulen and Cohen 1994).

In the context of the radio-galaxy/blazar unified scheme, an evolutionary
connection between FSRQ and BL~Lac objects has strong implications for the
parent populations. If (some) FSRQ evolve into BL~Lacs, then (some) FR~IIs
should evolve into FR~Is. Possible mechanisms for this transition exist, like
the deceleration of the high-velocity jets that give rise to an FR~II radio
source by an increase in the density of the intergalactic medium (De Young
1993) or a decrease in the mass accretion rate onto the central engine
(Baum \ea 1995).
The connection between FR~I and FR~II radio galaxies is still not fully
understood (Owen and Laing 1989; Owen and Ledlow 1994; Baum \ea 1995), but
it is interesting to note that studies of cluster environments of radio
sources do suggest both FR~IIs and radio quasars could evolve into FR~Is (Hill
and Lilly 1991; Yee and Ellingson 1993).

\subsubsection{Multiwavelength Spectral Continuity of BL~Lac Objects and
FSRQ} 
\label{sec:multi}

The multiwavelength spectra of HBL, LBL, and FSRQ form a continuous
sequence that suggests a common physical mechanism (Sambruna 1994; Sambruna \ea
1995; Maraschi \ea 1995).
The spectra of LBL (RBL) and FSRQ have relatively low peak wavelengths while
the HBL (XBL) have higher peak wavelengths. Very roughly, along the sequence
from HBL to LBL to FSRQ, the wavelength of the peak synchrotron emission
decreases\footnote{The peak wavelength is definitely longer for
LBL than HBL but the comparison for FSRQ and LBL has not yet
been done with sufficient statistics. Furthermore, the peak wavelength
lies in the far infrared band so that this question must really be
answered with {\it ISO}.},
the X-ray spectral index flattens, the ratio of gamma-ray to bolometric
luminosity increases, the bolometric luminosity itself increases, and the
mean redshift increases (luminosity and redshift are automatically correlated
in flux-limited samples).
These trends are illustrated in part by the multiwavelength spectra of a
typical
FSRQ (Fig.~\ref{fig:3c279_spec}) and a typical HBL and
LBL (Fig.~\ref{fig:multispec}).

Analyzing the spectra of these and other blazars
in terms of synchrotron models, Sambruna (1994) finds
that a simple transformation in terms of angle between HBL and LBL can not
explain the differences in their multiwavelength spectra. To account for the
shorter peak wavelength of the synchrotron emission in HBL compared to LBL,
it is necessary to
invoke both higher electron Lorentz factors (not to be confused with the bulk
Lorentz factor of the jet) and stronger magnetic fields in the X-ray-emitting
region. Similarly, the FSRQ have still lower electron energies and
weaker magnetic fields than the RBL. The lower electron energies mean
that in the X-ray band, the
synchrotron component is relatively less important than any Compton-scattered
emission (whether from scattered synchrotron photons or other ambient photons),
leading to the flatter X-ray spectra and perhaps the
stronger gamma-ray emission.

Thus it is plausible that all blazars are dominated by synchrotron plus
Compton-scattered continuum emission, with mean relativistic electron energies
and magnetic fields systematically declining with increasing bolometric
luminosity (Sambruna
1994). To establish this connection among BL~Lac objects and FSRQ
definitively, via analysis of the physical parameters in blazar jets, requires
extensive multiwavelength monitoring of complete samples of blazars.
Unfortunately it is difficult to correlate variability between gamma-ray and
other wavelengths because current detector sensitivity precludes variability
studies for all but a few of the brightest blazars (Kniffen et al. 1993;
Hunter \ea 1993).

\subsubsection{BL~Lac Objects as Gravitationally Micro-Lensed FSRQ} 
\label{sec:lensing}

The main differences between BL~Lacs and FSRQ are, to first order, that the
FSRQ are more distant, more luminous, and have stronger emission lines.
Ostriker and Vietri (1985, 1990) have suggested that gravitational
micro-lensing by stars in a foreground galaxy could turn an optically
violently variable (OVV) quasar (i.e., an FSRQ, in our terminology) into a
BL~Lac. Specifically, a
distant FSRQ with a compact optical/ultraviolet continuum region and an
extended
optical/ultraviolet emission-line region,
whose line of sight passes nearly through the center of
a foreground galaxy, would have the continuum preferentially amplified by
micro-lensing. A factor of $\simgt 10$ enhancement could, Ostriker and Vietri
estimated, reduce the equivalent widths of the BL~Lac emission lines
sufficiently. In addition, the BL~Lac would appear to lie in the (approximate)
center of the (nearby) lensing galaxy. Its redshift, in practice often
inferred from the ``host'' galaxy spectrum, would refer to the lensing galaxy
rather than the FSRQ, so the derived luminosity of the BL~Lac object would be
artificially low.

In this picture, BL~Lacs and FSRQ are intrinsically the same objects, and
there is no separate low-luminosity blazar class to be unified with FR~I radio
galaxies. Note that the relativistic jet is probably still required to explain
the rapid variability, polarization, and other blazar characteristics of FSRQ,
although micro-lensing could account for some low-amplitude, approximately
achromatic, variability in BL~Lac objects (Urry \ea 1993; Edelson \ea 1995).

There are a number of arguments against the micro-lensing hypothesis, albeit
none definitive. These include the following.
\begin{itemize}
\item
The optical amplification factors derived from numerical simulations under the
micro-lensing hypothesis, basically comparing the optical continuum fluxes of
BL~Lacs and FSRQ in the 1~Jy and 2~Jy samples, respectively, are much smaller
than the factor of 10 required by Ostriker and Vietri to swamp the emission
lines of quasars (Padovani 1992b).
\item
The observed optical number counts of BL~Lacs (Padovani and Urry 1991; Hawkins
\ea 1991) do not appear to flatten at $B \sim 16 - 18$, as predicted by
Ostriker and Vietri, although the faint counts are still quite uncertain.
\item
When BL~Lac ``host'' galaxies are detected (e.g., Abraham \ea 1991), their
nuclei are almost always well-centered on the galaxy (see Stocke \ea 1995 for
evidence of de-centering in MS~0205.7+3509, an EMSS BL~Lac), whereas the
micro-lensing
scenario would allow the background FSRQ to be well off the center of the
micro-lensing galaxy. Existing limits on de-centering may already be
sufficient to rule out the presence of even a small number of micro-lensed
sources (Merrifield 1992) and {\it HST} imaging of complete samples have the
potential to decide this issue finally.
\item
At least five of the low-redshift 1~Jy RBL have emission lines and absorption
lines at the same redshift (Stickel \ea 1993), indicating that in a substantial
number of cases, the absorption line redshifts are from the host galaxy rather
than the lensing galaxy.
\item
The systematic differences in VLBI polarization structure for BL~Lac objects
and
FSRQ, notably the $90^\circ$ difference in mean magnetic field orientation,
should not exist if one is simply an amplified version of the other
(Gabuzda \ea 1992). Micro-lensing changes the image position randomly with
respect to the direction of polarization (which is unchanged), since the lens
mass distribution does not ``know'' about the orientation of the background
jet.
Thus the VLBI polarization in BL~Lacs should be relatively uncorrelated with
jet
morphology compared to FSRQ, whereas they are parallel and perpendicular to
the jet, respectively.
\item
Micro-lensing should cause the radio-to-optical continuum of
the ``low-luminosity'' BL~Lac objects (basically, the HBL)
to be flatter than that of FSRQ, which is more or less as observed
(Stocke \ea 1985; Sambruna 1994); however, the observed difference is
larger than expected from micro-lensing. To shift the wavelength
of the peak synchrotron emission in FSRQ, generally in the infrared,
to the ultraviolet/EUV/soft X-ray, would require amplifying the
short-wavelength continuum by several orders
magnitude (see Fig.~\ref{fig:multispec}).
\end{itemize}

The evidence for micro-lensing in a few individual cases remains unclear. One
of the most promising micro-lensing candidates, 0235+164 (Stickel \ea 1988a),
which has multiple absorption systems (Yanny \ea 1989), a de-centered
foreground group of galaxies (Abraham \ea 1993), and an extremely high
superluminal velocity ($\beta_{\rm a} \sim 90$; B\aa\aa th 1984), is
almost certainly weakly amplified by macro-lensing (Abraham \ea 1993) but no
one has calculated the effects of micro-lensing by stars in the foreground
lens. (Abraham \ea 1993 argued that by analogy to a normal quasar,
micro-lensing could cause significant variability only on time scales of
years; however, the relevant source-lens velocity for FSRQ is the superluminal
jet velocity rather than the typical stellar velocity in a galaxy core
[Gopal-Krishna and Subramanian 1991].
Therefore, the variability time scales can be much shorter, quite comparable
to those observed in 0235+164.) In the case of PKS~0537$-$441, another lensing
candidate, Falomo \ea (1992) failed to detect the foreground galaxy the
presence of which had been suggested by Stickel \ea (1988b).

In summary, there are a number of arguments against micro-lensing in large
numbers of
BL~Lac objects, and there are no clear cases of micro-lensing in any one
BL~Lac. Still, there are no data that actually falsify the micro-lensing
hypothesis. We suggest a clear and incontrovertible test. As the {\it HST} Key
Project on quasar absorption lines has demonstrated (Bahcall \ea 1993),
the ultraviolet spectrum
of a distant FSRQ should show multiple absorption lines due to intervening
Ly$\alpha$ clouds. Suppose we have a BL~Lac object that
appears to lie in an elliptical galaxy at a redshift of, say, $z=0.25$, and we
want to test whether it is actually a quasar at a redshift of, say, $z=1.9$
(or less) that is micro-lensed by that galaxy. With an ultraviolet spectrum
from $\sim1500$ to 3500~\AA\ (with sufficient signal-to-noise ratio to detect
absorption lines with equivalent widths of a few hundred milliangstroms)
the presence or absence of unidentified (i.e., Ly$\alpha$) absorption features
answers this question unambiguously. This project is ideal for {\it
HST}.

\section{The Viability of Unified Schemes} 
\label{sec:prob_comp}

On the whole, unified schemes for radio-loud AGN are very successful. But
there remain some potential problems, including disturbing trends in
the linear sizes of radio galaxies and quasars, a dependence on redshift of
the ratio of quasars to radio galaxies, the lack of superluminal motion in
radio galaxies, and the un-FR-I-like radio morphologies of some BL Lac objects.
A number of complications also exist, like the need to include
evolution in unified schemes (the critical angle separating different AGN
classes or the ratio of $L_{\rm jet}/{\cal L}_{\rm ext}$ may well depend on
redshift),
even as the relatively simple beaming models adopted so far are
underconstrained by existing data. In this section we review known problems
for unified schemes and anticipate complications that should be addressed
when there are more complete data for larger samples of radio-loud AGN.

\subsection{Potential Problems with Unification} 
\label{sec:prob}

\subsubsection{Linear Sizes of Blazars and Radio Galaxies} 
\label{sec:sizes}

In principle, the linear sizes of radio sources can be used to test
unification because blazars oriented at small angles to the line of sight
should have systematically smaller large-scale radio structures than radio
galaxies in the plane of the sky. Right at the start, however, we must say
that conclusions regarding the relative linear sizes of blazars and radio
galaxies are extremely murky, having been the subject of quite a number of
papers, with often contradictory results (Gopal-Krishna and Kulkarni 1992;
Barthel
1989; Kapahi 1987; Kapahi 1989; Kapahi 1990; Barthel and Miley 1988; Onuora
1989; Onuora 1991; Hough and Readhead 1989; Nilsson \ea 1993, and
references therein). It turns out that the linear size depends, not
unexpectedly, on several factors like radio power and redshift. Even matching
these variables, the expected difference in mean linear size is quite modest,
only a factor of 2 for an orientation of $30^{\circ}$ compared to 90$^{\circ}$,
while the scatter in intrinsic sizes must be many times that large.

In a recent work, Singal (1993a; see also Singal 1988) studied the linear
sizes of a large, heterogeneous sample of
radio galaxies (about half of which have redshifts estimated from the
magnitude-redshift relation) and quasars. His main result is that both
size-luminosity and size-redshift correlations are significantly different for
the two classes. In particular, luminosity and size are directly correlated
for radio galaxies and inversely correlated for quasars, while the size of
radio galaxies falls strongly with redshift and not at all for quasars.
According to Singal, the only redshift range for which unification works is
$0.5 < z < 1$, the interval Barthel (1989) used in his original analysis of
the 3CR catalog ostensibly because of the better statistics there. Indeed,
for the lowest luminosity bin and $z \leq 0.4$, Singal (1993a) found that the
median linear size of quasars is larger than that of radio galaxies,
completely opposite to the predictions of the unified scheme.

These results, however, are uncertain and/or contradict other comparable work.
One problem with determining linear sizes is that often the radio source
morphology is unclear or the source is poorly resolved. In a similar study
limited to sources with clear FR~II double structures, which by definition
have sharp outer boundaries, Nilsson \ea (1993) found
quite different results for the dependence of linear size on redshift and
luminosity; in
direct contrast to Singal (1993a), the median values of the linear size
for quasars are (marginally) smaller than for galaxies in all luminosity bins.

Another issue is the use of complete flux-limited samples. The 3CR catalog
is excellent for this purpose because it is largely unbiased by beamed
flux. From the luminosity function analysis we found quasars
(effectively SSRQ) and radio galaxies are divided at
$\theta_{\rm c} \sim 38^\circ$ (Table~3);
the mean angle for SSRQ (between $14^\circ$ and $\sim 38^\circ$)
is therefore $28^\circ$ and the mean angle for FR~IIs is $66^\circ$.
The ratio of quasar linear size to radio galaxy linear size,
1.9 for these mean angles,
is both small and a very weak function of critical angle.
Estimating the critical angle from the numbers of quasars and radio
galaxies in the 3CR gives $\theta_{\rm c} = 44.4 ^\circ$
(Barthel 1989), with a $1~\sigma$ uncertainty of $\pm 6.6^\circ$;
this gives a foreshortening factor of $1.9^{+0.3}_{-0.2}$.
The foreshortening factor would be much larger were one to compare
only FSRQ and FR~IIs, but there are too few FSRQ in the 3CR and using
the 2~Jy sample would mean incorporating the selection bias of beaming, which
is nontrivial.

Another complication is that even modest misalignments between large-scale
radio structure and the obscuring torus are enough to confuse the expected
linear size trend (Gopal-Krishna \ea 1994). For example, the distributions
of observed linear size and bend angle for a large, heterogeneous sample of
quasars and radio galaxies are completely consistent
with random orientations within
and without a cone angle of $\sim50^\circ$, respectively, assuming only that
the jets have small intrinsic bend angles ($\simlt 25^\circ$) and modest
intrinsic arm-length differences (Lister \ea 1994a). In sum, considering the
many problems and potential complications, it is remarkable that the analysis
of
linear sizes of quasars and FR~IIs gives any sensible results at all.

The comparison of linear sizes of FR~I galaxies and BL~Lac objects is even
more difficult because the amorphous FR~I radio structures have no clear outer
boundaries. The largest angular sizes of the EMSS BL~Lacs are similar to those
of B2 FR~Is and of a heterogeneous sample of radio-selected BL~Lacs (Perlman
and Stocke 1993), in apparent contradiction to their unification. This may be
due to substantial misalignments between the parsec and kiloparsec scale jets
(Perlman and Stocke 1993), which are quite common in BL~Lacs (Kollgaard
\ea 1992). In that case, a small angle to the line of sight of the VLBI jet
does not necessarily imply a very small radio linear size.

In addition, if the bulk Lorentz factors in BL~Lac objects are small
(\S~\ref{sec:independent}), the critical angle separating them from FR~Is can
be quite large. For example, for $\gamma\sim2$ and $f\sim0.05$,
$\theta_{\rm c}\sim 19^\circ$ (Table~3).
The corresponding ratio of projected lengths of BL~Lacs
and FR~Is, evaluated at the mean angle (ignoring beaming)
is $\sim3$; for the XBL fit in Table~3 it is $\sim2.5$.

In the end, consideration of linear sizes of radio sources is not a compelling
test of unified schemes. Even the oft-stated problem of large deprojected
radio sizes --- wherein the lengths of superluminal sources and/or one-sided
jets as large as $\sim1$~Mpc in length (like NGC~6251) are thought to be
uncomfortably large when deprojected by their orientation angles --- vanishes
when one thinks about the numbers. Neither superluminal motion nor
one-sidedness require a particularly small angle (Fig.~\ref{fig:beta_app}
and Eq.~\ref{eq:beta_app}, Eq.~\ref{eq:jcj}), and in fact the expected
deprojection correction is only a factor of 2 or so.

\subsubsection{Dependence of Quasar Fraction on Redshift}
\label{sec:q_z}

In his original unification paper, Barthel (1989) used the ratio of quasars to
radio galaxies in the 3CR catalog (essentially SSRQ to FR~II) to determine a
critical angle separating the two classes.
However, Singal (1993b) found that the quasar fraction in the 3CR appears to
increase with redshift\footnote{In a related study with a largely overlapping
sample, Lawrence (1991) found the fraction of broad-line objects increased
with radio power and possibly {\it decreased} with redshift.}
and noted that Barthel's result depended on choosing the redshift interval
$0.5<z<1$. Given the small numbers involved, Singal's number ratios
are actually within $2~\sigma$ of no redshift dependence
assuming Poisson statistics (Gehrels 1986). Moreover, once low-excitation
3CR FR~II galaxies are excluded (\S~\ref{sec:distinction}),
the quasar/radio-galaxy fraction is independent of redshift (Laing \ea 1994).

The number ratios of quasars to radio galaxies binned by redshift
imply a set of critical angles which in turn correspond to predicted ratios of
mean linear sizes for the two classes. These size ratios agree
with the observed values to within $\sim 2~\sigma$,
contrary to Singal's conclusions, once the
associated uncertainties in the predicted values are taken into account
(Saikia and Kulkarni 1994).

Even taking Singal's numbers at face value, any inconsistencies can be
explained easily by allowing for a moderate misalignment ($\sim 20^{\circ} -
30^{\circ}$) between the radio axis and the axis of the optically thick torus
hiding the broad emission line region in radio galaxies (Gopal-Krishna \ea
1994).
Finally, as Singal (1993b) noted, a modest dependence of quasar fraction on
redshift might be explained by an evolution in the opening angle of the
obscuring torus, as it might by anything causing a correlation between opening
angle and luminosity (Lawrence 1991; \S~\ref{sec:torus}).

\subsubsection{Absence of Superluminal Motion in Radio Galaxies}
\label{sec:sup_fri}

The subrelativistic kiloparsec-scale jet velocities in FR~I galaxies,
typically of the order of $1000 - 10,000$ km s$^{-1}$ (Bicknell \ea
1990), used to represent a problem for the FR~I/BL~Lac unification, which
requires relativistic speeds at least on small scales. It was not clear that
FR~Is were relativistic even on the smallest scales, and if they were, it was
not clear how they were decelerated or what the observational
signature of that deceleration would be.
Now, new observational (Venturi \ea 1993; Feretti \ea
1993; Giovannini \ea 1994) and theoretical results (Laing 1994; Bicknell 1994)
supporting the presence of relativistic motion on parsec scales
in these sources have changed our understanding of FR~Is. Physically
reasonable models have also been proposed for the required deceleration (Laing
1994; Bicknell 1994).

If FR~Is are relativistic on small scales, they should show
superluminal motion, even near the plane of the sky. A jet with a Lorentz
factor
of 5 would have apparent transverse speeds $\beta_{\rm a} \simeq 2.3$ and 1
for viewing angles of 45 and 90 degrees, respectively (Eq.~\ref{eq:beta_app}).
The few measurements of jet proper motions with VLBI, however, suggest that
presently studied FR~I radio galaxies display relativistic but still
subluminal speeds. Using the data
compiled by Vermeulen and Cohen (1994), we obtain an average value
$\langle \beta_{\rm a} \rangle
\sim 0.5$ (converting to our adopted value of $H_0=50$~km~s$^{-1}$~Mpc$^{-1}$)
for four FR~Is (NGC~315, M87, Centaurus A, and NGC~6251). The speeds for
FR~IIs seem to be no different: $\beta_{\rm a} \simeq 0.5 - 1.0$ for Cygnus A
(Carilli \ea 1994).

These results may imply that radio galaxies have smaller
Lorentz factors than BL~Lacs and radio quasars. It is important to remember,
however, that detection of VLBI components in radio galaxies is
hampered by relativistic deamplification and dilution by unbeamed
emission. For $\gamma = 5$, for example, jets are deamplified for orientation
angles $\theta \simgt 35^{\circ}$ (Appendix A), which would include basically
all radio galaxies (see Table 3); for larger Lorentz factors,
the angle is even smaller and the deamplification larger.
(The jet-to-counterjet ratio remains significantly larger than unity
even at large angles; Fig.~\ref{fig:jcj}.)
Similarly, at these angles the ratio of
transverse jet luminosity to unbeamed luminosity is $\simlt 10^{-2}$ for
$\gamma=5$ and $f=0.01$ (Eq.~\ref{eq:r}) and equals $\sim 10^{-4}$ at
$90^\circ$ (Eq.~\ref{eq:rt}). Thus the bulk of the emission may well appear
to be stationary even if a transverse relativistic jet is present.
Another consideration
is that significant beaming can be reconciled with subluminal motion of knots
if these are reverse shocks advected by the jet; their motion would then give
a misleading indication of the flow velocity (Bicknell 1994).

In the end, either FR~Is are intrinsically different from BL~Lacs or they
will exhibit superluminal motion on small scales. Surveys of FR~Is with the
Very Long Baseline Array (VLBA)
should help decide this question. It is extremely
interesting that in one nearby FR~I galaxy, M87, which has been
extremely well mapped in the radio, superluminal motion has been detected
on kiloparsec scales, with $\beta_{\rm a}$ up to 2.5 (Biretta \ea 1995).
At least one FR~I, then, must
have an appreciable bulk flow on both parsec and kiloparsec scales.

\subsubsection{The Parent Population of BL Lac Objects} 
\label{sec:bll_parents}

There are some potential mismatches in the properties of FR~I radio galaxies
and BL~Lac objects, although these are far from demonstrated conclusively.
First, the narrow emission line strengths of BL~Lacs, where measured,
barely overlap with the distribution for FR~Is (\S~\ref{sec:narrow_low},
Fig.~\ref{fig:oiii}), although this is complicated by the lack of complete
samples
and uniform coverage, not to mention the extreme difficulty of detecting
very weak lines (and the lack of published upper limits) in BL~Lac objects.

Second, broad \MGII~lines are quite strong in some high-redshift BL~Lacs
(\S~\ref{sec:low_highzbl}, Fig.~\ref{fig:mgii})
while there is little evidence for similarly strong broad lines in nearby
FR~Is.
In part this could be due to the lack of sensitive
UV measurements of FR~Is or to a correlation with continuum luminosity
(and therefore redshift); both effects would limit the observation of
\MGII~in existing, very local, samples of FR~Is.

Third, the diffuse radio emission of BL~Lac objects displays FR~II
morphology in some cases (\S~\ref{sec:ext_low}) and sometimes quite
high unbeamed luminosity (Fig.~\ref{fig:pext_low}). There may also be a deficit
of twin-jet morphologies in BL~Lacs, depending on the true value of the
critical angle.
It may well be that the parent population of BL~Lac objects includes at
least some FR~II radio galaxies, presumably those with low-excitation optical
spectra. We expect that the beaming calculation under this hypothesis would
show this scheme to be viable and that the derived Lorentz factor(s) will
not change much.
Certainly this would be an interesting calculation to do once the optical
spectra of the 2~Jy sample are sufficient to separate the low-excitation
and high-excitation FR~IIs.

\subsection{Possible Complications for Unification}
\label{sec:comp}

\subsubsection{Properties of the Obscuring Torus}
\label{sec:torus}

The opening angle of the obscuring torus may well be a function of
source power. For example, the inner radius of the torus could be
determined by the evaporation temperature of dust,
${\cal R}_{\rm in} \sim 0.06 (L_{\rm bol}/10^{45})^{0.5}$~pc (Lawrence 1991;
Netzer and Laor 1994). Since more distant quasars are more luminous,
this would produce a positive dependence of the quasar fraction
on redshift. There are claims in the literature that the ratio
of broad- to narrow-line objects depends on redshift (Lawrence 1991;
Singal 1993b), but at least in the 3CR sample this trend disappears
once the low-excitation FR~IIs are excluded
(\S~\ref{sec:q_z}). The misclassification of some quasars as NLRG could also
confuse the issue (\S~\ref{sec:content}).

An interesting distinction between high- and low-luminosity unified
schemes for radio-loud AGN is that the high-luminosity AGN include
transition objects, the SSRQ, which are relatively unbeamed in terms of radio
emission but ``beamed'' (i.e., unobscured) in terms of optical spectrum.
The analog of SSRQ for the low-luminosity AGN --- broad-line radio galaxies
with FR~I morphologies --- is not seen (with the exception of the BLRG 3C~120);
surveys find either
narrow-line FR~I or BL~Lacs (with occasional broad lines).
If FR~I galaxies do have broad emission lines at some level,
the lack of broad lines could be explained if the opening angle
of the torus depended on luminosity
(Falcke \ea 1995; \S~\ref{sec:narrow_low}).
This would mean the torus opening angle in FR~Is is much smaller
than in FR~IIs. To eliminate broad-line FR~Is, it must be comparable to or
smaller
than the radio beaming angle, whereas in quasars the torus opening angle
is much larger than the critical angle for relativistic beaming of the
radio emission.
Sensitive infrared spectroscopy and spectropolarimetry of FR~Is will
address this issue by detecting or putting interesting limits
on the broad-line luminosities of FR~I radio galaxies.
{\it ISO} observations of radio galaxies, BL~Lac objects, and quasars will
constrain the relative amounts of obscuring matter in each class.

Direct mapping of the torus may be possible with high spatial resolution
spectroscopic observations of water masers.
Recent VLBI imaging of water masers in NGC~4258, a nearby galaxy
with low-level AGN activity, provides evidence of rotating molecular
gas in a region
less than 0.13 pc from the center (Miyoshi \ea 1995). This is important direct
evidence for molecular gas in the nuclei of galaxies, as implied by the
obscuring torus hypothesis. The case of NGC~4258 is not yet the desired
demonstration of a molecular torus in AGN, however, because
the molecular gas is distributed in a very thin disk rather than a torus.
Using the results of Miyoshi \ea (1995) we find
the ratio of disk height to inner radius is
$\simlt 0.02$, corresponding to an opening angle of $\sim 90^{\circ}$,
which will clearly not obscure the central source very effectively.
In addition, masers do not automatically signal the presence of a torus;
in NGC~1068, there are masers in the southern radio component with
locations and velocities approximately commensurate with rotation, but there
is also a maser to the north that apparently arises in molecular gas shocked
by the radio jet (Gallimore \ea 1995).

\subsubsection{Cosmic Evolution of Radio-Loud AGN} 
\label{sec:evol}

Cosmic evolution could be an important aspect of unified schemes.
Realistically,
one should probably expect evolution in the Lorentz factor ($\gamma$),
in the ratio of beamed to unbeamed flux ($R$), and in the intrinsic
fraction of luminosity radiated by the jet ($f$). But evolution
has been largely ignored in unified schemes
(cf. Vagnetti and Spera 1994) because of the poor statistics when
existing samples are divided into multiple redshift bands.

In most cases, one finesses the evolution of AGN by parameterizing it
with a smooth function and extrapolating evolving properties to zero
redshift. For example, comparisons between beamed and observed LFs
were done assuming exponential luminosity evolution (\S~\ref{sec:statistical}).
The preferred approach, statistics permitting, is to derive the LFs at
different redshifts and apply the beaming formalism to the single-epoch LFs.
This would allow, for example, the study of evolutionary trends of the
beaming parameters.

The well-identified radio-selected samples currently available --- the
3CR, the 1~Jy, the 2~Jy samples --- are quite shallow. For a
typical radio-loud AGN spectrum, optical and X-ray surveys are much deeper,
in many cases the equivalent of 1~mJy or fainter at 5~GHz.
Deep radio surveys, optically identified and classified morphologically
in the radio, would greatly improve our understanding of unified schemes
but represent a substantial technical challenge.

Another important aspect is the evolution of individual sources
and the potential consequences for their radio morphologies.
For example, the observed correlation between core-to-extended radio
flux and linear size (e.g., Lister \ea 1994b) can be explained by a
model wherein the radio source starts out as a luminous core that fades
with time as the lobes brighten (Hutchings \ea 1988).
If this evolutionary effect dominates over beaming and orientation, it could
complicate considerably the use of this flux ratio as a beaming indicator.

\subsubsection{Parameterization of Relativistic Beaming} 
\label{sec:param}

The adopted beaming model (\S~\ref{sec:effect}; Appendix~B)
is undoubtedly too simple. Jets are
almost certainly complicated, more so than usually assumed in the statistical
tests (Lind and Blandford 1985; Bridle \ea 1994). They are likely to be
bent, have variable bulk Lorentz factors, have hot spots at local shocks,
and so on, rather than being a well-behaved, ideal confluence of particles,
photons, and fields in equipartition. This is borne out by the complex
morphologies and trajectories of superluminal components in blazars.
Jet velocities probably decrease or
increase with increasing distance from the nucleus, whereas we have
assumed they are fixed; observed superluminal motion in well-monitored
sources is definitely complex (e.g., Biretta \ea 1986).
Finally, there is always the problem of flow versus
pattern velocity (\S~\ref{sec:superluminal}), with only the former causing
Doppler boosting.

Moreover, it is not clear how jets form and propagate. By what physical process
is matter accelerated to Lorentz factors $\sim 5$,
and sometimes as much as an order of magnitude higher? How is the balance
between kinetic and radiative energy determined? Jet formation and motion
is almost certainly affected by the local gas environment, which might
explain in part the differences between FR~I and FR~II radio galaxies
(De Young 1993; Bicknell 1994), and perhaps why the parsec-scale magnetic
field configurations differ in quasars
and BL~Lac objects. Jet motion is probably also
affected by competition between gravitational force and
radiation pressure, which moderates the accretion rate. If so,
the Lorentz factor could depend on the source Eddington ratio
($\dot m_{\rm Edd} \propto L/M$, where $M$ is the mass of the central
black hole; Abramowicz 1992).

Unified schemes for radio-loud objects can also be complicated if the beaming
is wavelength-dependent. Although this is probably not the case for BL~Lacs
(Sambruna 1994), it could be important in quasars (Impey \ea 1991, but see
also Hough and Readhead 1989). The dependence of beaming on wavelength
occurs naturally for an obscuring torus because its transparency
is wavelength dependent or for an inhomogeneous jet model with variable
Lorentz factor along its length.

\subsubsection{Compact Steep-Spectrum and Gigahertz Peaked-Spectrum Sources}
\label{sec:CSS}

The place of Compact Steep-Spectrum Sources (CSS) and the possibly related
Gigahertz Peaked-Spectrum Sources (GPS) in the unified scheme for
high-power radio sources is an open question. CSS and GPS include both
quasars and radio galaxies. Their most striking feature is their compact
radio structure ($\simlt 15$ and $\simlt 1$~kpc, respectively;
see Fanti \ea 1990a and O'Dea \ea 1991 for reviews of their properties).
In terms of the definitions in \S~\ref{sec:Introduction} (Table~1),
we include them in the SSRQ/FSRQ and NLRG classes, as appropriate.

The CSS and GPS sources constitute a non-negligible part of radio samples; for
example, CSS represent approximately 13\% of the 3CR sample (Fanti \ea 1990a)
and $\sim20$\% of the 2~Jy sample (Morganti \ea 1993).
For our statistical analysis in \S~\ref{sec:statistical},
we removed the galaxies (because they were not classified as either FR~I or
FR~II by their radio morphologies) but included the quasars; this might have
increased the critical angles slightly.

Statistical considerations exclude the possibility that most
CSS and GPS are intrinsically large radio sources seen at small angles
to the line of sight. Instead, they are probably inherently compact
(Fanti \ea 1990b). At the same time, the size distribution and relative
numbers of CSS quasars and galaxies are consistent with the former being
the latter viewed at angles smaller than $\sim 45^{\circ}$
(Fanti \ea 1990b), although the different radio morphologies of the
two classes are not entirely explained by orientation.
CSS and GPS could represent an early stage of radio source evolution (Fanti \ea
1990b) and if so, probably should be included in
unification. Alternatively, they may lie in unusual environments
(O'Dea \ea 1991). In
either case, unified schemes bear on their origin, and it is important
to understand how they fit into the present paradigm.

\subsubsection{Selection Effects in the Identification of Quasars} 
\label{sec:selec}

If quasar unification schemes are correct, all quasars are oriented at a
preferred angle from which the broad-line region is visible. This
undoubtedly has strong effects on other quasar properties. For example,
it could explain why no Ly$\alpha$ edges are seen in the ultraviolet
spectra of quasars (Koratkar \ea 1992; Kinney 1994) even though they
are expected from accretion disks, an important element in the current
AGN paradigm (Fig.~\ref{fig:cartoon}). That is, assuming that the axes
of accretion disks and obscuring matter coincide, no edge-on disks
would be seen and so no Ly$\alpha$ edges should be seen.

Other secondary effects are also possible. If the accretion disks in
quasars are thick, as expected for high-luminosity AGN, then their
radiation patterns can be anisotropic independent of the obscuring torus
(Madau 1988). Because its
radiation pattern is narrower than the opening angle of the obscuring
torus, the thick disk can by itself introduce mild selection effects
(Urry \ea 1991b) wherein quasars are more face-on than implied by their
emission line properties (assuming the axes of the thick disk and torus
are aligned). In that case, the optical-through-ultraviolet
continuum emission from quasars will appear bluer and more luminous
than if it were emitted isotropically.

\subsubsection{Extended Continuum Emission in Type~2 AGN} 
\label{sec:ext_cont}

Spectropolarimetry has indicated an interesting complication in a number of
Type~2 AGN; specifically, the scattered broad-line flux
is sometimes more polarized than the continuum (Tran 1995).
This implies that in
addition to the nuclear power-law continuum source there must be an
extended continuum source in the scattering region which dilutes
the polarized flux of the scattered nuclear continuum.
The {\it in situ} extended featureless continuum must be similar
to the nuclear power law,
and in some ways the match between the nuclear and extended properties
is a bit mysterious. The details of the spectral shapes depend on
very high signal-to-noise spectropolarimetric data, however,
available only for a few of the brightest sources,
and so remain to be determined for the Type~2 AGN as a class.

The presence of an extended optical continuum component
can also be inferred from the uniformity of $H\beta$ equivalent widths in
AGN types from Seyferts to radio-loud Type~2 AGN
to quasars
(Binette \ea 1993). According to unification, the intensity of the
narrow $H\beta$ line should be relatively orientation independent while
the broad line and the nuclear continuum are not, meaning the equivalent
width should differ strongly between Type~1 and Type~2 AGN.
The lack of a difference can be explained if appreciable
continuum emission is generated in or near near the line-emitting cloud,
roughly in proportion to the incident photoionizing continuum,
in which case the equivalent width will not depend strongly on orientation.

Locally-generated continuum, such as shock-excited emission from a jet
interacting with the interstellar medium
(Sutherland \ea 1993), could also explain the similar reddening
seen in lines and continuum (Binette \ea 1993), as well as the association
of optical emission line gas with extended radio structure
(\S~\ref{sec:polarim}). Extended optical/UV continuum associated with
emission-line regions has in fact been seen in some nearby Seyfert~2
galaxies (Pogge and De Robertis 1993) and in Cygnus~A (Pierce and Stockton
1986; Antonucci \ea 1994).
To first order, an extended continuum source should mitigate
the selection effects of beaming in the optical band; the degree to which
this is the case depends critically on its luminosity relative to the
nuclear continuum.

\section{The Ten Most Important Questions} 
\label{sec:ten_quest}

Strong anisotropy in AGN is well established
(\S~\ref{sec:anis_obs} and \S~\ref{sec:anis_rel}).
This means some form of unification through orientation must be valid.
The first step in understanding AGN, then, is to unify the principal
classes of radio-loud AGN as described in this
review. The next step is to untangle the effects of cosmic
evolution, which causes the luminosity, morphology, and/or number of AGN
to change with time (redshift). With an ensemble of AGN for which orientation
and evolutionary effects have been removed, we will finally be able to
approach the more physically interesting issues of black hole mass,
black hole spin, accretion rate, formation and evolution of
the individual radio source, etc.

For radio-loud AGN --- radio galaxies, quasars, and blazars ---
relativistic beaming is likely to be very important, on both small and large
scales. This has an enormous effect on the observed properties. Two schemes,
one unifying low-luminosity (FR~I) radio galaxies with weak-lined blazars
(BL~Lac objects) and the other unifying high-luminosity (FR~II) radio galaxies
with strong-lined blazars (OVV, HPQ, FSRQ), and SSRQ, are both consistent with
essentially all available data (\S\S~\ref{sec:basis}, \ref{sec:statistical},
and \ref{sec:q_bl}).
There are some remaining problems or complications that require further
analysis (\S~\ref{sec:prob_comp})
but these are far from falsifying the concept of unification.

{}From today's vantage point, we identify the ten most important
issues for unification and ultimately for understanding AGN.
That unification works quite well with very simple models means we
do understand something important about AGN. As we learn more and more about
radio sources, and the simplicity of present unified schemes inevitably gives
way to increasing complexity, we will doubtless discover even more about what
it is that we do not know.
As an anonymous person (who was surely a scientist) said:

\begin{quote}
{\it The greater the sphere of our knowledge, the larger is the surface of its
contact with the infinity of our ignorance.}\footnote{Anonymous quotation in
{\it A Short History of Astronomy}, A. Berry (New York, Dover).}
\end{quote}
\noindent
In that spirit, recognizing that the questions we ask today,
which we believe will lead to fundamental understanding of AGN,
may not be the questions asked or answered tomorrow,
we put forward the following list:
\bigskip

\begin{itemize}
\begin{enumerate}

\item Is there evidence for BL~Lacs or obscured quasars in all radio galaxies?

\item What is the relation between HBL, LBL and FSRQ?

\item Are the observed distributions of $\beta_{\rm a}$, $R$,
and jet/counter-jet ratios commensurate with beaming?

\item Is the Lorentz factor higher in the high-luminosity radio sources
(quasars and FR~II radio galaxies) than in low-luminosity radio sources
(BL~Lac objects and FR~I radio galaxies)?

\item Do FR~Is have broad emission line regions?

\item What is the relation between FR~Is and FR~IIs?

\item How do jets form and propagate?

\item What is the physical cause of the radio loud -- radio quiet
distinction?

\item Where are the narrow-line (Type~2) radio-quiet quasars?

\item What are the fundamental parameters governing the central engine, and is
it powered by a black hole?

\end{enumerate}
\end{itemize}

\acknowledgments
We thank many of our colleagues for helpful comments and discussions, including
Ski Antonucci,
Stefi Baum,
John Biretta,
Howard Bond,
Sperello di Serego Alighieri,
James Dunlop,
Mike Eracleous,
Carla Fanti,
Roberto Fanti,
Bob Fosbury,
Gabriele Ghisellini,
Paolo Giommi,
Bob Goodrich,
Gopal-Krishna,
Ann Gower,
Tim Heckman,
John Hutchings,
Buell Jannuzi,
Ken Kellermann,
Ron Kollgaard,
Robert Laing,
Andy Lawrence,
Laura Maraschi,
Raffaella Morganti,
Chris O'Dea,
Frazer Owen,
Joe Pesce,
Eric Perlman,
Rita Sambruna,
Chris Simpson,
Ashok Singal,
John Stocke,
Steve Unwin,
and Paul Wiita.
We further thank
Jim Braatz, James Dunlop, John Hutchings, Raffaella Morganti, Frazer Owen,
Geoffrey Taylor, and Andrew Wilson for providing data, images, and figures;
Joe Pesce, John Godfrey, and Dave Paradise for help producing figures;
Sarah Stevens-Rayburn for library assistance;
and the Space Telescope Science Institute, the STScI Visitor Program,
Christina Padovani, and Andrew Szymkowiak for support of the collaborative
visits which made completion of this work possible.

\appendix
\section{Relativistic Beaming Parameters}

We discuss here the various beaming parameters and how they are related to one
another. Many of the equations can be found in Ghisellini \ea
(1993).

The kinematic Doppler factor of a moving source is defined as
\begin{equation}
\delta \equiv [\gamma(1 -\beta \cos\theta)]^{-1},
\end{equation}
where $\beta$ is its bulk velocity in units of the speed of light,
$\gamma=(1-\beta^2)^{-1/2}$ is the corresponding Lorentz factor, and $\theta$
is the angle between the velocity vector and the line of sight. The Doppler
factor has a strong dependence on the viewing angle (as shown in
Fig.~\ref{fig:delta}), which gets stronger for larger Lorentz factors. For
$0^{\circ} \leq \theta \leq 90^{\circ}$, $\delta$ ranges between $\delta_{\rm
min} = \delta(90^{\circ}) = \gamma^{-1}$ and $\delta_{\rm max} =
\delta(0^{\circ}) = (1+\beta)\gamma$ $\sim 2\gamma$ for $\gamma \gg 1$.
Moreover, $\delta = 1$ for $\theta_{\delta} = \arccos
\sqrt{(\gamma-1)/(\gamma+1)}$ (e.g., for an angle $\theta_{\delta} \simeq
35^{\circ}$ if $\gamma=5$), and for decreasing $\theta_{\delta}$ with
increasing
$\gamma$ (Fig.~\ref{fig:delta}); for angles larger than $\theta_{\delta}$
relativistic {\it deamplification} takes place.

Given a value of $\delta$, a lower limit to the Lorentz
factor is given by the condition $\delta \le \delta_{\rm
max}$; that is,
\begin{equation}
\gamma\, \ge\, \di {\strut 1 \ov 2} \left(\delta + \di
{\strut 1 \ov \delta}
\right).
\end{equation}
When $\delta$ is a lower limit, as in the SSC case, this expression is valid
only for $\delta > 1$, since for $\delta < 1$, $\delta + 1/\delta$ decreases
for increasing $\delta$. It can also be shown that for any value of $\gamma$,
\begin{equation}
\sin \theta \le \di {\strut 1 \ov \delta} ,
\end{equation}
which gives a useful upper limit to $\theta$ if $\delta > 1$.

In the relativistic beaming model, the observed transverse
velocity of an emitting blob, $v_{\rm a} = \beta_{\rm a} c$,
is related to its true velocity, $v = \beta c$, and the angle
to the line of sight by
\begin{equation}
\beta_{\rm a} = {\beta \sin\theta \over 1 - \beta \cos\theta}.
\label{eq:beta_app}
\end{equation}
It can be shown that if $\beta > 1/\sqrt{2} \simeq 0.7$, then for some
orientations superluminal motion is observed. The maximum value of the
apparent velocity, $\beta_{\rm a,max} = \sqrt{\gamma^2 -1}$, occurs when $\cos
\theta = \beta$ or $\sin \theta = \gamma^{-1}$; for this angle, $\delta =
\gamma$. This implies a minimum value for the Lorentz factor $\gamma_{\rm min}
= \sqrt{\beta_{\rm a}^2 + 1}$ (see Fig.~\ref{fig:beta_app}). For example, if
one
detects superluminal motion in a source with $\beta_{\rm a} \sim 5$,
the Lorentz factor responsible for it has to be at least 5.1. It is
also apparent from Fig.~\ref{fig:beta_app} that superluminal speeds are
possible even for large angles to the line of sight; sources oriented at
$\theta \sim 50^{\circ}$, have $\beta_{\rm a} \simgt 2$ if $\gamma
\simgt 5$, and sources in the plane of the sky ($\theta = 90^{\circ}$) have
$\beta_{\rm a} = \beta \sim 1$ for $\gamma \simgt 3$.

The apparent velocity in terms of $\gamma$ and $\delta$ is
\begin{equation}
\beta_{\rm a} = \sqrt{2 \delta \gamma - \delta^2 - 1}.
\end{equation}
We find from equations (A1) and (A4), $\beta_{\rm a} = \delta \gamma \beta
\sin \theta$, and for the angle that maximizes the apparent velocity, $\sin
\theta = \gamma^{-1}$, $\beta_{\rm a} = \delta \beta$ $= \gamma \beta$ $=$
$\sqrt{\gamma^2 -1}$ $\simeq$ $\gamma$ $\simeq$ $\delta$.

With a measurement of superluminal velocity and an independent estimate of the
Doppler factor (for example from an SSC calculation), one can combine Eqs.
(A1) and (A4) to obtain two equations in four unknowns. That is, under the
hypothesis that the ``bulk'' and ``pattern'' speeds are the same, one can
derive the value of the Lorentz factor and the angle to the line of sight:
\begin{equation}
\gamma = {\beta_{\rm a}^2 + \delta^2 +1 \over 2 \delta}
\label{eq:gamma_bd}
\end{equation}
\begin{equation}
\tan\theta = {2\beta_{\rm a}\over\beta_{\rm a}^2 + \delta^2
-1}.
\label{eq:tan_theta}
\end{equation}

Note that $\gamma$ reaches its minimum value when $\delta = \sqrt{\beta_{\rm
a}^2 +1} \equiv \gamma_{\rm min}$. If $\delta$ is a lower limit (as when it is
derived from an SSC calculation)
and $\delta < \sqrt{\beta_{\rm a}^2 +1}$, then the $\gamma$ estimated
from Eq.~(\ref{eq:gamma_bd}) is an upper limit
(of course always bound to be $\geq
\gamma_{\rm min}$), while if $\delta > \sqrt{\beta_{\rm a}^2 +1}$, it is a
lower limit. For $\delta \gg \sqrt{\beta_{\rm a}^2 +1}\,\,$, $\gamma \simeq
\delta/2$, while if $\delta \ll \sqrt{\beta_{\rm a}^2 +1}\,\,$, $\gamma \simeq
\gamma_{\rm min}/2 \delta$. As for Eq.~(\ref{eq:tan_theta}),
when $\delta$ is a lower limit, the inferred $\theta$ is always an upper limit.

The predicted jet/counter-jet ratio (i.e., the ratio between the approaching
and
receding jets), can be expressed in terms of $\delta$ and $\beta_{\rm a}$ as
\begin{eqnarray}
\label{eq:jcj}
J &=& \left( {1 + \beta \cos\theta \over 1 - \beta \cos
\theta} \right)^p \\
 &=& (\beta_{\rm a}^2 + \delta^2)^p .
\end{eqnarray}
In the simplest cases, $p = 2 +\alpha$ or $3 + \alpha$
(Appendix~B). Figure~\ref{fig:jcj} shows the jet/counter-jet
ratio as a function of the viewing angle for various values of the Lorentz
factors and for $p=2$ (which minimizes the effect, as $p$ is likely to be
larger). The dependence on orientation is very strong,
since $J \approx \delta^{2p}$. From the jet/counter-jet ratio alone
(Eq.~\ref{eq:jcj}), we obtain
\begin{equation}
\beta \cos\theta = {J^{1/p}-1 \over J^{1/p} + 1},
\label{eq:beta_cos_theta}
\end{equation}
from which an upper limit to $\theta$ (since $\beta \le 1$)
and a lower limit to $\beta$ (since $\cos\theta \le 1$) can
be derived.

It is useful to calculate the angular parameters relevant to tests
of unified schemes. For sources randomly oriented within the angular range
$\theta_1$ to $\theta_2$, the mean orientation angle is
\begin{eqnarray}
\langle \theta \rangle &\equiv & {\int_{\theta_1}^{\theta_2} \theta
\sin\theta d\theta \over \int_{\theta_1}^{\theta_2} \sin\theta d\theta} \\
	& = & {\theta_1 \cos\theta_1 - \theta_2 cos\theta_2 + \sin\theta_2
-\sin\theta_1 \over \cos\theta_1 - \cos\theta_2} .
\label{eq:mean_theta}
\end{eqnarray}
The linear size of extended sources is proportional to $\sin\theta$,
for which the mean value is
\begin{eqnarray}
\langle \sin\theta \rangle & \equiv & {\int_{\theta_1}^{\theta_2} \sin^2\theta
 d\theta \over \int_{\theta_1}^{\theta_2} \sin\theta d\theta} \\
	& = & {\theta_2 -\theta_1+\cos\theta_1 \sin\theta_1 - \cos\theta_2
\sin\theta_2 \over 2(\cos\theta_1 - \cos\theta_{2})} .
\label{eq:mean_stheta}
\end{eqnarray}
Finally, the mean value of $\cos\theta$ is given by
\begin{eqnarray}
\langle \cos\theta \rangle & \equiv &
	\int_{\theta_1}^{\theta_2} \cos\theta \sin\theta d\theta \over
	\int_{\theta_1}^{\theta_2} \sin\theta d\theta \\
	& = & \frac{1}{4}
	{\cos(2\theta_1) - \cos(2\theta_2) \over
	\cos\theta_1 - \cos\theta_2 } ~.
\label{eq:mean_ctheta}
\end{eqnarray}

\setcounter{equation}{0}
\section{Doppler Enhancement}

The Doppler factor, $\delta$ (Appendix~A), relates {\it intrinsic} and {\it
observed} flux for a source {\it moving} at relativistic speed $v =\beta c$.
For an approaching source, time intervals measured in the observer frame are
shorter than in the rest frame (even allowing for time dilation) because the
emitter ``catches up'' to its own photons:
\begin{equation}
t = \delta^{-1} t^\prime,
\label{eq:t_dt}
\end{equation}
where primed quantities refer to the rest frame of the source.
Since the number of wavefronts per unit time is constant, the
emission is blue-shifted (essentially the inverse relation):
\begin{equation}
\nu = \delta \nu^\prime.
\label{eq:nu_dnu}
\end{equation}

The intensity enhancement (``Doppler boosting'') is an even more dramatic
effect. Because $I_{\nu}/\nu^3$ is a relativistic invariant, the
transformation of specific intensity is:
\begin{equation}
I_{\nu}(\nu) = \delta^3 I^{\prime}_{\nu^{\prime}}(\nu^{\prime})
\label{eq:boost3}
\end{equation}
(Rybicki and Lightman 1979). One power of $\delta$ comes from the compression
of the time interval (Eq.~\ref{eq:t_dt}) and two come from the transformation
of the solid angle, $d \Omega = \delta^2 d \Omega^\prime$.

If the emission is isotropic in the source rest frame (i.e.,
$I^{\prime}_{\nu^{\prime}}$ is not a function of angle), the flux density,
$F_\nu$, transforms in the same way as the specific intensity. For a power-law
spectrum of the form $F^{\prime}_{\nu^{\prime}} \propto
(\nu^{\prime}){}^{-\alpha}$, Eq.~(\ref{eq:boost3}) becomes
\begin{equation}
F_{\nu}(\nu) = \delta^{3+\alpha} F^{\prime}_{\nu^{\prime}}(\nu);
\label{eq:boost3a}
\end{equation}
$\delta^\alpha$ is just the ratio of the intrinsic power-law fluxes at
the observed and emitted frequencies.

Broad-band fluxes are obtained from integrating Eq.~(\ref{eq:boost3}) over
frequency, and since $d\nu = \delta d\nu^{\prime}$ (Eq.~\ref{eq:nu_dnu}),
these are boosted by another factor of
$\delta$:
\begin{equation}
F = \delta^{4} F^{\prime}.
\label{eq:boost4}
\end{equation}

The degree of variability in AGN is frequently measured by the change in
flux over a given period of time, which from Eqs.~(\ref{eq:t_dt}) and
(\ref{eq:boost4}) is:
\begin{equation}
{\Delta F \over \Delta t}= \delta^5 {\Delta F^{\prime} \over \Delta
t^{\prime}}.
\label{eq:boost5}
\end{equation}

Equations~(\ref{eq:boost3a}), (\ref{eq:boost4}), and (\ref{eq:boost5})
assume that the emission comes from a moving source, so they
apply to the case of discrete, essentially point-like, components. For a
smooth, continuous jet, the observed emitting volume is decreased by one power
of the Doppler factor because of Lorentz contraction, so that the exponents
in Eqs.~(\ref{eq:boost3a}), (\ref{eq:boost4}), and (\ref{eq:boost5})
become $2+\alpha$, 3, and 4, respectively (Begelman \ea 1984; Cawthorne 1991;
Ghisellini \ea 1993).

Relaxing the assumption of isotropic emission in the rest frame can also
change the relationship between intrinsic and observed flux. If the source is
an optically thin jet with magnetic field parallel to its axis, then in the
rest frame $F^{\prime}_{{\nu}^{\prime}}(\theta^{\prime}) \propto (\sin
\theta^{\prime})^{1+\alpha}$. Since $\sin \theta^{\prime} = \delta \sin
\theta$, Eq.~(\ref{eq:boost3a}) becomes $F_{\nu}(\nu) = \delta^{(3+2\alpha)}
(\sin
\theta)^{1+\alpha} F^{\prime}_{\nu^\prime}(\nu)$ (Cawthorne 1991; Begelman
1993).

So far we have assumed power-law emission in the rest-frame. Strictly
speaking, for synchrotron emission this would mean the jet is either
completely optically thin ($F_{\nu} \propto \nu^{-\alpha}$, $\alpha \simgt
0.5$)
or completely optically thick ($F_{\nu} \propto \nu^{5/2}$). Real jets are
probably inhomogeneous and have flat spectra caused by the superposition of
individual synchrotron components with different self-absorption frequencies.
Relativistic beaming distorts these components differentially because of the
dependence of optical depth and $F_\nu$ on $\delta$, so the overall spectral
shape should change. For a standard conical jet with tangled magnetic field,
the integrated flux transforms approximately as $\delta^{2+\alpha}$, where
$\alpha$ refers to the observed integrated spectrum rather than the spectral
index of the local emission (Cawthorne 1991).

Additional complications in the evaluation of the amplification factor include
the lifetimes of the emitting components, the radial dependence of the their
emissivities, and the presence of shocks (Lind and Blandford 1985). In the
following we will hide all these possibilities in a single parameter, $p$, by
assuming that the observed luminosity, $L_j$, of a relativistic jet is related
to its intrinsic luminosity, $\l_{j}$, via
\begin{equation}
L_j = \delta^p \l_j,
\label{eq:lj_lu}
\end{equation}
with $p = 3+\alpha$ for a moving, isotropic source and $p = 2+\alpha$ for a
continuous jet (other values are certainly possible).

The recent detection of superluminal motion within our own Galaxy (Mirabel and
Rodr\'\i guez 1994) permits for the first time a direct estimate of $p$. Since
proper motions are measured for both the approaching and receding components,
$\beta \cos\theta$ is uniquely determined to be $0.323 \pm 0.016$ (where
$\beta$ refers to the pattern speed; Mirabel and Rodr\'\i guez 1994). Using
the observed jet/counter-jet ratio, $F_j/F_{cj} = 8 \pm 1$, and assuming the
bulk speed in Eq.~(\ref{eq:jcj}) is equal to the pattern speed,
we find $p=3.10\pm0.25$.
One might expect $p=3+\alpha$ because the components are discrete blobs,
whereas our estimate implies $p \approx 2+\alpha$ since the measured spectral
index is $\alpha=0.8$. Alternatively, $p\sim3.8$ is allowed if the bulk speed
is actually lower than the pattern speed, with a ratio $\beta_{\rm
bulk}/\beta_{\rm pattern} \sim 0.8$ (cf. Bodo and Ghisellini 1995).

Regardless of the precise value of $p$, relativistic beaming has a very strong
effect on the observed luminosity. A relativistic jet has $\gamma\simlt\delta
\simlt 2\gamma$ for $\theta \simlt \arcsin \gamma^{-1}$ (Appendix~A), meaning
a modest bulk Lorentz factor of $\gamma = 10$ amplifies the intrinsic power by
$2-5$ orders of magnitude (depending on $p$). The Doppler boosted radiation is
strongly collimated and sharply peaked: at $\theta \sim 1/\gamma \sim
6^{\circ}$, the observed jet power is already $\sim 4 - 16$ times fainter than
at $\theta = 0^\circ$ (for $p = 2 - 4$). At $90^{\circ}$, the reduction is
huge, a factor of $\sim 10^4 - 10^8$. Although this is a very large ratio, it
is actually much smaller than the inferred extinction at optical wavelengths
caused by an obscuring torus, which can be up to a factor of $10^{20}$
(Djorgovski \ea 1991).

\setcounter{equation}{0}
\section{Ratio of Core- to Extended-Flux}

The ratio between core and extended flux is an important beaming indicator.
We use the observed luminosity ratio, $R = L_{\rm core}/L_{\rm ext}$, which is
related to the observed flux ratio via a $K$-correction:
\begin{eqnarray}
R & = & {{F_{\rm core}} \over {F_{\rm ext}}}
	(1+z)^{ \alpha_{\rm core} - \alpha_{\rm ext} } \\
 & \sim & {F_{\rm core} \over F_{\rm ext}} (1+z)^{-1} .
\label{eq:kcorr}
\end{eqnarray}
We do not correct for the fact that $F_{\rm ext}$ diminishes faster with
redshift than $F_{\rm core}$, as this depends on the source morphology. (In
any case it is not a large effect. The corrections calculated by Perlman and
Stocke 1993 for 14 BL~Lac objects with redshifts from 0.2 to 0.5 are
roughly proportional to $[1+z]^{1.2}$.)

We associate the core with the relativistically beamed jet and the extended
power with the unbeamed emission. In terms of the beaming formalism,
\begin{equation}
R = L_j/\l_u = f \delta^p ,
\label{eq:rdef}
\end{equation}
where $L_j$ is the observed jet luminosity (Eq.~\ref{eq:lj_lu}), $\l_u$ is the
unbeamed luminosity, $f \equiv \l_j /\l_u$ is the ratio of intrinsic jet to
unbeamed luminosity, $\delta$ is the Doppler factor, and $p$ is the
appropriate exponent (Appendix~B). We call sources ``beamed'' when $R >
R_{\rm min} \sim 1$.

The largest angle between the jet and the line of sight for beamed objects is
a critical angle, $\theta_{\rm c}$, defined by the condition $R_{\rm min}
\equiv f\delta^p_{\rm min}$, where $\delta_{\rm min} = \delta(\theta_{\rm
c})$. That is,
\begin{equation}
\theta_{\rm c} = \arccos \left \{ {1\over\beta} - {1 \over\beta\gamma}
	\left ( {f \over R_{\rm min}} \right )^{1/p} \right \} ~.
\label{eq:thetac}
\end{equation}
Conversely, the largest ratio, $R_{\rm max} \equiv f\delta^p_{\rm max}$,
will occur at the smallest angle, $\theta_{\rm min}$. If $\theta_{\rm min} =
0^{\circ}$, as is usually the case, then
\begin{equation}
f = {R_{\rm max} \over [\gamma (1+\beta) ]^p}.
\label{eq:fraction}
\end{equation}
If there is a range of Lorentz factors, $R_{\rm max}$ is evaluated at
$\gamma_{\rm max}$. If $\theta_{\rm min} \ne 0^{\circ}$, as is the case for
SSRQ, $R$ will be maximum for $\gamma = 1/\sin \theta_{\rm min}$.

For large angles ($\theta \simgt \arccos [0.5/\beta]$), emission from the
receding jet is no longer negligible and Eq.~(\ref{eq:rdef}) is replaced by
\begin{equation}
R = f \{[\gamma (1 -\beta \cos \theta)]^{-p} +
	[\gamma (1 +\beta \cos \theta)]^{-p} \}.
\label{eq:r}
\end{equation}

The relationship between
our parameter $f$ and $R_{\rm T} \equiv R(90^{\circ})$ used
by other authors (e.g., Orr and Browne 1982) is given by
\begin{equation}
R_{\rm T} = {2 f \over \gamma^p}.
\label{eq:rt}
\end{equation}
For $\theta_{\rm min} = 0^\circ$,
$R_{\rm max} = f[\gamma(1-\beta)]^{-p} \simeq f(2\gamma)^p$,
so that
$R_{\rm max} / R_{\rm T} \simeq 2^{p-1}\gamma^{2p}$.
This implies
\begin{equation}
\gamma \simeq [(R_{\rm max}/R_{\rm T})/2^{p-1}]^{1/2p}.
\label{eq:gamma_rmax_rt}
\end{equation}
Combining the maximum value of $R$ for a set of beamed objects, $R_{\rm
max,b}$,
and the minimum value of $R$ for the parent population, $R_{\rm min,unb}$,
gives
a lower limit to the value of $\gamma$ (since $R_{\rm max}/R_{\rm T} \simgt
R_{\rm max,b}/R_{\rm min,unb}$). In the case of a distribution of $\gamma$s,
the Lorentz factor derived from this argument is the largest one,
since that will be responsible both for $R_{\rm max}$ ($\propto \gamma^p$) and
$R_{\rm T}$ ($\propto \gamma^{-p}$).

\setcounter{equation}{0}
\section{Glossary of Acronyms}

Here we list all the acronyms defined in the text and the section in which
they are first found.

\begin{itemize}

\item AGN: Active Galactic Nuclei, \S~\ref{sec:Introduction}

\item BAL: Broad Absorption Line (Quasars), \S~\ref{sec:AGNprop}

\item BLRG: Broad-Line Radio Galaxies, \S~\ref{sec:AGNprop}

\item CDQ: Core-Dominated Quasars, \S~\ref{sec:AGNprop}

\item CSS: Compact-Steep Spectrum, \S~\ref{sec:content}

\item EMSS: ({\it Einstein} Observatory) Extended Medium Sensitivity Survey,
\S~\ref{sec:ext_low}

\item FR I: Fanaroff-Riley Type I, \S~\ref{sec:AGNprop}

\item FR II: Fanaroff-Riley Type II, \S~\ref{sec:AGNprop}

\item FSRQ: Flat Spectrum Radio Quasars, \S~\ref{sec:AGNprop}

\item FWHM: Full Width Half Maximum, \S~\ref{sec:narrow_low}

\item GPS: Gigahertz-Peaked Spectrum, \S~\ref{sec:content}

\item HBL: High (-Energy Cutoff) BL Lacs, \S~\ref{sec:terms}

\item HGLS: (EXOSAT) High Galactic Latitude Survey, \S~\ref{sec:bl_samples}

\item HPQ: Highly Polarized Quasars, \S~\ref{sec:AGNprop}

\item KS: Kolmogorov-Smirnov, \S~\ref{sec:ext_high}

\item LASS: ({\it HEAO}-1) Large Area Sky Survey, \S~\ref{sec:ext_low}

\item LBL: Low (-Energy Cutoff) BL Lacs, \S~\ref{sec:terms}

\item LF: Luminosity Function, \S~\ref{sec:effect}

\item NELG: Narrow-Emission-Line X-ray Galaxies, \S~\ref{sec:AGNprop}

\item NLRG: Narrow-Line Radio Galaxies, \S~\ref{sec:AGNprop}

\item OVV: Optically Violently Variable (Quasars), \S~\ref{sec:AGNprop}

\item RBL: Radio (Selected) BL Lacs, \S~\ref{sec:ext_low}

\item SSC: synchrotron-self Compton, \S~\ref{sec:gammaray}

\item SSRQ: Steep Spectrum Radio Quasars, \S~\ref{sec:AGNprop}

\item VLBA: Very Long Baseline Array, \S~\ref{sec:sup_fri}

\item VLBI: Very Long Baseline Interferometry, \S~\ref{sec:superluminal}

\item XBL: X-ray (Selected) BL Lacs, \S~\ref{sec:ext_low}

\end{itemize}

\newpage
\noindent
{\bf Figure Captions}
\vskip 0.01truecm\nobreak
\begin{figure}
\caption{A schematic diagram of the current paradigm for radio-loud AGN (not
to scale). Surrounding the central black hole is a luminous accretion
disk. Broad emission lines are produced in clouds orbiting above the disk and
perhaps by the disk itself. A thick dusty torus (or warped disk) obscures the
broad-line region from transverse lines of sight; some continuum and
broad-line emission can be scattered into those lines of sight by hot
electrons that pervade the region. A hot corona above the accretion disk may
also play a role in producing the hard X-ray continuum. Narrow lines are
produced in clouds much farther from the central source. Radio jets, shown
here as the diffuse jets characteristic of low-luminosity, or FR~I-type, radio
sources, emanate from the region near the black hole, initially at
relativistic speeds. For a $10^8$\sm~black hole,
the black hole radius is $\sim3\times10^{13}$~cm,
the accretion disk emits mostly from $\sim 1 - 30 \times 10^{14}$~cm,
the broad-line clouds are located within $\sim2 - 20 \times 10^{16}$~cm
of the black hole, and
the inner radius of the dusty torus is perhaps $\sim10^{17}$~cm.
The narrow-line region extends approximately from $10^{18} - 10^{20}$~cm,
and radio jets have been detected on scales from $10^{17}$
to several times $10^{24}$~cm, a factor of ten larger than the
largest galaxies.
\label{fig:cartoon}
}
\end{figure}

\begin{figure}
\caption{Radio images of the two types of radio galaxies: (a) at low
luminosity, an FR~I radio galaxy, 1231+674,
with diffuse, approximately symmetric jets whose surface brightness
falls off away from the center, and (b) at high luminosity,
an FR~II radio galaxy, 1232+414, with sharp-edged
lobes and bright hot spots; the jets in this case are often too faint to see.
(Courtesy of Frazer Owen and Mike Ledlow.)
\label{fig:rad_image}
}
\end{figure}

\newpage
\begin{figure}
\caption
{{\it HST} WFPC image of the nearby Seyfert 2 galaxy NGC~5728 taken in the
light of \OIII~with resolution 0.1 arcsec (Wilson et al. 1993). The bi-conical
structure suggests that a hidden nuclear source with quasar-like ultraviolet
luminosity is photoionizing gas in the narrow-line region. (Copyright
American Astronomical Society, reprinted with permission.)}
\label{fig:ion_cones}
\end{figure}

\begin{figure}
\caption{Multiwavelength spectra of the gamma-ray-bright
superluminal quasar 3C279 (Maraschi \ea 1994a) at two epochs,
a high state in June 1991 and a low state in January 1993. In the high state,
the gamma-ray luminosity is ten times the luminosity in the synchrotron
component seen at lower energies, while in the low state the two are
comparable.
(Copyright American Astronomical Society, reproduced with permission.)
\label{fig:3c279_spec}
}
\end{figure}

\begin{figure}
\caption{Gamma-ray light curve of the superluminal quasar 3C279 during
the bright outburst in June 1991 (Kniffen \ea 1993). The combination of
rapid variability and high gamma-ray luminosity strongly suggests the
emission is relativistically beamed.
(Copyright American Astronomical Society, reproduced with permission.)
\label{fig:3c279_lc}
}
\end{figure}

\clearpage
\begin{figure}
\caption{A comparison of the extended radio powers at 5 GHz for FR~II radio
galaxies ({\it dashed line}), steep-spectrum radio quasars
({\it dot-dashed line}), and flat-spectrum radio quasars ({\it
solid line}) from the 2~Jy sample (Wall and Peacock 1985; di Serego
Alighieri \ea 1994). Data for the FR~IIs come from Morganti \ea (1993), those
for FSRQ have been derived from the $R$ values (Appendix~C)
given by Padovani (1992b),
updated with lower limits from Ulvestad \ea (1981), while extended radio
powers for SSRQ have been collected from the literature (mostly from Browne
and Murphy 1987). Upper limits on extended power for FSRQ are indicated
with the symbol ``$<$''. Radio
powers have been de-evolved using the best-fit evolution for each
class (Table~2).
\label{fig:pext_high}
}
\end{figure}

\begin{figure}
\caption{The \OII~emission line luminosity for quasars ({\it filled circles})
and powerful (FR II) radio galaxies ({\it open squares}) in the 3CR sample,
plotted versus redshift to allow comparisons at similar radio luminosity
(which is well correlated with redshift in a radio-flux-limited sample)
and/or observed wavelength of \OII.
The narrow emission line luminosities
of radio galaxies and quasars, which should be unaffected by obscuration
or relativistic beaming, span the same values over the full
redshift range. (Data from Hes \ea 1993.)
\label{fig:oii}
}
\end{figure}

\begin{figure}
\caption{Optical image of the radio-loud quasar 2135$-$147 obtained with
the CFHT with resolution 0.5~arcsec. The host galaxy is clearly
visible, as are two close companions $\sim2$~arcsec and $\sim5$~arcsec
to the east (Hutchings and Neff 1992; copyright
American Astronomical Society, reproduced with permission).
}
\label{fig:qhost}
\end{figure}

\begin{figure}
\caption{
The distribution of radio galaxies and the host galaxies of radio-loud quasars
on the $K$-band $\mu_{1/2}$, log$R_{1/2}$ plane, where $\mu_{1/2}$ is
the $K$-band surface brightness at the half-light radius, $R_{1/2}$,
for samples matched statistically in redshift and radio power
to minimize selection effects. Both radio galaxies and
quasar hosts are very luminous and have large scale lengths;
the best fit correlation for the combined dataset (dotted line) is
very similar, in both slope and normalization, to those for
brightest cluster galaxies (solid lines; from Schneider, Gunn and Hoessel 1983,
shifted to the K-band by the observed mean zero-redshift color of the radio
galaxies in the sample, $B-K = 3.75$). The properties of
radio galaxies and quasar hosts on this diagram are statistically
indistinguishable, lending strong support to unified schemes, at least
for $z < 0.4$ (Taylor \ea 1995; figure courtesy of James Dunlop).}
\label{fig:host_prop}
\end{figure}

\begin{figure}
\caption{A comparison of the extended radio powers at 5 GHz for FR~I radio
galaxies ({\it dashed line}) from the 2 Jy sample (Wall and Peacock 1985;
di Serego Alighieri \ea 1994b), radio-selected BL~Lac objects ({\it solid
line}) from the 1~Jy sample (Stickel \ea 1991) and X-ray-selected BL~Lacs
({\it dot-dashed line}) from the EMSS sample (Stocke \ea 1991). Data for the
FR~Is come from Morganti \ea (1993); those for RBL have been derived from the
$R$ values (Appendix~C) given by Padovani (1992a), updated with data from
Kollgaard \ea (1992) and Murphy \ea (1993) plus lower limits from Ulvestad \ea
(1981); data for the XBL are taken from Perlman and Stocke
(1993). Upper limits on extended power ($<$) and lower limits on redshift
($>$) for BL~Lac objects are indicated.
\label{fig:pext_low}
}
\end{figure}

\begin{figure}
\caption{The \OIII~emission line luminosity for the
1~Jy BL~Lacs ({\it filled circles}), 2~Jy FR~Is ({\it open circles}),
and 2~Jy flat-spectrum radio quasars ({\it open squares}), including
some upper limits for the FR~Is, plotted versus redshift to allow
comparisons at similar radio luminosity (which is well correlated with redshift
in radio-flux-limited samples) and/or observed wavelength.
At a given redshift, the line
luminosities of FR~Is are generally smaller than those of BL~Lacs,
although the highest FR~I \OIII~luminosities are comparable.
The line luminosities of quasars are larger than those
of BL~Lacs by more than a factor of ten. (The one quasar located in the BL~Lac
region, PKS~0521$-$365, is discussed in the text;
\S~7.1.)
This argues against the idea that the high-redshift, high-luminosity
BL~Lac objects are more closely related to flat-spectrum radio quasars
than to low-redshift, low-luminosity BL~Lacs, although the BL~Lac data extend
only to $z \sim 0.4$.
\label{fig:oiii}
}
\end{figure}

\begin{figure}
\caption{
The central part of an NTT R-band image of the BL~Lac object PKS~0548$-$322
(center) and its environment (Falomo \ea 1995). The host galaxy has the
$r^{1/4}$ profile typical of an elliptical. Numerous other galaxies
visible in the field have the same redshift as the BL~Lac ($z=0.069$),
indicating this BL~Lac object lies in a rich cluster (Abell richness class
2; Falomo \ea 1995).
A number of these galaxies are distorted or otherwise unusual; for
example, the bright galaxy at the left edge of the image has a small jet-like
structure extending to the southeast.
(Copyright American Astronomical Society, reproduced with permission.)
\label{fig:bll_host}
}
\end{figure}

\begin{figure}
\caption{The effect of relativistic beaming on observed luminosity functions.
(a) Simple case where all emitted flux comes from a randomly oriented
relativistic jet. For a delta-function intrinsic luminosity
({\it thin dashed line}), the observed luminosity function after beaming
({\it thick dashed line}) is a flat power law of differential slope
$1 \simlt (p+1)/p \simlt 1.5$ (Eq.~4). A power-law intrinsic
luminosity function ({\it thin solid line}), after integration
of Eq.~(5), gives rise to an observed luminosity function that is
a double power law ({\it thick solid line}), with flat slope $(p+1)/p$ at low
luminosities and the same slope as the intrinsic power law at high
luminosities. (b) Case where the intrinsic luminosity in the jet is a fraction
$f=0.001$, 0.01, 0.1, or 1 of the unbeamed luminosity. As in (a), the beamed
luminosity function has slope $(p+1)/p$ below the break and the parent
luminosity function slope above the break. In both (a) and (b), the high
luminosity objects (those in the steep part of the beamed luminosity function)
are oriented close to the line of sight ($\theta \sim 0^\circ$).
\label{fig:fake_lf}
}
\end{figure}

\begin{figure}
\caption{Local differential radio luminosity functions for high-luminosity
radio sources: flat spectrum radio quasars ({\it filled circles}),
steep spectrum radio quasars ({\it open triangles}), and FR~II galaxies
({\it open squares}).
The error bars represent the sum in quadrature of the $1~\sigma$ Poisson errors
(Gehrels 1986) and the variations of the number density associated with a
$1~\sigma$ change in the evolutionary parameter $\tau$ (see Padovani and Urry
1992 for details).
The luminosity functions of FR~IIs and SSRQ have
similar slopes and extend to the same luminosity at the bright end,
although the FR~IIs extend a decade lower at the faint end. The FSRQ
luminosity function is distinctly flatter and extends to higher luminosities.
This agrees well with the predictions of a beaming model calculated
for FSRQ ({\it solid line}) and SSRQ ({\it dot-dashed line});
see \S~6.1.3 (Table~3) for details of the model parameters.
}
\label{fig:rlf_high}
\end{figure}

\begin{figure}
\caption{Multiwavelength spectra of a low-energy cutoff BL Lac (LBL)
(PKS~0537$-$441; {\it filled circles}) and a high-energy cutoff BL Lac (HBL)
(Mrk~421; {\it open squares}). The peak synchrotron luminosity occurs at
infrared/optical wavelengths for the LBL and at soft X-ray wavelengths for the
HBL. In this example, the HBL is both an XBL and an RBL, and (as in most cases)
the LBL is an RBL. A value of $q_0 = 0.5$ has been assumed. (Following Maraschi
\ea 1994b; figure courtesy of Rita Sambruna.)
}
\label{fig:multispec}
\end{figure}

\begin{figure}
\caption{The local differential X-ray luminosity functions of low-luminosity
radio-loud AGN. The observed LF of FR~I galaxies is
represented by a broken power law ({\it dashed line}).
The observed LFs for the EMSS XBL sample (no evolution:
{\it filled circles}; anti-evolution: {\it open squares}) are fitted by
beaming models for the non-evolving case ({\it solid line}) and
the anti-evolving case ({\it dash-dotted line});
data and evolution estimate from Wolter \ea (1994).
Error bars correspond to $1~\sigma$ Poisson errors (Gehrels 1986).
\label{fig:xlf_low}
}
\end{figure}

\begin{figure}
\caption{The local differential radio luminosity functions of
radio-selected BL~Lac objects ({\it filled circles}; Stickel \ea 1991)
and FR~I radio galaxies ({\it open squares}; \S~6.2.4),
compared to the fitted beaming model ({\it solid line}; first set of
RBL parameters in Table~3).
Error bars correspond to $1~\sigma$ Poisson errors (Gehrels 1986).
\label{fig:rlf_low}
}
\end{figure}

\begin{figure}
\caption{The ``approximate bolometric'' luminosity function for 1 Jy LBL ({\it
filled circles}) and EMSS HBL ({\it open squares}). This is the bivariate
peak luminosity function, where the peak luminosity has been defined as the
maximum in $\nu L_{\nu}$ from the data of Giommi \ea (1995) and no evolution
has been assumed. Error bars correspond to $1~\sigma$ Poisson errors (Gehrels
1986). The HBL appear to be more numerous in the range of overlapping
luminosity but selection effects inherent in the 1~Jy and EMSS samples
are already present in these luminosity functions.
\label{fig:bol_lf}
}
\end{figure}

\begin{figure}
\caption{The available \MGII~emission line luminosities for 1~Jy BL~Lac objects
({\it filled circles}) and 2~Jy flat-spectrum radio quasars
({\it open squares}),
plotted versus redshift so that comparisons at similar luminosity (which is
well
correlated with redshift in flux-limited samples) and/or observed wavelength
are
possible.
BL~Lacs and FSRQ line luminosities differ by more than one order of magnitude
up to $z \sim 0.8$, above which some of the high-redshift BL~Lacs have line
luminosities comparable to the lower-luminosity quasars.
\label{fig:mgii}
}
\end{figure}

\begin{figure}
\caption{The dependence of the Doppler factor on the angle to the line of
sight. Different curves correspond to different Lorentz factors: from the top
down, $\gamma = 15$, 10, 5, 2. The expanded scale on the inset shows
the angles for which $\delta=1$.
\label{fig:delta}
}
\end{figure}

\begin{figure}
\caption{The apparent velocity relative to the speed of light versus angle to
the line of sight for an emitter approaching at relativistic speed.
Different curves correspond to different Lorentz factors: from the top down,
$\gamma = 15$, 10, 5, 2. The dotted line corresponds to $\beta_{\rm a} = 1$.
Note that $\beta_{\rm a}$ is essentially independent of
$\gamma$ at large angles.
\label{fig:beta_app}
}
\end{figure}

\begin{figure}
\caption{The jet to counter-jet ratio, $J$, versus angle to the line of sight
for
$p=2$. Different curves correspond to different Lorentz factors: from the top
down, $\gamma = 15$, 10, 5, 2. Note that the ratio is essentially independent
of $\gamma$ at large angles.
\label{fig:jcj}
}
\end{figure}
\vspace{5in}


\begin{references}

\reference Abraham, R. G., Crawford, C. S., and McHardy, I. M. 1992, \apj,
	401, 474
\reference Abraham, R. G., Crawford, C. S., Merrifield, M. R., Hutchings, J.
	B., and McHardy, I. M. 1993, \apj, 415, 101
\reference Abraham, R. G., McHardy, I. M., and Crawford, C. S. 1991, \mnras,
	252, 482
\reference Abramowicz, M. A. 1992, in Extragalactic Radio Sources: From Beams
	to Jets, ed. J. Roland, H. Sol, and G. Pelletier (Cambridge, Cambridge
	Univ. Press), p. 206
\reference Angel, J. R. P., and Stockman, H. S. 1980, AR\aap, 8, 321
\reference Antonucci, R. 1984, \apj, 278, 299
\reference Antonucci, R. 1993, AR\aap, 31, 473
\reference Antonucci, R., and Barvainis, R. 1990, \apj, 363, L17
\reference Antonucci, R., Hurt, T., and Kinney, A. 1994, Nature, 371, 313
\reference Antonucci, R., and Miller, J. S. 1985, \apj, 297, 621
\reference Antonucci, R., and Ulvestad, J. S. 1985, \apj, 294, 158
\reference Arnaud, K. A., Johnstone, R. M., Fabian, A. C., Crawford, C. S.,
	Nulsen, P. E. J., Shafer, R. A., and Mushotzky, R. F. 1987, \mnras,
	227, 241
\reference Avni, Y., and Bahcall, J. N. 1980, \apj, 235, 694
\reference Axon, D. J., Pedlar, A., Unger, S. W., Meurs, E. J. A., and
	Whittle, D. M. 1989, Nature, 341, 631
\reference B\aa\aa th, L. B. 1984, in VLBI and Compact Radio Sources, ed. R.
	Fanti et al. (Dordrecht, Kluwer), p. 127
\reference Bade, N., Fink, H. H., and Engels, D. 1994, \aap, 286, 381
\reference Bahcall, J. N., et al. 1993, \apjs, 87, 1
\reference Bahcall, J. N., Kirhakos, S., and Schneider, D. P. 1995, \apj, in
	press
\reference Bailey, J., Sparks, W., Hough, J., and Axon, D. 1986, Nature, 322,
	150
\reference Barthel, P. D. 1989, \apj, 336, 606
\reference Barthel, P. D., and Miley, G. K. 1988, Nature, 333, 319
\reference Baum, S. A., and Heckman, T. M. 1989, \apj, 336, 681
\reference Baum, S. A., Zirbel, E. L., and O'Dea, C. P. 1995, \apj, 451
	(Sep 20), in press
\reference Begelman, M. C. 1993, in Jets in Extragalactic Radio Sources, ed.
	H.-J. R\"oser, K. Meisenheimer, R. A. Perley, and P. A. G. Scheuer
	(Berlin, Springer), p. 145
\reference Begelman, M. C., Blandford, R. D., and Rees, M. J. 1984, Rev. Mod.
	Phys., 56, 255
\reference Bicknell, G. V. 1994, \apj, 422, 542
\reference Bicknell, G. V., de Ruiter, H. R., Parma, P., Morganti, R., and
	Fanti, R. 1990, \apj, 354, 98
\reference Binette, L., Fosbury, R. A. E., and Parker, D. 1993, PASP, 105, 1150
\reference Biretta, J. A., Moore, R. L., and Cohen, M. H. 1986, \apj, 308, 93
\reference Biretta, J. A., Zhou, F., and Owen, F. N. 1995, \apj, in press
\reference Blandford, R. D. 1990, in Active Galactic Nuclei, ed. T. J.-L.
	Courvoisier, M. Mayor (Saas-Fee Advanced Course 20) (Berlin, Springer),
	p. 161
\reference Blandford, R. D., and Rees, M. J. 1978, in Pittsburgh Conference on
	BL~Lac Objects, ed. A. N. Wolfe (Pittsburgh, Univ. of Pitt. Press),
	p. 328
\reference Bodo, G., and Ghisellini, G. 1995, \apj, 441, L69
\reference Bridle, A. H. 1992, in Testing the AGN Paradigm, ed. S. S. Holt,
	S. G. Neff, and C. M. Urry (New York, AIP), p. 386
\reference Bridle, A. H., Hough, D. H., Lonsdale, C. J., Burns, J. O., and
	Laing, R. A. 1994, \aj, 108, 766
\reference Bridle, A. H., and Perley, R. A. 1984, AR\aap, 22, 319
\reference Browne, I. W. A. 1983, \mnras, 204, 23p
\reference Browne, I. W. A. 1989, in BL~Lac Objects, ed. L. Maraschi, T.
	Maccacaro, and M.-H. Ulrich (Berlin, Springer), p. 401
\reference Browne, I. W. A., and March\~a, M. J. M. 1993, \mnras, 261, 795
\reference Browne, I. W. A., and Murphy, D. W. 1987, \mnras, 226, 601
\reference Burbidge, G., and Hewitt, A. 1989, in BL~Lac Objects, ed. L.
	Maraschi, T. Maccacaro, and M.-H. Ulrich (Berlin, Springer), p. 412
\reference Carilli, C. L., Bartel, N., and Diamond, P. 1994, \aj, 108, 64
\reference Cawthorne, T. V. 1991, in Beams and Jets in Astrophysics, ed. P. A.
	Hughes (Cambridge, Cambridge Univ. Press), p. 187
\reference Celotti, A., Maraschi, L., Ghisellini, G., Caccianiga, A., and
	Maccacaro T. 1993, \apj, 416, 118
\reference Chambers, K. C., Miley, G. K., and van Breugel, W. 1987, Nature,
	329, 604
\reference Cimatti, A., di Serego Alighieri, S., Fosbury, R. A. E., Salvati,
	M., and Taylor, D. 1993, \mnras, 264, 421
\reference Della Ceca, R. 1993, Ph.D. Thesis, University of Bologna
\reference Della Ceca, R., Zamorani, G., Maccacaro, T., Wolter, A., Griffiths,
	R., Stocke, J., and Setti, G. 1994, \apj, 430, 533
\reference de Ruiter, H. R., Parma, P., Fanti, C., and Fanti, R. 1990, \aap,
	227, 351
\reference De Young, D. S. 1993, \apj, 405, L13
\reference di Serego Alighieri, S., Binette, L., Courvoisier, T. J.-L.,
	Fosbury, R. A. E., and Tadhunter, C. N. 1988, Nature, 334, 591
\reference di Serego Alighieri, S., Cimatti, A., and Fosbury, R. A. E. 1994a,
	\apj, 431, 123
\reference di Serego Alighieri, S., Danziger, J., Morganti, R., and Tadhunter,
	C. 1994b, \mnras, 269, 998
\reference di Serego Alighieri, S., Fosbury, R. A. E., Quinn, P. J., and
	Tadhunter, C. N. 1989, Nature, 341, 307
\reference Djorgovski, S., Weir, N., Matthews, K., and Graham, J. R. 1991,
	\apj, 372, L67
\reference Dondi, L., and Ghisellini, G. 1995, \mnras, 273, 583
\reference Dunlop, J. S., and Peacock, J. A. 1990, \mnras, 247, 19
\reference Dunlop, J. S., Taylor, G. L., Hughes, D. H., and Robson, E. I.
	1993, \mnras, 264, 455
\reference Economou, F., Lawrence, A., Ward, M. J., and Blanco, P. R. 1995,
	\mnras, 272, L5
\reference Edelson, R. A., et al. 1995, \apj, 438, 120
\reference Elvis, M., Plummer, D., Schachter, J., and Fabbiano, G. 1992, \apjs,
	80, 257
\reference Evans, I. N., Ford, H. C., Kinney, A. L., Antonucci, R. R. J.,
	Armus, L., and Caganoff, S. 1991, \apj, 369, L27
\reference Fabbiano, G., Miller, L., Trinchieri, G., Longair, M. S., and Elvis,
	M. 1984, \apj, 277, 115
\reference Fabbiano, G., Willner, S. P., Carleton, N. P., and Elvis, M. 1986,
	\apj, 304, L37
\reference Falcke, H., Gopal-Krishna, and Biermann, P. L. 1995, \aap, in press
\reference Falomo, R., Melnick, J., and Tanzi, E. G. 1992, \aap, 255, L17
\reference Falomo, R., Pesce, J. E., and Treves, A. 1993, \apj, 411, L63
\reference Falomo, R., Pesce, J. E., and Treves, A., 1995, \apj, 438, L9
\reference Fanaroff, B. L., and Riley, J. M. 1974, \mnras, 167, 31p
\reference Fanti, C., Fanti, R., de Ruiter, H. R., and Parma, P. 1987, \aaps,
	69, 57
\reference Fanti, C., Fanti, R., O'Dea, C. P., and Schilizzi, R. T. (eds.)
	1990a, Compact Steep-Spectrum and GHz-Peaked Radio Sources (Bologna,
	Istituto di Radioastronomia-CNR)
\reference Fanti, R., Fanti, C., Schilizzi, R. T., Spencer, R. E., Nan Rendong,
	Parma, P., van Breugel, W. J. M., and Venturi, T. 1990b, \aap, 231, 333
\reference Feretti, L., Comoretto, G., Giovannini, G., Venturi, T., and
	Wehrle, A. E. 1993, \apj, 408, 446
\reference Feretti, L., Giovannini, G., Gregorini, L., Parma, P., and Zamorani,
	G. 1984, \aap, 139, 55
\reference Fleming, T. A., Green, R. F., Jannuzi, B. T., Liebert, J., Smith,
	P. S., and Fink, H. 1993, \aj, 106, 1729
\reference Franceschini, A., Danese, L., De Zotti, G., and Toffolatti, L. 1988,
	\mnras, 233, 157
\reference Fried, J. W., Stickel, M., and K\"uhr, H. 1993, \aap, 268, 53
\reference Fugmann, W. 1989, \aap, 205, 86
\reference Gabuzda, D. C., Cawthorne, T. V., Roberts, D. H., and Wardle, J. F.
	C. 1992, \apj, 388, 40
\reference Gabuzda, D. C., Kollgaard, R. I., Roberts, D. H., and Wardle, J. F.
	C. 1993, \apj, 410, 39
\reference Gabuzda, D. C., Mullan, C. M., Cawthorne, T. V., Wardle, J. F. C.,
	and Roberts, D. H. 1994, \apj, 435, 140
\reference Gallimore, J. F., Baum, S. A., O'Dea, C. P., Brinks, E., and Pedlar,
	A. 1995, in preparation
\reference Garrington, S. T., and Conway, R. G. 1991, \mnras, 250, 198
\reference Gehrels, N. 1986, \apj, 303, 336
\reference Ghisellini, G. 1987, Ph.D. Thesis, SISSA, Trieste
\reference Ghisellini, G., and Maraschi, L. 1989, \apj, 340, 181
\reference Ghisellini, G., Padovani, P., Celotti, A., and Maraschi, L. 1993,
	\apj, 407, 65
\reference Giommi, P., Ansari, S. G., and Micol, A. 1995, \aaps, 109, 267
\reference Giommi, P., \ea 1991, \apj, 378, 77
\reference Giommi, P., and Padovani, P. 1994, \mnras, 268, L51
\reference Giovannini, G., Feretti, L., Gregorini, L., and Parma, P. 1988,
\aap,
	199, 73
\reference Giovannini, G., Feretti, L., Venturi, T., Lara, L., Marcaide,
	J., Rioja, M., Spangler, S. R., and Wehrle, A. E. 1994, \apj, 435, 116
\reference Goodrich, R. W., and Cohen, M. H. 1992, \apj, 391, 623
\reference Gopal-Krishna, and Kulkarni, V. K. 1992, \aap, 257, 11
\reference Gopal-Krishna, Kulkarni, V. K., and Mangalam, A. V. 1994, \mnras,
268
\reference Gopal-Krishna, and Subramanian, K. 1991, Nature, 349, 766
\reference Gopal-Krishna, and Wiita, P. J. 1993, Nature, 363, 142
\reference Grandi, S. A., and Osterbrock, D. A. 1978, \apj, 220, 783
\reference Green, R. F., Schmidt, M., and Liebert, J. 1986, \apjs, 61, 305
\reference Halpern, J. P., Impey, C. D., Bothun, G. D., Tapia, S., Skillman, E.
	D., Wilson, A. S., and Meurs, E. J. 1986, \apj, 302, 711
\reference Harms, R. J., \ea 1994, \apj, 435, L35
\reference Hawkins, M. R. S., V\'eron, P., Hunstead, R. W., and Burgess, A. M.
	1991, \aap, 248, 421
\reference Heckman, T. M., O'Dea, C. P., Baum, S. A., and Laurikainen, E. 1994,
	\apj, 428, 65
\reference Hes, R., Barthel, P. D., and Fosbury, R. A. E. 1993, Nature, 362,
326
\reference
\reference Hill, G. J., Goodrich, R. W., and DePoy, D. L. 1995, \apj, submitted
\reference Hill, G. J., and Lilly, S. J. 1991, \apj, 367, 1
\reference Hine, R. G., and Longair, M. S. 1979, \mnras, 188, 111
\reference Hjellming, R. M., and Rupen, M. P. 1995, preprint
\reference Holt, S. S., Neff, S. G., and Urry, C. M. (eds.) 1992, Testing the
	AGN Paradigm (New York, AIP)
\reference Hough, J. H., Brindle, C. B., Axon, D. J., Bailey, J., and Sparks,
W.
	B. 1987, \mnras, 224, 1013
\reference Hough, J. H., Brindle, C. B., Wills, B. J., Wills, D., and Bailey,
J.
	1991, \apj, 372, 478
\reference Hough, D. H., and Readhead, A. C. S. 1989, \aj, 98, 1208
\reference Hoyle, F. R. S., Burbidge, G. R., and Sargent, W. L. W. 1966,
	Nature, 209, 751
\reference Hunter, S. D., \ea 1993, \apj, 409, 134
\reference Hutchings, J. B. 1987, \apj, 320, 122
\reference Hutchings, J. B., Holtzman, J., Sparks, W. B., Morris, S. C.,
	Hanisch, R. J., and Mo, J. 1994, \apj, 429, L1
\reference Hutchings, J. B., and Morris, S. L. 1995, preprint
\reference Hutchings, J. B., and Neff, S. G. 1992, \aj, 104, 1
\reference Hutchings, J. B., Price, R., and Gower, A. C. 1988, \apj, 329, 122
\reference Impey, C. D., Lawrence, C. R., and Tapia, S. 1991, \apj, 375, 46
\reference Inglis, M. D., Brindle, C., Hough, J. H., Young, S., Axon, D. J.,
	Bailey, J. A., and Ward, M. J. 1993, \mnras, 263, 895
\reference Isobe, T., Feigelson, E. D., Akritas, M. G., and Babu, G. J. 1990,
	\apj, 364, 104
\reference Jackson, N., and Browne, I. W. A. 1990, Nature, 343, 43
\reference Jackson, N., and Tadhunter, C. N. 1993, \aap, 272, 105
\reference Jannuzi, B. T., Smith, P. S., and Elston, R. 1994, \apj, 428, 130
\reference Jones, P. A., McAdam, W. B., and Reynolds, J. E. 1994,
	\mnras, 268, 602
\reference Jones, T. W., O'Dell, S. L., and Stein, W. A. 1974, \apj, 188, 353
\reference Kapahi, V. K. 1987, in Observational Cosmology, ed. A. Hewitt,
	G. Burbidge, and L. Z. Fang (Dordrecht, Reidel), p. 251
\reference Kapahi, V. K. 1989, \aj, 97, 1
\reference Kapahi, V. K. 1990, in Parsec Scale Radio Jets, ed.
	J.A. Zensus and T.J. Pearson (Cambridge, Cambridge Univ. Press), p. 304
\reference Kapahi, V. K., Athreya, R. M., Subrahmanya, C. R., McCarthy, P. J.,
	van Breugel, W., Hunstead, R. W., and Baker, J. C. 1994, in Astronomy
	Posters of the XXII IAU, ed. H. van Woerden (Sliedrecht, Twin Press),
	p. 195
\reference Kellermann, K. I., and Pauliny-Toth, I. I. K. 1969, \apj, 155, L71
\reference Kellermann, K. I., Sramek, R., Schmidt, M., Shaffer, D. B., and
	Green, R. 1989, \aj, 98, 1195
\reference Kinney, A. L. 1994, in The Physics of Active Galaxies, ed. G. V.
	Bicknell, M. A. Dopita, and P. J. Quinn, ASP Conf. Series, 54, p. 61
\reference Kniffen, D. A., \ea 1993, \apj, 411, 133
\reference Kollgaard, R. I. 1994, Vistas in Astronomy, 38, 29
\reference Kollgaard, R. I., Wardle, J. F. C., and Roberts, D. H. 1990, \aj,
	100, 1057
\reference Kollgaard, R. I., Wardle, J. F. C., Roberts, D. H., and Gabuzda, D.
	C. 1992, \aj, 104, 1687
\reference Koratkar, A., Kinney, A. L., and Bohlin, R. C. 1992, \apj, 400, 435
\reference Kristian, J., Sandage, A. R., and Westphal, J. A. 1978, \apj, 221,
	383
\reference Kurfess, J. D., Johnson, W. N., and McNaron-Brown, K. 1994, in The
	Gamma Ray Sky Seen with CGRO and SIGMA, in press
\reference La Franca, F., Gregorini, L., Cristiani, S., de Ruiter, H.,
	and Owen, F. 1994, AJ, 108, 1548
\reference Laing, R. A. 1988, Nature, 331, 149
\reference Laing, R. A. 1994, in The Physics of Active Galaxies, ed. G. V.
	Bicknell, M. A. Dopita, and P. J. Quinn, ASP Conf. Series, 54, p. 227
\reference Laing, R. A., Jenkins, C. R., Wall, J. V., and Unger, S. W. 1994, in
	The Physics of Active Galaxies, ed. G. V. Bicknell, M. A. Dopita,
	and P. J. Quinn, ASP Conf. Series, 54, p. 201
\reference Laing, R. A., Jenkins, C. R., Wall, J. V., and Unger, S. W. 1995,
	in preparation
\reference Laing, R. A., Riley, J. M., and Longair, M. S. 1983, \mnras, 204,
151
\reference Laurent-Muehleisen, S. A., Kollgaard, R. I., Moellenbrock, G. A.,
and
	Feigelson, E. D. 1993, \aj, 106, 875
\reference Lawrence, A. 1987, PASP, 99, 309
\reference Lawrence, A. 1991, \mnras, 252, 586
\reference Lawrence, A. 1993, in The Nearest Active Galaxies, ed. J. Beckman,
	L. Colina, and H. Netzer (Consejo Superior de Investigationes
	Cientificas), p. 3
\reference Lawrence, A., and Elvis, M. 1982, \apj, 256, 410
\reference Ledden, J. E., and O'Dell, S. L. 1985, \apj, 298, 630
\reference Lehnert, M. D., Heckman, T. M., Chambers, K. C., and Miley, G. K.
	1992, \apj, 393, 68
\reference Lind, K. R., and Blandford, R. D. 1985, \apj, 295, 358
\reference Lister, M., Hutchings, J. B., and Gower, A. C. 1994a, \apj, 427, 125
\reference Lister, M., Gower, A. C., and Hutchings, J. B. 1994b, \aj, 108, 821
\reference Madau, P. 1988, \apj, 327, 116
\reference Madau, P., Ghisellini, G., and Persic, M. 1987, \mnras, 224, 257
\reference Madejski, G. M., and Schwartz, D. A. 1983, \apj, 275, 467
\reference Maraschi, L., \ea 1994a, \apj, 435, L91
\reference Maraschi, L., Fossati, G., Tagliaferri, G., and Treves, A. 1995,
	\apj, 443, 578
\reference Maraschi, L., Ghisellini, G., and Boccasile, A. 1994b, in
	The Nature of Compact Objects in AGN, ed. A. Robinson and R. J.
	Terlevich (Cambridge, Cambridge Univ. Press), p. 381
\reference Maraschi, L., Ghisellini, G., and Celotti, A. 1992, \apj, 397, L5
\reference Maraschi, L., Ghisellini, G., Tanzi, E., and Treves A. 1986, \apj,
	310, 325
\reference Maraschi, L., and Rovetti, F. 1994, \apj, 436, 79
\reference Marscher, A. P., Marshall, F. E., Mushotzky, R. F., Dent, W. A.,
	Balonek, T. J., and Hartman, M. F. 1979, \apj, 233, 498
\reference McCarthy, P. J. 1989, Ph.D. Thesis, Univ. Cal. Berkeley
\reference McCarthy, P. J., van Breugel, and Kapahi, V. K. 1991, \apj, 371, 478
\reference McCarthy, P. J., Spinrad, H., van Breugel, W., Liebert, J.,
	Dickinson, M., Djorgovski, S., and Eisenhardt, P. 1990, \apj, 365, 487
\reference McCarthy, P. J., van Breugel, W., Spinrad, H., and Djorgovski, S.
	1987, \apj, 321, L29
\reference McHardy, I. M., Merrifield, M. R., Abraham, R. G., and Crawford, C.
	S. 1994, \mnras, 268, 681
\reference Merrifield, M. R. 1992, \aj, 104, 1306
\reference Miller, J. S., Goodrich, R. W., and Mathews, W. G. 1991, \apj, 378,
	47
\reference Mirabel, I. F., and Rodr\'\i guez, L. F. 1994, Nature, 371, 46
\reference Miyoshi, M., Moran, J., Herrnstein, J., Greenhill, L., Nakai, N.,
	Diamond, P., and Inoue, M. 1995, Nature, 373, 127
\reference Morganti, R., Fosbury, R. A. E., Hook, R. N., Robinson, A., and
	Tsvetanov, Z. 1992, \mnras, 256, 1p
\reference Morganti, R., Killeen, N. E. B., and Tadhunter, C. N. 1993, \mnras,
	263, 1023
\reference Morganti, R., Oosterloo, T. A., Fosbury, R. A. E., and Tadhunter, C.
	N. 1995, \mnras, in press
\reference Morganti, R., Robinson, A., Fosbury, R. A. E., di Serego Alighieri,
	S., Tadhunter, C. N., and Malin, D. F. 1991, \mnras, 249, 91
\reference Morris, S. L., Stocke, J. T., Gioia, I. M., Schild, R. E., Wolter,
	A., Maccacaro, T., and Della Ceca, R. 1991, \apj, 380, 49
\reference Mulchaey, J. S., Mushotzky, R. F., and Weaver, K. A. 1992,
	\apj, 390, L69
\reference Murphy, D. W., Browne, I. W. A., and Perley, R. A. 1993, \mnras,
264,
	298
\reference Mushotzky, R. F., Serlemitsos, P. J., Boldt, E. A., Holt, S. S.,
	and Becker, R. H. 1978, \apj, 220, 790
\reference Mushotzky, R. F. 1982, \apj, 256, 92
\reference Netzer, H., and Laor, A. 1994, \apj, 404, L51
\reference Nilsson, K., Valtonen, M. J., Kotilainen, J., and Jaakkola, T. 1993,
	\apj, 413, 453
\reference O'Dea, C. P., Baum, S. A., and Stanghellini, C. 1991, \apj, 380, 660
\reference Onuora, L. I. 1989, ApSpSci, 162, 343
\reference Onuora, L. I. 1991, \apj, 377, 36
\reference Orr, M. J. W., and Browne, I. W. A. 1982, \mnras, 200, 1067
\reference Osterbrock, D. E. 1989, Astrophysics of Gaseous Nebulae and Active
	Galactic Nuclei (Mill Valley, University Science Books)
\reference Ostriker, J. P., and Vietri, M. 1985, Nature, 318, 446
\reference Ostriker, J. P., and Vietri, M. 1990, Nature, 344, 45
\reference Owen, F. N., and Laing, R. A. 1989, \mnras, 238, 357
\reference Owen, F. N., and Ledlow, M. J. 1994, in The Physics of Active
	Galaxies, ed. G. V. Bicknell, M. A. Dopita, and P. J. Quinn, ASP Conf.
	Series, 54, p. 319
\reference Owen, F. N., Ledlow, M. J., and Keel, W. C. 1995, \apj, in press
\reference Owen, F. N., and White, R. A. 1991, \mnras, 249, 164
\reference Padovani, P. 1992a, \aap, 256, 399
\reference Padovani, P. 1992b, \mnras, 257, 404
\reference Padovani, P. 1993, \mnras, 263, 461
\reference Padovani, P., Ghisellini, G., Fabian, A. C., and Celotti, A. 1993,
	260, L21
\reference Padovani, P., and Giommi, P. 1995, \apj, 444, 567
\reference Padovani, P., and Urry, C. M. 1990, \apj, 356, 75
\reference Padovani, P., and Urry, C. M. 1991, \apj, 368, 373
\reference Padovani, P., and Urry, C. M. 1992, \apj, 387, 449
\reference Parma, P., et al. 1992, in Astrophysical Jets, Poster Papers from
	the Space Telescope Science Institute Symposium, ed. Denis Burgarella,
	Mario Livio, and Chris O'Dea (Baltimore, Space Telescope Science
	Institute), p. 30
\reference Parma, P., Fanti, C., Fanti, R., Morganti, R., and de Ruiter, H. R.
	1987, \aap, 181, 244
\reference Peacock, J. A. 1987, in Astrophysical Jets and
	Their Engines, ed. W. Kundt (Dordrecht, Reidel), p. 185
\reference Pearson, T. J., and Readhead, A. C. S. 1988, \apj, 328, 114
\reference P\'erez-Fournon, I., and Biermann, P. 1984, \aap, 130, L13
\reference Perlman, E. S., and Stocke, J. T. 1993, \apj, 406, 430
\reference Perlman, E. S., Stocke, J. T., Shaffer, D. B., Carilli, C. L., and
	Ma, C. 1994, \apj, 424, L69
\reference Perlman, E. S., Stocke, J. T., Schachter, J. F., Elvis, M.,
	Ellingson, E., Urry, C. M., Impey, C. D., Smith, P. S., and
	Kolchinsky, P. 1995, \apj, submitted
\reference Pesce, J. E., Falomo, R., and Treves, A. 1994, \aj, 107, 494
\reference Phillips, M. M., Jenkins, C. R., Dopita, M. A., Sadler, E. M., and
	Binette, L. 1986, \aj, 91, 1062
\reference Piccinotti, G., Mushotzky, R. F., Boldt, E. A., Holt, S. S.,
	Marshall, F. E., Serlemitsos, P. J., and Shafer, R. A. 1982,
	\apj, 253, 485
\reference Pier, E. A., and Krolik, J. H. 1992, \apj, 399, L23
\reference Pier, E. A., and Krolik, J. H. 1993, \apj, 418, 673
\reference Pierce, M. J., and Stockton, A. 1986, \apj, 305, 204
\reference Pogge, R. 1988, \apj, 328, 519
\reference Pogge, R. W., and De Robertis, M. M. 1993, \apj, 404, 563
\reference Polatidis, A. G., Wilkinson, P. N., Xu, W., Readhead, A. C. S.,
	Pearson, T. J., Taylor, G. B., and Vermeulen, R. C. 1995, \apjs, 98, 1
\reference Prestage, R. M., and Peacock, J. A. 1988, \mnras, 230, 131
\reference Quirrenbach, A., Witzel, A., Krichbaum, T. P., Hummel, C. A.,
	Wegner, R., Schalinski, C., Ott, M., Alberdi, A., and Rioja, M.
	1992, \aap, 258, 279
\reference Rawlings, S., and Saunders, R. 1991, Nature, 349, 138
\reference Rawlings, S., Saunders, R., Eales, S. A., and Mackay, C. D. 1989,
	\mnras, 240, 701
\reference Readhead, A. C. S. 1994, \apj, 426, 51
\reference Rees, M. J. 1966, Nature, 211, 468
\reference Robinson, A., Binette, L., Fosbury, R. A. E., and Tadhunter, C. N.
	1987, \mnras, 227, 97
\reference Romanishin, W. 1992, \apj, 401, L65
\reference Rowan-Robinson, M. 1968, \mnras, 138, 445
\reference Rowan-Robinson, M. 1977, \apj, 213, 638
\reference Rybicki, G., and Lightman, A. 1979, Radiation Processes in
	Astrophysics (New York, Wiley)
\reference Saikia, D. J., and Kulkarni, V. K. 1994, \mnras, 270, 897
\reference Sambruna, R. 1994, Ph.D. Thesis, SISSA, Trieste
\reference Sambruna, R. \ea 1995, in preparation
\reference Sanders, D. B., Phinney, E. S., Neugebauer, G., Soifer, B. T., and
	Matthews, K. 1989, \apj, 347, 29
\reference Sanders, D. B., Soifer, B. T., Elias, J. H., Madore, B. F.,
	Matthews, K., Neugebauer, G., and Scoville, N. Z. 1988, \apj, 325, 74
\reference Sarazin, C. L., and Wise, M. W. 1993, \apj, 411, 55
\reference Scarrott, S. M., Rolph, C. D., and Tadhunter, C. N. 1990, \mnras,
	243, 5p
\reference Schachter, J. F., et al. 1993, \apj, 412, 541
\reference Scheuer, P. A. G. 1987, in Superluminal Radio Sources, ed. J. A.
	Zensus and T. J. Pearson (Cambridge, Cambridge Univ. Press), p. 104
\reference Scheuer, P. A. G., and Readhead, A. C. S. 1979, Nature, 277, 182
\reference Schmidt, M. 1968, \apj, 151, 393
\reference Schneider, D. P., Gunn, J. E., and Hoessel, J. G. 1983, \apj,
	268, 476
\reference Schwartz, D. A., Brissenden, R. J. V., Tuohy, I. R., Feigelson, E.
	D., Hertz, P. L., and Remillard, R. A. 1989, in BL~Lac Objects, ed. L.
	Maraschi, T. Maccacaro, and M.-H. Ulrich (Berlin, Springer), p. 64
\reference Schwartz, D. A., and Ku, W. H.-M. 1983, \apj, 266, 459
\reference Singal, A. K. 1988, \mnras, 233, 87
\reference Singal, A. K. 1993a, \mnras, 263, 139
\reference Singal, A. K. 1993b, \mnras, 262, L27
\reference Singal, A. K., and Gopal-Krishna 1985, \mnras, 215, 383
\reference Smith, E. P., Heckman, T. M., Bothun, G. D., Romanishin, W., and
	Balick, B. 1986, \apj, 306, 64
\reference Smith, E. P., and Heckman, T. M. 1989, \apj, 341, 658
\reference Smith, E. P., and Heckman, T. M. 1990, \apj, 348, 38
\reference Smith, E. P., O'Dea, C. P., and Baum, S. A. 1995, \apj, 441, 113
\reference Sparks, W. B., Fraix-Burnet, D., and Macchetto, F. 1992, Nature,
355,
	804
\reference Spinrad, H., Djorgovski, S., Marr, J., and Aguilar, L. 1985, PASP,
	97, 932
\reference Stiavelli, M., Biretta, J., M\o ller, P., and Zeilinger, W. W. 1992,
	Nature, 355, 802
\reference Stickel, M., Fried, J. W., and K\"uhr, H. 1988a, \aap, 198, L13
\reference Stickel, M., Fried, J. W., and K\"uhr, H. 1988b, \aap, 206, L30
\reference Stickel, M., Fried, J. W., and K\"uhr, H. 1993, \aaps, 98, 393
\reference Stickel, M., and K\"uhr, H. 1993, \aaps, 97, 483
\reference Stickel, M., and K\"uhr, H. 1994, \aaps, 103, 349
\reference Stickel, M., Meisenheimer, K., and K\"uhr H. 1994, \aaps, 105, 211
\reference Stickel, M., Padovani, P., Urry, C. M., Fried, J. W., and K\"uhr, H.
	1991, \apj, 374, 431
\reference Stocke, J. T., Liebert, J., Schmidt, G., Gioia, I. M., Maccacaro,
T.,
	Schild, R. E., Maccagni, D., and Arp, H. C. 1985, \apj, 298, 619
\reference Stocke, J. T., Morris, S. L., Gioia, I. M., Maccacaro, T., Schild,
	R., Wolter, A., Fleming, T. A., and Henry, J. P. 1991, \apjs, 76, 813
\reference Stocke, J. T., Wurtz, R., and Perlman, E. S. 1995, \apj, submitted
\reference Stocke, J. T., Wurtz, R., Wanf, Q., Elston, R., and Jannuzi, B. T.
	1992, \apj, 400, L17
\reference Sutherland, R. S., Bicknell, G. V., and Dopita, M. A. 1993,
	\apj, 414, 510
\reference Tadhunter, C. N., Fosbury, R. A. E., and di Serego Alighieri, S.
	1989, in BL~Lac Objects, ed. L. Maraschi, T. Maccacaro, and M.-H. Ulrich
	(Berlin, Springer), p. 79
\reference Tadhunter, C. N., Morganti, R., di Serego Alighieri, S., Fosbury, R.
	A. E., and Danziger, I. J. 1993, \mnras, 263, 999
\reference Tadhunter, C. N., Scarrott, S. M., and Rolph, C. D. 1990, \mnras,
	246, 163
\reference Tadhunter, C. N., and Tsvetanov, Z. 1989, Nature, 341, 422
\reference Taylor, G. B., Vermeulen, R. C., Pearson, T. J., Readhead, A. C. S.,
	Henstock, D. R., Browne, I. W. A., and Wilkinson, P. N.
	1994, \apjs, 95, 345
\reference Taylor, G. L., Dunlop, J. S., and Hughes, D. H. 1995, \mnras,
	in preparation
\reference Ter\"asranta, H., and Valtaoja, E. 1994, \aap, 283, 51
\reference Thakkar, D. D., Xu, W., Readhead, A. C. S., Pearson, T. J., Taylor,
	G. B., Vermeulen, R. C., Polatidis, A. G., and Wilkinson, P. N.
	1995, \apjs, 98, 33
\reference Tingay, S., \ea 1995, Nature, 374, 141
\reference Tran, H. D. 1995, \apj, 440, 597
\reference Tribble, P. C. 1992, \mnras, 256, 281
\reference Turnshek, D. A. 1984, \apj, 280, 51
\reference Turnshek, D. A. 1988, in QSO Absorption Lines: Probing the Universe,
	ed. J. C. Blades, D. A. Turnshek, and C. A. Norman (Cambridge,
	Cambridge Univ. Press), p. 17
\reference Ueno, S., Koyama, K., Nishida, M., Yamauchi, S., and Ward, M. J.
	1994, \apj, 431, L1
\reference Ulrich, M.-H. 1981, \aap, 103, L1
\reference Ulrich, M.-H. 1989, in BL~Lac Objects, ed. L. Maraschi, T.
Maccacaro,
	and M.-H. Ulrich (Berlin, Springer), p. 45
\reference Ulvestad, J., Johnston, K., Perley, R., and Fomalont, E. 1981, \aj,
	86, 1010
\reference Urry, C. M. 1984, Ph.D. Thesis, The Johns Hopkins University
\reference Urry, C. M., et al. 1993, \apj, 411, 614
\reference Urry, C. M., et al. 1995, \apj, submitted
\reference Urry, C. M., Marziani, P., and Calvani, M. 1991b, \apj, 371, 510
\reference Urry, C. M., and Padovani, P. 1991, \apj, 371, 60
\reference Urry, C. M., Padovani, P., and Stickel, M. 1991a, \apj, 382, 501
\reference Urry, C. M., and Shafer, R. A. 1984, \apj, 280, 569
\reference Vagnetti, F., Giallongo, E., and Cavaliere, A. 1991, \apj, 368, 366
\reference Vagnetti, F., and Spera, R. 1994, \apj, 436, 611
\reference Valtaoja, E., Ter\"asranta, H., Urpo, S., Nesterov, N. S., Lainela,
	M., and Valtonen, M. 1992, \aap, 254, 80
\reference Venturi, T., Giovannini, G., Feretti, L., Comoretto, G., and
	Wehrle, A. E. 1993, \apj, 408, 81
\reference Vermeulen, R. C., and Cohen, M. H. 1994, \apj, 430, 467
\reference V\'eron-Cetty, M.-P., and Woltjer, L. 1990, \aap, 236, 69
\reference V\'eron-Cetty, M.-P., and V\'eron, P. 1993, A Catalogue of Quasars
	and Active Nuclei, ESO Scientific Report n. 13, 1993
\reference von Montigny, C., et al. 1995, \apj, 440, 525
\reference Wall, J. V., and Peacock, J. A. 1985, \mnras, 216, 173
\reference Ward, M. J., Blanco, P. R., Wilson, A. S., and Nishida, M. 1991,
	\apj, 382, 115
\reference Wardle, J. F. C., Moore, R. L., and Angel, J. R. P. 1984, \apj, 279,
	93
	1991, \apj, 373, 23
\reference Wilkes, B. J. 1986, \mnras, 218, 331
\reference Wills, B. J., Wills, D., Breger, M., Antonucci, R., and Barvainis,
	R. 1992a, \apj, 398, 454
\reference Wills, B. J., Wills, D., Evans, N. J., Natta, A., Thompson, K. L.,
	Breger, M., and Sitko, M. L. 1992b, \apj, 400, 96
\reference Wilson, A. S., Braatz, J., Heckman, T. M., Krolik, J. H., and
	Miley, G. K. 1993, \apj, 419, L61
\reference Wilson, A. S., and Colbert, E. J. M. 1995, \apj, 438, 62
\reference Wolter, A., Caccianiga, A., Della Ceca, R., and Maccacaro, T. 1994,
	\apj, 433, 29
\reference Worrall, D. M., and Birkinshaw, M. 1994, \apj, 427, 134
\reference Worrall, D. M., and Wilkes, B. J. 1990, \apj, 360, 396
\reference Wurtz, R., Ellingson, E., Stocke, J. T., and Yee, H. K. C. 1993,
\aj,
	106, 869
\reference Yanny, B., York, D. G., and Gallagher, J. S. 1989, \apj, 338, 735
\reference Yates, M. G., Miller, L., and Peacock, J. A. 1989, \mnras, 240, 129
\reference Yee, H. K. C., and Ellingson, E. 1993, \apj, 411, 43
\reference Zirbel, E. L., and Baum, S. A. 1995, \apj, in press

\end{references}
\end{document}